\documentclass[aps,prd,twocolumn,amssymb,showpacs,superscriptaddress,nofootinbib,longbibliography,floatfix]{revtex4-1}

\usepackage{amsfonts}
\usepackage{latexsym}
\usepackage{yfonts}
\usepackage{amsmath}
\usepackage{amssymb}
\usepackage{amsthm}
\usepackage{mathrsfs}
\usepackage{upgreek}
\usepackage{bm}
\usepackage{dcolumn}
\usepackage{epsfig}
\usepackage{graphicx}
\usepackage{latexsym}
\usepackage{rotating}
\usepackage{color}
\usepackage{soul}
\usepackage[english]{babel}
\soulregister\cite7
\soulregister\ref7
\soulregister\pageref7

\usepackage{lmodern}
\usepackage[T1]{fontenc}
\usepackage[utf8]{inputenc}

\usepackage{times}
\usepackage{setspace}

\usepackage{color}
\usepackage[unicode=true,pdfusetitle,
 bookmarks=true,bookmarksnumbered=false,bookmarksopen=false,
 breaklinks=false,pdfborder={0 0 0},backref=false,colorlinks=true]
 {hyperref}
\hypersetup{linkcolor=[rgb]{0.1,0.4,0.1}, linktoc=page}
\hypersetup{citecolor=[rgb]{0.6,0,0.1}}
\hypersetup{urlcolor=[rgb]{0.1,0,0.6}}

\DeclareTextFontCommand{\emph}{\sl}

\usepackage{setspace}
\usepackage[hang]{footmisc}
\usepackage{lipsum}

\makeatletter
\renewcommand{\overset}[3][0ex]{%
  \mathrel{\mathop{#3}\limits^{
    \vbox to#1{\kern-2\ex@
    \hbox{$\scriptstyle#2$}\vss}}}}
\makeatother

\definecolor{capri}{rgb}{0.0, 0.75, 1.0}

\definecolor{ggreen}{rgb}{0.05,0.45,0.1}

\def \nn   {\!\!\!\!\!}

\AtBeginDocument{%
    \newwrite\bibnotes
    \def\bibnotesext{Notes.bib}
    \immediate\openout\bibnotes=\jobname\bibnotesext
    \immediate\write\bibnotes{@CONTROL{REVTEX41Control}}
    \immediate\write\bibnotes{@CONTROL{%
    apsrev41Control,author="08",editor="1",pages="1",title="0",year="1"}}
     \if@filesw
     \immediate\write\@auxout{\string\citation{apsrev41Control}}%
    \fi
}%

\renewcommand{\emph}[1]{\textit{#1}}

\begin{document}

\title{Motion of localized sources in general relativity:\\
gravitational self-force from quasilocal conservation laws}

\author{Marius Oltean}
\email[]{oltean@ice.cat}

\affiliation{\mbox{Institute of Space Sciences (ICE, CSIC), Campus Universitat Aut\`{o}noma de Barcelona,}\\
\mbox{Carrer de Can Magrans s/n, 08193 Cerdanyola del Vall\`{e}s (Barcelona), Spain}}

\affiliation{\mbox{Institute of Space Studies of Catalonia (IEEC),}\\
\mbox{Carrer del Gran Capit\`{a}, 2-4, Edifici Nexus, despatx 201, 08034 Barcelona, Spain}}

\affiliation{\mbox{Departament de F\'isica, Facultat de Ci\`{e}ncies, Universitat Aut\`{o}noma de Barcelona,}\\
\mbox{Edifici C, 08193 Cerdanyola del Vall\`{e}s (Barcelona), Spain}}

\affiliation{\mbox{Observatoire des Sciences de l'Univers en r\'{e}gion Centre (OSUC), Universit\'{e} d'Orl\'{e}ans,}\\
\mbox{1A rue de la F\'{e}rollerie, 45071 Orl\'{e}ans, France}}

\affiliation{Laboratoire de Physique et Chimie de l'Environnement et de l'Espace (LPC2E), Centre National de la Recherche Scientifique (CNRS), 3A Avenue de la Recherche Scientifique, 45071 Orl\'{e}ans, France}

\author{Richard J. Epp}

\affiliation{Department of Physics and Astronomy, University of Waterloo, 200 University Avenue West, Waterloo, Ontario N2L 3G1, Canada}

\author{Carlos F. Sopuerta}

\affiliation{\mbox{Institute of Space Sciences (ICE, CSIC), Campus Universitat Aut\`{o}noma de Barcelona,}\\
\mbox{Carrer de Can Magrans s/n, 08193 Cerdanyola del Vall\`{e}s (Barcelona), Spain}}

\affiliation{\mbox{Institute of Space Studies of Catalonia (IEEC),}\\
\mbox{Carrer del Gran Capit\`{a}, 2-4, Edifici Nexus, despatx 201, 08034 Barcelona, Spain}}

\author{Alessandro D.A.M. Spallicci}

\affiliation{\mbox{Observatoire des Sciences de l'Univers en r\'{e}gion Centre (OSUC), Universit\'{e} d'Orl\'{e}ans,}\\
\mbox{1A rue de la F\'{e}rollerie, 45071 Orl\'{e}ans, France}}

\affiliation{Laboratoire de Physique et Chimie de l'Environnement et de l'Espace (LPC2E), Centre National de la Recherche Scientifique (CNRS), 3A Avenue de la Recherche Scientifique, 45071 Orl\'{e}ans, France}

\affiliation{\mbox{P\^{o}le de Physique, Collegium Sciences et Techniques (CoST), Universit\'{e} d'Orl\'{e}ans,}\\
\mbox{Rue de Chartres, 45100  Orl\'{e}ans, France}}

\author{Robert B. Mann}

\affiliation{Department of Physics and Astronomy, University of Waterloo, 200 University Avenue West, Waterloo, Ontario N2L 3G1, Canada}

\affiliation{Perimeter Institute for Theoretical Physics, 31 Caroline Street North, Waterloo, Ontario N2L 2Y5, Canada}

\date{\today}

\begin{abstract}

An idealized ``test'' object in general relativity moves along a geodesic.
However, if the object has a finite mass, this will create additional
curvature in the spacetime, causing it to deviate from geodesic motion. If
the mass is nonetheless sufficiently small, such an effect is usually treated
perturbatively and is known as the gravitational self-force due to the object. This issue is still an open problem in gravitational physics
today, motivated not only by basic foundational interest, but also by the need for its direct
application in gravitational-wave astronomy. In particular, the
observation of extreme-mass-ratio inspirals by the future
space-based detector LISA will rely crucially on an accurate
modeling of the self-force driving the orbital evolution and
gravitational wave emission of such systems. 

In this paper, we present a novel derivation,
based on conservation laws, of the basic equations of motion for this
problem. They are formulated with the use of a quasilocal (rather than
matter) stress-energy-momentum tensor---in particular, the Brown-York
tensor---so as to capture gravitational effects in the  momentum
flux of the object, including the self-force. Our formulation and resulting equations of motion are independent of the choice of the perturbative gauge. We show that, in addition to the usual gravitational self-force term, they also lead to an additional ``self-pressure'' force not found in previous analyses, and also that our results correctly recover known formulas under appropriate conditions. Our approach thus offers a fresh geometrical
picture from which to understand the self-force fundamentally, and
potentially useful new avenues for computing it practically.
\end{abstract}

\maketitle


\section{Introduction}

\subsection{Gravitational waves and extreme-mass-ratio inspirals}
The recent advent of gravitational wave astronomy---propelled
by the ground-based direct detections achieved by the LIGO/Virgo collaboration (see~\cite{abbott_et_al._gwtc-1:_2018} for the detections during the O1 and O2 observing runs),
the success of the LISA Pathfinder mission as a proof of principle
for future space-based interferometric detectors \cite{armano_et_al._sub-femto-g_2016,armano_et_al._beyond_2018},
and the subsequent approval of the LISA mission for launch in the
2030s \cite{amaro-seoane_et_al._gravitational_2013,amaro-seoane_et_al._laser_2017}---has brought a multitude 
of both practical and foundational problems to the foreground of gravitational
physics today. While a plethora of possibilities for gravitational wave sources are
actively being investigated theoretically and anticipated to become accessible
observationally, both on the Earth as well as in space, the most ubiquitous
class of such sources has manifestly been---and foreseeably will
remain---the coalescence of compact object binaries \cite{colpi_gravitational_2016,celoria_lecture_2018}.
These are two-body systems consisting of a pair of compact objects, say of masses $M_{1}$
and $M_{2}$, orbiting and eventually spiraling into each other.
Each of these is, usually, either a \emph{black hole} (BH) or a neutron star. There are also more general possibilities being investigated, including that of having a brown dwarf as one of the objects \cite{amaro-seoane_x-mris:_2019}.

The LIGO/Virgo detections during the first scientific runs~\cite{abbott_et_al._gwtc-1:_2018}, O1 and O2, have all involved binaries of \emph{stellar-mass
compact objects} (SCOs) located in our local neighbourhood. These have
comparable masses, of the order of a few tens of solar masses each
($M_{1}\sim M_{2}\sim10^{0-2}M_{\odot}$). 
In addition second- and third- generation terrestrial detectors can also eventually see \emph{intermediate-mass-ratio
inspirals}, binaries consisting of an intermediate-mass BH, of $10^{2-4}M_{\odot}$, and an SCO. While there is as yet no direct evidence for the existence of the former sorts of objects, there are good reasons to anticipate their detection (through gravitational waves) most likely at the centers of globular clusters, and their study provides an essential link to the strongly perturbative regime of compact object binary dynamics.

\begin{figure}
\begin{centering}
\includegraphics[scale=0.6]{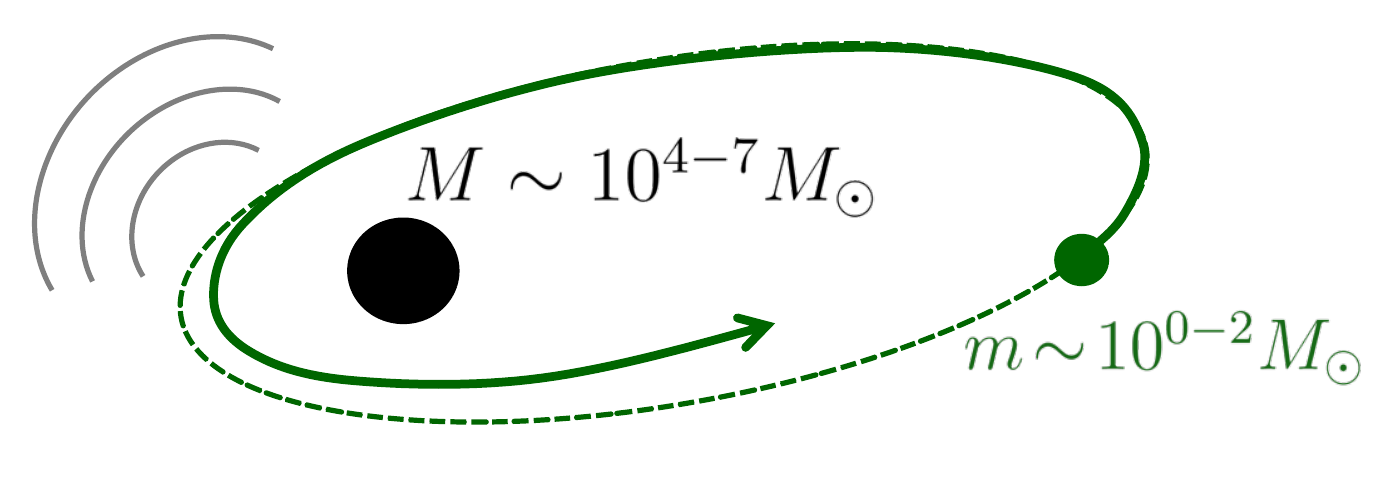}
\par\end{centering}
\caption{Sketch of an extreme-mass-ratio inspiral (EMRI), a two-body system consisting of a stellar-mass compact object (SCO), usually a stellar-mass black hole, of mass $m\sim10^{0-2}M_{\odot}$, orbiting and eventually spiralling into a (super-) massive black hole (MBH), of mass $M\sim10^{4-7}M_{\odot}$, and emitting gravitational waves in the process.}\label{fig-emri}
\end{figure}

It is even further in this direction that future space-based
detectors such as LISA are anticipated to take us. In particular, LISA is expected to see \emph{extreme-mass-ratio
inspirals} (EMRIs) \cite{amaro-seoane_relativistic_2018}, compact binaries where $M_{1}\gg M_{2}$. An elementary sketch is depicted in Figure \ref{fig-emri}. The
more massive object could be a (\emph{super}-) \emph{massive}\textit{
}black hole (MBH) of mass $M_{1}=M\sim10^{4-7}\,M_{\odot}$ located
at a galactic center, with the significantly less massive object---effectively
orbiting and eventually spiraling into the MBH---being an SCO: either
a stellar-mass black hole or a neutron star, with $M_{2}=m\sim10^{0-2}M_{\odot}$.

Average estimates indicate that LISA will be able to see on the order
of hundreds of EMRI events per year~\cite{babak_science_2017}, with an expectation of observing, for each,
thousands of orbital cycles over a period on the order of one year before the final plunge \cite{barack_self-force_2018}.
The trajectories defining these cycles and the gravitational wave signals produced
by them will generally look much more complex than the relatively
generic signals from mergers of stellar-mass black holes of comparable masses as observed, for
example, by LIGO/Virgo.

EMRIs will therefore offer an ideal experimental milieu for strong
gravity: the complicated motion of the SCO around the MBH will effectively
``map out'' the geometry---that is, the gravitational field---around
the MBH, thus presenting us with an unprecedented opportunity for
studying gravity in the very strong regime~\cite{babak_science_2017,berry_unique_2019}. In particular, among the
possibilities offered by EMRIs are the measurement of the mass and
spin of the MBH to very high accuracy, testing the validity of the
Kerr metric as the correct description of BHs within general relativity
(GR), and testing GR itself as the correct theory of gravity.

\begin{figure}
\begin{centering}
\includegraphics[scale=0.47]{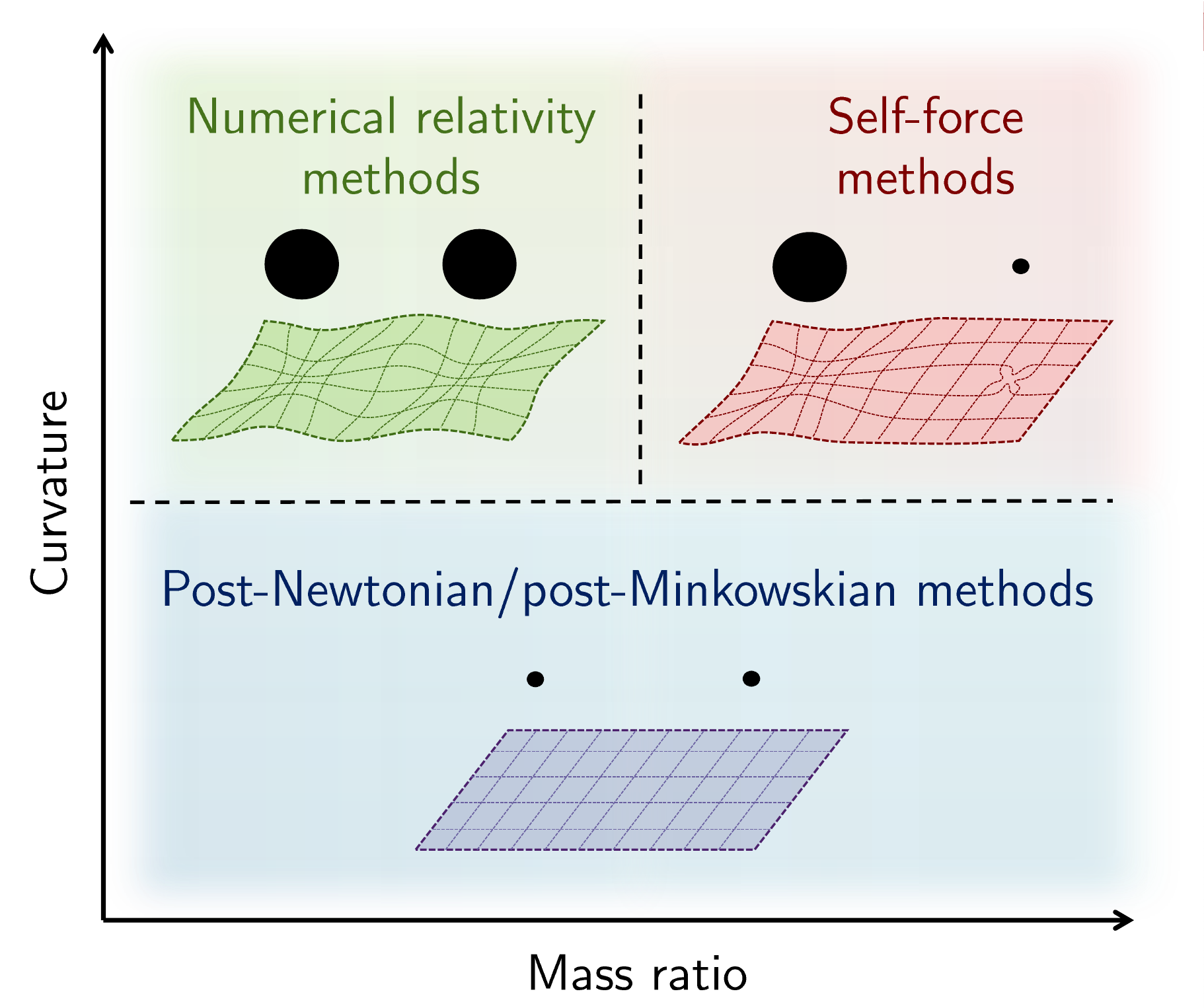}
\par\end{centering}
\caption{The main approaches used in practice for the modeling of compact object binaries as a function of the mass ratio (increasing from $1$) and the spacetime curvature involved. For low curvature (high separation between the bodies), post-Newtonian and post-Minkowskian methods are used. For high curvature (low separation) and low mass ratio, numerical relativity is used. For high curvature and extreme mass ratios, as the scale of a numerical grid would have to span orders of magnitude thus rendering it impracticable,  perturbation theory must be used---in particular, self-force methods.}\label{fig-pn-nr-sf}
\end{figure}

Yet, the richness of the observational opportunities presented by EMRIs
comes with an inexorable cost: that is, a significant and as yet ongoing
technical challenge in their theoretical modeling. This is all the
more pressing as the EMRI signals expected from LISA are anticipated
to be much weaker than the instrumental noise of the detector. Effectively,
what this means is that extremely accurate models are necessary in
order to produce the waveform templates that can be used to extract
the relevant signals from the detector data stream. At the theoretical
level, the problem of EMRI modeling cannot be tackled directly with
numerical relativity (used for the LIGO/Virgo detections), simply
due to the great discrepancy in (mass/length) scales; however, for
the same reason, the approach that readily suggests itself is perturbation
theory. See Figure \ref{fig-pn-nr-sf} for a graphic depicting the main methods used for compact object binary modeling in the different regimes. In particular, modeling the strong gravity, extreme mass ratio regime turns out to be equivalent to a general
and quite old problem which can be posed in any (not just gravitational)
classical field theory: the so-called \emph{self-force} problem.

\subsection{The self-force problem} 

Suppose we are dealing with a theory for a field $\psi(x)$ in some
spacetime. If the theory admits a Lagrangian formulation, we can usually
assume that the field equations have the general form 
\begin{equation}
L[\psi\left(x\right)]=S\left(x\right)\,,\label{eq:general_field_eqn}
\end{equation}
where $L$ is a (partial, possibly nonlinear and typically second-order)
differential operator, and we refer to $S$ as the \emph{source} of
the field $\psi$. Broadly speaking, the problem of the self-force
is to find solutions $\psi(x)$ satisfying (\ref{eq:general_field_eqn})
when $S$ is ``localized'' in spacetime. Intuitively, it is the
question of how the existence of a dynamical (field-generating) ``small
object'' (a mass, a charge etc.) backreacts upon the total field
$\psi$, and hence in turn upon its own future evolution subject to
that field. Thus, an essential part of any detailed self-force analysis
is a precise specification of what exactly it means for $S$ to be
localized. In standard approaches, one typically devises a perturbative
procedure whereby $S$ ends up being approximated as a distribution,
usually a Dirac delta, compactly supported on a worldline---that
is, the \emph{background} (zeroth perturbative order) worldline of
the small object. However, this already introduces a nontrivial mathematical
issue: if $L$ is non-linear (in the standard PDE sense), then the
problem (\ref{eq:general_field_eqn}) with a distributional source $S$ is
mathematically ill-defined, at least within the classical theory
of distributions \cite{schwartz_theorie_1957} where products of distributions
do not make sense \cite{schwartz_sur_1954}\footnote{Nonlinear theories of distributions are being actively investigated
by mathematicians \cite{li_review_2007,colombeau_nonlinear_2013,bottazzi_grid_2019}. Some work has been done to apply these to the electromagnetic self-force problem \cite{gsponer_classical_2008} and to study their general applicability in GR \cite{steinbauer_use_2006}, however at this point, to our knowledge, their potential usefulness
for the gravitational self-force problem has not been contemplated to any significant extent.}. 

One might therefore worry that nonlinear physical theories, such as
GR, would a priori not admit solutions sourced by distributions, and
we refer the interested reader to Ref. \cite{geroch_strings_1987}
for a classic detailed discussion of this topic. The saving point
is that, while the full field equation (in this case, the Einstein
equation) may indeed be generally non-linear, if we devise a perturbative
procedure (where the meaning of the perturbation is prescribed
in such a way as to account for the presence of the small object itself),
then the first-order field equation is, by construction, linear in
the (first-order) perturbation $\delta\psi$ of $\psi$. Thus, assuming
the background field is a known exact solution of the theory,
it always makes sense to seek solutions $\delta\psi$ to
\begin{equation}
\delta L[\delta\psi\left(x\right)]=S\left(x\right)\,,\label{eq:general_perturbed_field_eqn}
\end{equation}
for a distributional source $S$, where $\delta L$ indicates the
first-order part of the operator $L$ in the full field equation (\ref{eq:general_field_eqn}).
As this only makes sense for the (linear) first-order problem, such
an approach becomes again ill defined if we begin to ask about the
(nonlinear) second- or any higher-order problem. Additional technical
constructions are needed to deal with these, the most common of which
for the gravitational self-force has been the so-called ``puncture'' (or ``effective source'')
method \cite{barack_scalar-field_2007,barack_m-mode_2007,vega_regularization_2008,barack_self-force_2018}; similar ideas have proven to be very useful also in numerical relativity \cite{campanelli_accurate_2006,baker_gravitational-wave_2006}. For work on the second-order equation of motion for the gravitational self-force problem, see \textit{e.g.} Refs. \cite{pound_second-order_2012,gralla_second-order_2012,pound_nonlinear_2017}. For now, we assume that
we are interested here in the first-order self-force problem (\ref{eq:general_perturbed_field_eqn}) only.

\begin{figure}
\begin{centering}
\includegraphics[scale=0.6]{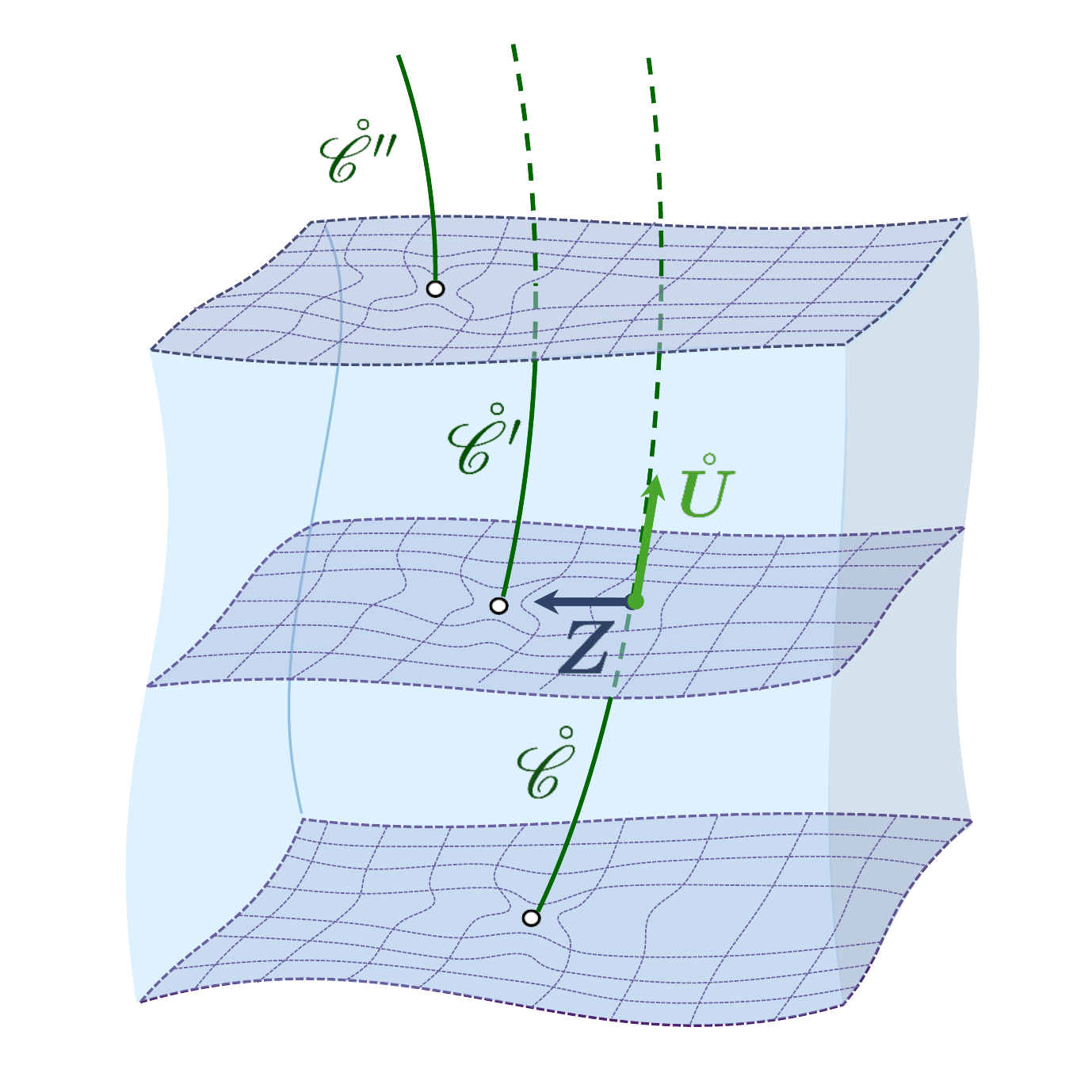}
\par\end{centering}
\caption{A depiction of the perturbative problem for the gravitational self-force (GSF). In particular, this represents one of the most popular conceptions of a so-called ``self-consistent'' approach \cite{gralla_rigorous_2008}: at a given step (on a given Cauchy surface) in the time evolution problem, one computes the ``correction to the motion'' away from geodesic ($\mathring{\mathscr{C}}$) in the form of a deviation vector $Z^{a}$, determined by the GSF. Then, at the next time step, one begins on a new (``corrected'') geodesic ($\mathring{\mathscr{C}}'$), computes the new deviation vector, and so on.}\label{fig-sf}
\end{figure}

Now concretely, in GR, our physical field $\psi$ is simply the spacetime metric $g_{ab}$ (where Latin letters from the beginning of the alphabet indicate spacetime indices), and following standard convention we
denote a first-order perturbation thereof by $\delta g_{ab}=h_{ab}$.
The problem (\ref{eq:general_perturbed_field_eqn}) is then just the
first-order Einstein equation,
\begin{equation}
\delta G_{ab}[h_{cd}]=\kappa T_{ab}^{\textrm{PP}}\,,\label{eq:first_order_EFE}
\end{equation}
where $G_{ab}$ is the Einstein tensor, $\kappa=8\pi$ (in geometrized
units $c=G=1$) is the Einstein constant, and $T_{ab}^{\textrm{PP}}$ the energy-momentum tensor 
of a ``point particle'' (PP) compactly supported on a given worldline $\mathring{\mathscr{C}}$. We will return later
to discussing this more precisely, but in typical approaches, $\mathring{\mathscr{C}}$
turns out to be a geodesic---that is, the ``background motion''
of the small object, which is in this case a small mass\footnote{The problem of deriving geodesic motion for appropriately defined non-dynamical ``test particles'' from the Einstein equation (in
lieu of postulating it as an independent axiom of GR additional
to the Einstein equation) is a long-standing and interesting issue
in its own right. Einstein was involved in some of the earliest work on this \cite{einstein_gravitational_1938}, and over the decades various proofs
have been put forward outside of the context of the gravitational self-force problem. See Refs. \cite{geroch_motion_1975,ehlers_equation_2004} for some of the most famous such proofs. 
See also Ref. \cite{weatherall_geometry_2018} for a recent general
review of the most widely used approaches as well as an interesting
novel proposal. We consider later in this paper in detail one approach
to the gravitational self-force which also proves geodesic motion as the background motion of point particles in GR.}. Thus, simply solving (\ref{eq:first_order_EFE}) for $h_{ab}$ assuming
a fixed $\mathring{\mathscr{C}}$ for all time, though mathematically
well-defined, is by itself physically meaningless: it would simply
give us the metric perturbations caused by a small object eternally
moving on the same geodesic. Instead what we would ultimately like
is a way to take into account how $h_{ab}$ modifies the motion of
the small object itself. Thus in addition to the field equation (\ref{eq:first_order_EFE}),
any self-force analysis must be supplemented by an \emph{equation
of motion} (EoM) telling us, essentially, how to move from a given
background geodesic $\mathring{\mathscr{C}}$ at one step in the (ultimately
numerical) time evolution problem to a new background geodesic $\mathring{\mathscr{C}}'$
at the next time step---with respect to which the field equation
(\ref{eq:first_order_EFE}) is solved anew, and so on. This is sometimes referred to as a ``self-consistent'' approach. See Fig. \ref{fig-sf} for a visual depiction. 

The first proposal for an EoM for the \emph{gravitational self-force} (GSF)
problem was put forward in the late 1990s, since known as the MiSaTaQuWa
equation after its authors \cite{mino_gravitational_1997,quinn_axiomatic_1997}.
On any $\mathring{\mathscr{C}}$, it reads:
\begin{equation}
\ddot{Z}^{a}=-\mathring{E}_{b}\,^{a}Z^{b}+F^{a}[h_{cd}^{\textrm{tail}};\mathring{U}^{e}]\,.\label{eq:intro_MiSaTaQuWa}
\end{equation}
The LHS is a second (proper) time derivative of a \emph{deviation
vector} $Z^{a}$ on $\mathring{\mathscr{C}}$ pointing in the direction
of the ``true motion'' (away from $\mathring{\mathscr{C}}$), to
be defined more precisely later. On the RHS, $\mathring{E}_{ab}$
is the electric part of the Weyl tensor on $\mathring{\mathscr{C}}$,
such that the first term is a usual ``geodesic deviation'' term.
The second term on the RHS is the one usually understood as being responsible for self-force effects: $F^{a}[\cdot;\cdot]$ is a four-vector functional of a symmetric
rank-$2$ contravariant tensor and a vector, to which we refer in general (for any
arguments) as the \emph{GSF functional}. In any spacetime with a given
metric $\mathring{g}_{ab}$ and compatible derivative operator $\mathring{\nabla}_{a}$,
it is explicitly given by the following simple action of a first-order
differential operator:
\begin{multline}
F^{a}[H_{bc};V^{d}]\\
=-\left(\mathring{g}^{ab}+V^{a}V^{b}\right)\left(\mathring{\nabla}_{c}H_{bd}-\frac{1}{2}\mathring{\nabla}_{b}H_{cd}\right)V^{c}V^{d}\,.\label{eq:intro_GSF_functional}
\end{multline}
While this is easy enough to calculate once one knows the arguments,
the main technical challenge in using the MiSaTaQuWa equation (\ref{eq:intro_MiSaTaQuWa})
lies precisely in the determination thereof: in particular, $h_{ab}^{\textrm{tail}}$
is not the \emph{full} metric perturbation $h_{ab}$ which solves the field
equation (\ref{eq:first_order_EFE}), but instead represents what
is called the ``tail'' integral of the Green functions of $h_{ab}$ \cite{poisson_motion_2011}.
This quantity is well defined, but difficult to calculate in practice
and usually requires the fixing of a perturbative gauge---typically
the \emph{Lorenz} gauge, $\mathring{\nabla}^{b}(h_{ab}-\frac{1}{2}h_{cd}\mathring{g}^{cd}\mathring{g}_{ab})=0$. Physically, $h_{ab}^{\textrm{tail}}$ can be thought of as the part of the full perturbation $h_{ab}$ which is scattered back by the spacetime curvature. (In this way, $h_{ab}$ can be regarded as the sum of $h_{ab}^{\textrm{tail}}$ and the remainder, what is sometimes called the ``instantaneous'' or ``direct'' part $h_{ab}^{\textrm{direct}}$, responsible for waves radiated to infinity \cite{spriao14}.)

An alternative, equivalent GSF EoM was proposed by Detweiler and Whiting
in the early 2000s \cite{detweiler_self-force_2003}. It relies upon
a regularization procedure for the metric perturbations, \textit{i.e.} a choice
of a decomposition for $h_{ab}$ [the full solution of the field equation
(\ref{eq:first_order_EFE})] into the sum of two parts: one
which diverges---in fact, one which contains \emph{all} divergent
contributions\emph{\emph{---}}on $\mathring{\mathscr{C}}$, denoted
$h_{ab}^{\textrm{S}}$ (the so-called ``singular'' field, related to the ``direct'' part of the metric perturbation), and a
remainder which is finite, $h_{ab}^{\textrm{R}}$ (the so-called ``regular''
field, related to the ``tail'' part), so that one writes $h_{ab}=h_{ab}^{\textrm{S}}+h_{ab}^{\textrm{R}}$.
An analogy with the self-force problem in electromagnetism gives some
physical intuition behind how to interpret the meaning of this decomposition
\cite{barack_self-force_2018}, with $h_{ab}^{\textrm{S}}\sim m/r$
having the heuristic form of a ``Coulombian self-field.'' However,
no procedure is known for obtaining the precise expression of $h_{ab}^{\textrm{S}}$
in an arbitrary perturbative gauge, and moreover, once a gauge is
fixed (again, usually the Lorenz gauge), this splitting is not unique \cite{barack_self-force_2018}.
Nevertheless, if and when such an $h_{ab}^{\textrm{S}}$ is obtained
(from which we thus also get $h_{ab}^{\textrm{R}}=h_{ab}-h_{ab}^{\textrm{S}}$),
the Detweiler-Whiting EoM for the GSF reads:
\begin{equation}
\ddot{Z}^{a}=-\mathring{E}_{b}\,^{a}Z^{b}+F^{a}[h_{cd}^{\textrm{R}};\mathring{U}^{e}]\,.\label{eq:intro_Detweiler-Whiting}
\end{equation}

The EoMs (\ref{eq:intro_MiSaTaQuWa}) and (\ref{eq:intro_Detweiler-Whiting})
are equivalent in the Lorenz gauge and form the basis of the two most popular
methods used today for the numerical computation of the GSF. Yet a
great deal of additional technical machinery is required for handling
gauge transformations. This is essential because, in the EMRI problem,
the background spacetime metric---that of the MBH---is usually assumed
to be Schwarzschild-Droste or Kerr. Perturbation theory for such spacetimes
has been developed and is most easily carried out in, respectively,
the so-called Regge-Wheeler and radiation gauges; in other words,
in practice, it is often difficult (though not infeasible---see, \textit{e.g.}, Ref. \cite{balo05}) to compute $h_{ab}$ directly
in the Lorenz gauge for use in (\ref{eq:intro_MiSaTaQuWa}) or (\ref{eq:intro_Detweiler-Whiting}).

A proposal for an EoM for the GSF problem that is valid in a wider class of
perturbative gauges was presented by Gralla in 2011 \cite{gralla_gauge_2011}.
It was therein formulated in what are called ``parity-regular''
gauges, \textit{i.e.} gauges satisfying a certain parity condition. This condition
ultimately has its origins in the Hamiltonian analysis of Regge and
Teitleboim in the 1970s \cite{regge_role_1974}, wherein the authors
introduce it in order to facilitate the vanishing of certain surface
integrals and thus to render certain general-relativistic Hamiltonian
notions, such as multipoles and ``center of mass,'' well-defined
mathematically. In parity-regular gauges (satisfying the Regge-Teitleboim
parity condition), the Gralla EoM---mathematically equivalent, in
the Lorenz gauge, to the MiSaTaQuWa and the Detweiler-Whiting EoMs---is:
\begin{equation}
\ddot{Z}^{a}=-\mathring{E}_{b}\,^{a}Z^{b}+\frac{1}{4\pi}\lim_{r\rightarrow0}\intop_{\mathbb{S}_{r}^{2}}\bm{\epsilon}^{}_{\mathbb{S}^{2}}\,F^{a}[h_{cd};\mathring{U}^{e}]\,.\label{eq:intro_Gralla}
\end{equation}
The GSF (last) term on the RHS is obtained in this approach by essentially
relating the deviation vector $Z^{a}$ (the evolution of which is expressed
by the LHS) with a gauge transformation vector and then performing
an ``angle average'' over a ``small'' $r$-radius two-sphere $\mathbb{S}_{r}^{2}$,
with $\bm{\epsilon}^{}_{\mathbb{S}^{2}}$ the volume form of the unit two-sphere, of the so-called ``bare''
GSF, $F^{a}[h_{bc};\mathring{U}^{d}]$. The latter is just the GSF
functional [Eq. (\ref{eq:intro_GSF_functional})] evaluated directly using the \emph{full} metric perturbatiuon $h_{ab}$ (\textit{i.e.} the ``tail'' plus ``direct'' parts, or equivalently, the ``regular'' plus ``singular'' parts),
around (rather than at the location of) the distributional source. 
The point therefore is that this formula never requires the evaluation
of $h_{ab}$ on $\mathring{\mathscr{C}}$ itself, where it is divergent
by construction; instead, away from $\mathring{\mathscr{C}}$ it is
always finite\footnote{This is true provided one does not transform to a perturbative gauge wherein any of the multipole moments of $\bm{h}$ diverge away from $\mathring{\mathscr{C}}$ as a consequence of the gauge definition. Examples of gauges leading to such divergences are the ``half-string'' and ``full-string'' radiation gauges of Ref. \cite{pound_gravitational_2014}, which exhibit string-like singularities in $\bm{h}$ along a radial direction. Nevertheless, in this work it was also shown that one can define a ``no-string'' radiation gauge which is in the parity-regular class, and where the singularities in $\bm{h}$ remain only on $\mathring{\mathscr{C}}$ thus rendering the integral (\ref{eq:intro_GSF_functional}) well-defined.}, and (\ref{eq:intro_Gralla}) says that it suffices
to compute the GSF functional (\ref{eq:intro_GSF_functional}) with
$h_{ab}$ directly in the argument (requiring no further transformations),
and integrate it over a small sphere.

The manifest advantage of (\ref{eq:intro_Gralla}) relative to (\ref{eq:intro_MiSaTaQuWa})
or (\ref{eq:intro_Detweiler-Whiting}) is that no computations of
tail integrals or regularizations of the metric perturbations are
needed at all. Yet, to our knowledge, there has thus far been no attempted
numerical computation of the GSF using this formula. One of the issues
with this remains that of the perturbative gauge: depending upon the detailed setup of the problem, one may still not easily be able to compute $h_{ab}$ directly in a parity-regular gauge (although manifestly, working in the parity-regular ``no-string'' radiation gauge \cite{pound_gravitational_2014} may be useful for a GSF calculation in Kerr), \textit{i.e.} a gauge in which (\ref{eq:intro_Gralla}) is strictly valid, and so further gauge transformations may be needed.  Aside from the practical issues with a possible numerical
implementation of this, there is also a conceptual issue: this formula
originates from an essentially mathematical argument---by a convenient ``averaging'' over the angles---so as to make it well-defined in a Hamiltonian setting via a relation to a canonical
definition of the center of mass. Yet its general form as a closed two-surface
integral suggestively hints at the possibility of interpreting it
not merely as a convenient mathematical relation, but as a real physical
flux of (some notion of) ``gravitational momentum''. We contend
and will demonstrate in this paper that indeed an even more general
version of (\ref{eq:intro_Gralla}), not restricted by any perturbative gauge choice (so long as one does not construct it in such a way that produces divergences in $\bm{h}$ away from $\mathring{\mathscr{C}}$), results from the consideration
of momentum conservation laws in GR.

\subsection{The self-force problem via conservation laws}

\begin{figure}
\begin{centering}
\includegraphics[scale=0.8]{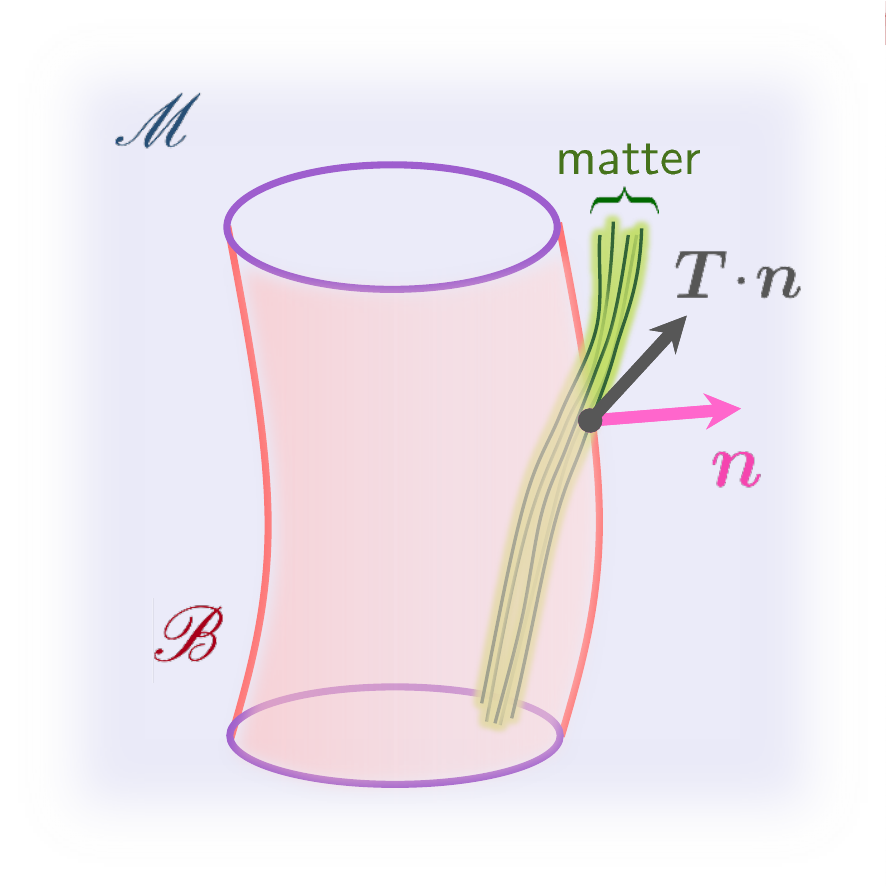}
\par\end{centering}
\caption{A worldtube boundary $\mathscr{B}$
(topologically $\mathbb{R}\times\mathbb{S}^{2}$) in $\mathscr{M}$, with (outward-directed) unit normal
$n^{a}$. The change in matter four-momentum between two constant time slices of this worldtube is given by the flux of the normal projection (in one index) of the matter stress-energy-momentum tensor $T_{ab}$ through the portion of $\mathscr{B}$ bounded thereby.\label{figure-worldtube-boundary}}
\end{figure}

The idea of using conservation laws for tackling the self-force problem
was appreciated and promptly exploited quite early on for the electromagnetic
self-force. In the 1930s \cite{dirac_classical_1938}, Dirac was the
first to put forward such an analysis in flat spacetime, and later
on in 1960 \cite{dewitt_radiation_1960}, DeWitt and Brehme extended
it to non-dynamically curved spacetimes\footnote{By this, we mean spacetimes with non-flat but fixed metrics, which do not evolve dynamically (gravitationally) in response to the matter stress-energy-momentum present therein.}. In such approaches, it can
be shown\footnote{See Ref. \cite{poisson_introduction_1999} for a basic and more contemporary
presentation.} that the EoM for the electromagnetic self-force follows from local
conservation expressions of the form
\begin{equation}
\Delta P^{a}=\intop_{\Delta\mathscr{B}}\bm{\epsilon}^{\,}_{\mathscr{B}}T^{ab}n_{b}\,,\label{eq:local_cons}
\end{equation}
where the LHS expresses the flux of \emph{matter} four-momentum $P^{a}$ (associated with $T_{ab}$) between
the ``caps'' of (\textit{i.e.} closed spatial two-surfaces delimiting) a
portion (or ``time interval'') of a thin worldtube boundary $\mathscr{B}$
(topologically $\mathbb{R}\times\mathbb{S}^{2}$), with natural volume
form $\bm{\epsilon}^{\,}_{\mathscr{B}}$ and (outward-directed) unit normal
$n^{a}$ (see Figure~\ref{figure-worldtube-boundary}). In particular, one takes a time derivative of (\ref{eq:local_cons})
to obtain an EoM expressing the time rate of change of momentum in
the form of a closed spatial two-surface integral (by differentiating the worldtube boundary integral). For the electromagnetic
self-force problem, the introduction of an appropriate matter stress-energy-momentum tensor
$T_{ab}$ into Eq.~(\ref{eq:local_cons}) and a bit of subsequent argumentation
reduces the integral expression to the famous Lorentz-Dirac equation;
on a spatial three-slice in a Lorentz frame and in the absence of
external forces, for example, this simply reduces to $\dot{P}^{i}=\frac{2}{3}q^{2}\dot{a}^{i}$ for a charge $q$. 

Along these lines, see also Ref. \cite{burko_self-force_1999} for a treatment of the scalar and electromagnetic self-forces based on the notion of a limit of an extended charge. Formulations of the scalar and electromagnetic self-forces using generalized Killing fields have more recently been put forward in \cite{harte_self-forces_2008,harte_electromagnetic_2009}. Then in Ref. \cite{harte_mechanics_2012}, the idea of generalized Killing fields was also used to study the gravitational self-force problem by employing definitions of momenta as bulk (volume) integrals (over the worldtube interior), involving the internal matter stress-energy-momentum tensor of the moving object (mass). This in fact proved the validity of the Detweiler-Whiting approach, i.e. viewing the ``point mass'' as moving on a geodesic in an effective (regularized) metric, to all multiple orders and also for the angular momentum. As we will shortly elaborate, we posit that gravitational momentum in GR cannot fundamentally be regarded as a bulk (local) density, and a more detailed analysis to reveal how the (locally-formulated) results of Ref. \cite{harte_mechanics_2012} relate to our work here would be very interesting.

The success of conservation law approaches for formulating the electromagnetic
self-force in itself inspires hope that the same may be done
in the case of the gravitational self-force (GSF) problem. In particular,
Gralla's EoM (\ref{eq:intro_Gralla})
strongly hints at the possibility of understanding the RHS
not just as a mathematical (``angle averaging'') device, but as a \emph{true, physical
flux of gravitational momentum} arising from a consideration of conservation
expressions.

Nevertheless, to our knowledge, there has thus far been no proposed
general treatment of the GSF following such an approach. This may, in large
part, be conceivably attributed to the notorious conceptual difficulties
surrounding the very question of the basic formulation of conservation
laws in GR. Local conservation laws, along the lines of Eq.~(\ref{eq:local_cons}) that can
readily be used for electromagnetism, no longer make sense fundamentally once gravity
is treated as dynamical. The reason has a simple explanation in the
equivalence principle \cite{misner_gravitation_1973}: one can always find a local frame of reference
with a vanishing local ``gravitational field'' (metric connection
coefficients), and hence a vanishing local ``gravitational energy-momentum'',
irrespective of how one might feel inclined to define the latter\footnote{It is worth remarking here that, in a perturbative setting, an approach that is sometimes taken is to work with an ``effective'' local gravitational stress-energy-momentum tensor, defined as the RHS of a suitably-rearranged (first-order) Einstein equation. This is a common tactic often used for studying, for example, the energy-momentum of gravitational waves, with some applicable caveats (see, \textit{e.g.}, Chapter 35 of \cite{misner_gravitation_1973}). In fact, one of the first formulations of the gravitational self-force---in particular, the derivation of the MiSaTaQuWa EoM (\ref{eq:intro_MiSaTaQuWa}) presented in Section III of \cite{mino_gravitational_1997}---made use of the local conservation (vanishing of the spacetime divergence) of a suitably-defined local tensor of such a sort (in analogy with the approach of DeWitt and Brehme to the electromagnetic self-force \cite{dewitt_radiation_1960}). We elaborate in the remainder of this subsection and at greater length in Section \ref{sec:qf} on why such a notion of gravitational conservation principles, while demonstrably useful for operational computations in some situations, cannot in general be expected to capture the fundamentally quasilocal (boundary density) nature of gravitational energy-momentum. See \cite{epp_momentum_2013} for a detailed discussion and comparison between these two (local and quasilocal) views of gravitational energy-momentum.}.

A wide variety of approaches have been taken over the decades towards
formulating sensible notions of gravitational energy-momentum, with still no general
consensus among relativists today on which to qualify as ``the best'' \cite{szabados_quasi-local_2004,jaramillo_mass_2011}. Often the preference
for employing certain definitions over others may simply come down
to context or convenience, but in any case, there exist agreements
between the most typical definitions in various limits. A very common
feature among them is the idea of replacing a local notion of gravitational
energy-momentum, \textit{i.e.} energy-momentum as a volume density, with what is known as a \emph{quasilocal}
energy-momentum, \textit{i.e.} energy-momentum as a boundary density. The typical Hamiltonian definitions
of the (total) gravitational energy-momentum for an asymptotically-flat spacetime,
for example, are of such a form. Among the most commonly used generalizations
of these definitions to arbitrary (finite) spacetime regions was proposed
in the early 1990s by Brown and York \cite{brown_quasilocal_1993}, and follow from what is now eponymously known
as the Brown-York stress-energy-momentum tensor. It is a quasilocal tensor, meaning it
is only defined on the boundary of an arbitrary spacetime region.
For example, using this, the total (matter plus gravitational) energy
inside a spatial volume is given up to a constant factor by
the closed two-surface (boundary) integral of the trace of the boundary extrinsic
curvature---precisely in agreement with the Hamiltonian definition of energy for the entire spacetime in the appropriate limit
(where the closed two-surface approaches a two-sphere
at asymptotically-flat spatial infinity) but, in principle, applicable
to any region in any spacetime. 

The formulation of general energy-momentum conservation laws in GR from the Brown-York tensor 
has been achieved in recent years with the use of a construction called \emph{quasilocal
frames} \cite{epp_momentum_2013}, a concept first proposed in Ref. \cite{epp_rigid_2009}. Essentially, the idea is that, in order to describe a (finite) gravitational system of interest and its associated boundary fluxes of gravitational energy-momentum, it does not suffice to merely
specify, as in the local matter conservation laws of the form of Eq.~(\ref{eq:local_cons}), a worldtube boundary $\mathscr{B}$ (the interior of which is to be regarded as containing that system, and across which its energy-momentum flux is measured) as an embedded submanifold of $\mathscr{M}$. What is in
fact required is the specification of a \emph{congruence} making up this
worldtube boundary, \textit{i.e.} a two-parameter family of timelike worldlines
with some chosen four-velocity field representing the motion of a
topological two-sphere's worth of \emph{quasilocal observers}. We will motivate this construction in greater amplitude shortly, but the reason for needing it is basically to be able to meaningfully define ``time-time'' and ``time-space'' directions on $\mathscr{B}$ for our conservation laws. A congruence of this sort is
what is meant by a quasilocal frame. 

The enormous advantage in using these quasilocal conservation laws
over other approaches lies in the fact that they hold in any arbitrary
spacetime. Thus the existence of Killing vector fields---a typical
requirement in other conservation law formulations---is in no way
needed here.

This idea has been used successfully in a number of applications so far \cite{mcgrath_quasilocal_2012,epp_existence_2012,epp_momentum_2013,mcgrath_post-newtonian_2014,mcgrath_rigid_2014,oltean_geoids_2016,oltean_geoids_2017}. These include the resolution of a variation of Bell's spaceship paradox\footnote{Proposed initially by E. Dewan and M. Beran \cite{dewan_note_1959} and later made popular by J.S. Bell's version \cite{bell_how_1976}.} in which a box accelerates rigidly in a transverse,
uniform electric field \cite{mcgrath_quasilocal_2012}, recovering under appropriate conditions the typical (but more limited) local matter conservation expressions of the form of Eq.~(\ref{eq:local_cons}) from the quasilocal ones \cite{epp_momentum_2013}, application to post-Newtonian theory \cite{mcgrath_post-newtonian_2014} and to relativistic geodesy \cite{oltean_geoids_2016,oltean_geoids_2017}. 

A similar idea to quasilocal frames, called \emph{gravitational screens}, was proposed more recently in Refs.  \cite{freidel_gravitational_2015,freidel_non-equilibrium_2015}. There, the authors also make use of quasilocal ideas to develop conservation laws very similar in style and form to those obtained via quasilocal frames. A detailed comparison between these two approaches has thus far not been carried out, but it would be very interesting to do so in future work. In particular, the notion of gravitational screens has been motivated more from thermodynamic considerations, and similarly casting quasilocal frames in this language could prove quite fruitful. For example, just as these approaches have given us operational definitions of concepts like the ``energy-momentum in an arbitrary spacetime region'' (and not just for special cases such as an entire spacetime), they may help to do the same for concepts like ``entropy in an arbitrary spacetime region'' (and not just for known special cases such as a black hole).

\subsection{Executive summary of the paper}

We now summarize the structure and main result of this paper. 

Section \ref{sec:qf} is entirely devoted to an overview of quasilocal frames and quasilocal conservation laws, in complete self-contained technical detail for our purposes here.  

In Section \ref{sec:general-analysis}, we prove the main general result of this paper from the quasilocal momentum conservation law: in particular, we show that for any localized gravitational system, the change between some initial and final time in its total linear momentum\footnote{This is written using lower-case typewriter
font ($\mathtt{p}$) to distinguish it from the purely matter four-momentum $P^{a}$ associated with $T_{ab}$.} $\mathtt{p}$,
\begin{equation}
\Delta\mathtt{p}^{(\bm{\phi})}=\mathtt{p}_{\textrm{final}}^{(\bm{\phi})}-\mathtt{p}_{\textrm{initial}}^{(\bm{\phi})}\,,
\end{equation}
along a spatial direction determined
by a vector $\bm{\phi}$ (the precise meaning of which is to be elaborated
later), is given by the following flux through the worldtube boundary interval
$\Delta\mathscr{B}$ bounded by these times (see Figure~\ref{figure-worldtube-boundary}): \begin{equation}
\Delta\mathtt{p}^{(\bm{\phi})}=-\frac{1}{4\pi}\intop_{\Delta\mathscr{B}}\bm{\text{\ensuremath{\epsilon}}}_{\mathscr{B}}^{\,}\,\frac{1}{r}\bm{\phi}\cdot\boldsymbol{\mathcal{F}}[h_{ab};\mathring{u}^{c}]+\mathcal{O}\left(r\right)\,.\label{eq:main_result}
\end{equation}
Here, $\mathcal{F}^{a}[\cdot;\cdot]$ is an \emph{extended} GSF functional.
In particular, it is the usual GSF functional $F^{a}$ [see Eq. (\ref{eq:intro_GSF_functional})]
plus a novel piece to which we refer as the \emph{gravitational self-pressure
force} $\wp^{a}$, arising from a quasilocal pressure effect
(also to be elaborated later): 
\begin{equation}
\mathcal{F}^{a}=F^{a}+\wp^{a}\,.\label{eq:main_result_GSF_functional}
\end{equation}
The first argument of this functional $\mathcal{F}^{a}$, just as
in Gralla's formula [our Eq.~(\ref{eq:intro_Gralla})], is the metric perturbation
$h_{ab}$ on $\mathscr{B}$. This avoids any potential singularities
in inside the worldtube, \textit{i.e.} within the spacetime volume of which $\mathscr{B}$ is the (exterior) boundary, and therefore the need for performing regularizations
or any further transformations. 
The second argument of $\mathcal{F}^{a}$
in our result [see Eq.~(\ref{eq:main_result})], \textit{i.e.} $\mathring{u}^{a}$,
is the four-velocity \emph{not} of any background geodesic contained inside the worldtube, as in the typical GSF EoMs discussed earlier---indeed, strictly speaking,
the main result [Eq. (\ref{eq:main_result})] holds without necessarily having
to even introduce any such geodesic, or more generally, without having
to say anything specific about the content of the worldtube interior---but
instead, that of the background quasilocal observers on $\mathscr{B}$
itself.

Manifestly, our formula [Eq. (\ref{eq:main_result})] bears significant
resemblance to that of Gralla [Eq. (\ref{eq:intro_Gralla})], and it is
the scope of Section \ref{sec:gralla-wald-analysis} to compare the two in the appropriate setting. For this, we introduce
the general setup of the Gralla-Wald approach to the GSF, and apply
our conservation law formula for two choices of quasilocal frames:
one which is inertial with the SCO in the perturbed spacetime (and
hence not inertial with the geodesic-following point particle in the
background), and one which is inertial with the geodesic-following
point particle in the background (and hence not inertial with the
SCO in the perturbed spacetime). We derive the EoMs in both of these
cases and discuss their correspondence to the known GSF EoMs.

Finally, Section \ref{sec:concl} concludes the paper.

\subsection{Notation and Conventions\label{ssec:notation}}

We work in the $(-,+,+,+)$ signature of spacetime. Script upper-case
letters ($\mathscr{A}$, $\mathscr{B}$, $\mathscr{C}$, ...) are
reserved for denoting mathematical spaces (manifolds, curves, etc.).
The $n$-dimensional Euclidean space is denoted as usual by $\mathbb{R}^{n}$,
the $n$-sphere of radius $r$ by $\mathbb{S}_{r}^{n}$, and the unit
$n$-sphere by $\mathbb{S}^{n}=\mathbb{S}_{1}^{n}$. For any two spaces
$\mathscr{A}$ and $\mathscr{B}$ that are topologically equivalent
(\textit{i.e.} homeomorphic), we indicate this by writing $\mathscr{A}\simeq\mathscr{B}$.

The set of $(k,l)$-tensors on any manifold $\mathscr{U}$ is denoted
by $\mathscr{T}^{k}\,_{l}(\mathscr{U})$. In particular, $T\mathscr{U}=\mathscr{T}^{1}\,_{0}(\mathscr{U})$
is the tangent bundle and $T^{*}\mathscr{U}=\mathscr{T}^{0}\,_{1}(\mathscr{U})$
the dual thereto. Any $(k,l)$-tensor in any $(3+1)$-dimensional
(Lorentzian) spacetime $\mathscr{M}$ is equivalently denoted either
using the (boldface) index-free notation $\bm{A}\in\mathscr{T}^{k}\,_{l}(\mathscr{M})$
following the practice of, \textit{e.g.}, Refs.~\cite{misner_gravitation_1973,hawking_large_1975},
or the abstract index notation $A^{a_{1}\cdots a_{k}}\,_{b_{1}\cdots b_{l}}\in\mathscr{T}^{k}\,_{l}(\mathscr{M})$
following that of, \textit{e.g.}, Ref.~\cite{wald_general_1984}; that is,
depending upon convenience, we equivalently write
\begin{equation}
\bm{A}=A^{a_{1}\cdots a_{k}}\,_{b_{1}\cdots b_{l}}\in\mathscr{T}^{k}\,_{l}(\mathscr{M})\,,\label{eq:tensor}
\end{equation}
with Latin letters from the beginning of the alphabet ($a$, $b$,
$c$, ...) being used for spacetime indices ($0,1,2,3$). The components of $\bm{A}$
in a particular choice of coordinates $\{x^{\alpha}\}_{\alpha=0}^{3}$
are denoted by $A^{\alpha_{1}\cdots\alpha_{k}}\,_{\beta_{1}\cdots\beta_{l}}$,
using Greek (rather than Latin) letters from the beginning of the
alphabet ($\alpha$, $\beta$, $\gamma$, ...). Spatial indices on
an appropriately defined (three-dimensional Riemannian spacelike)
constant time slice of $\mathscr{M}$ are denoted using Latin letters
from the middle third of the alphabet in Roman font: in lower-case
($i$, $j$, $k$, ...) if they are abstract, and in upper-case ($I$,
$J$, $K$, ...) if a particular choice of coordinates $\{x^{I}\}_{I=1}^{3}$
has been made.

For any $n$-dimensional manifold $(\mathscr{U},\bm{g}_{\mathscr{U}}^{\,},\bm{\nabla}_{\mathscr{U}}^{\,})$ with metric $\bm{g}_{\mathscr{U}}^{\,}$ and compatible derivative operator $\bm{\nabla}_{\mathscr{U}}^{\,}$,
we denote its natural volume form by
\begin{equation}
\bm{\epsilon}_{\mathscr{U}}^{\,}=\sqrt{\left|{\rm det}\left(\bm{g}_{\mathscr{U}}^{\,}\right)\right|}\;{\rm d}x^{1}\wedge\cdots\wedge{\rm d}x^{n}\,.\label{eq:vol_form}
\end{equation}

Let $\mathscr{S}\simeq\mathbb{S}^{2}$ be any
(Riemannian) closed two-surface that is topologically a two-sphere. Latin letters from the middle
third of the alphabet in Fraktur font ($\mathfrak{i}$, $\mathfrak{j}$,
$\mathfrak{k}$, ...) are reserved for indices of tensors in $\mathscr{T}^{k}\,_{l}(\mathscr{S})$.
In particular, for $\mathbb{S}^{2}$ itself, $\mathfrak{S}_{\mathfrak{ij}}$
is the metric, $\mathfrak{D}_{\mathfrak{i}}$ the associated derivative
operator, and $\epsilon_{\mathfrak{ij}}^{\mathbb{S}^{2}}$ the volume
form; in standard spherical coordinates $\{\theta,\phi\},$ the latter
is simply given by 
\begin{equation}
\bm{\epsilon}_{\mathbb{S}^{2}}^ {}=\sin\theta\,{\rm d}\theta\wedge{\rm d}\phi\,.\label{eq:S2_volume_form}
\end{equation}

Contractions are indicated in the usual way in the abstract index
notation: \textit{e.g.}, $U^{a}V_{a}$ is the contraction of $\bm{U}$ and
$\bm{V}$. Equivalently, when applicable, we may simply use the ``dot
product'' in the index-free notation, \textit{e.g.} $U^{a}V_{a}=\bm{U}\cdot\bm{V}$,
$A_{ab}B^{ab}=\bm{A}:\bm{B}$, etc. We must keep in mind that such
contractions are to be performed using the metric of the space on
which the relevant tensors are defined. Additionally, often we find it convenient to denote the component (projection) of a tensor in a certain direction by simply replacing its pertinent abstract index therewith: \textit{e.g.}, we equivalently write $U^{a}V_{b}=\bm{U}\cdot\bm{V}=U_{\bm{V}}=V_{\bm{U}}$,
$A_{ab}U^{a}=A_{\bm{U}b}$, $A_{ab}U^{a}V^{b}=A_{\bm{U}\bm{V}}$, etc.
For any $(0,2)$-tensor $A_{ab}$ that is not a metric, we indicate its trace (in
non-boldface) as $A=A_{a}\,^{a}={\rm tr}(\bm{A})$. (If it is a metric, this notation is reserved, as usual, for its determinant rather than its trace.)

Finally, let $\mathscr{U}$ and $\mathscr{V}$ be any two diffeomorphic
manifolds and let $f:\mathscr{U}\rightarrow\mathscr{V}$ be a map
between them. This naturally defines a map between tensors on the
two manifolds, which we denote by $f_{*}:\mathscr{T}^{k}\,_{l}(\mathscr{U})\rightarrow\mathscr{T}^{k}\,_{l}(\mathscr{V})$
and its inverse $(f^{-1})_{*}=f^{*}:\mathscr{T}^{k}\,_{l}(\mathscr{V})\rightarrow\mathscr{T}^{k}\,_{l}(\mathscr{U})$ if it exists.
We generically refer to any map of this sort as a \emph{tensor transport}
\cite{felsager_geometry_2012}. It is simply the generalization to
arbitrary tensors of the pushforward $f_{*}:T\mathscr{U}\rightarrow T\mathscr{V}$
and pullback $f^{*}:T^{*}\mathscr{V}\rightarrow T^{*}\mathscr{U}$,
the action of which is defined in the standard way—see, \textit{e.g.}, Appendix
C of Ref. \cite{wald_general_1984}. (Note that our convention
for sub-/super-scripting the star is the generally more common one
used in geometry \cite{felsager_geometry_2012,lee_introduction_2002};
it is sometimes opposite to and sometimes congruous with that used
in the physics literature, \textit{e.g.}, Refs.~\cite{wald_general_1984} and~\cite{carroll_spacetime_2003} respectively).

\section{Setup: quasilocal conservation laws\label{sec:qf}}

Let $(\mathscr{M},\bm{g},\bm{\nabla})$ be any $(3+1)$-dimensional
spacetime such that, given any matter stress-energy-momentum
tensor $T_{ab}$, the Einstein equation, 
\begin{equation}
\bm{G}=\kappa\,\bm{T}\enspace\textrm{in}\enspace\mathscr{M}\,,\label{eq:Einstein_eqn}
\end{equation}
 holds.  In what follows, we introduce the concept of quasilocal frames \cite{epp_rigid_2009,epp_existence_2012,mcgrath_quasilocal_2012,epp_momentum_2013,mcgrath_post-newtonian_2014,mcgrath_rigid_2014,oltean_geoids_2016,oltean_geoids_2017} and describe the basic steps for their construction, as well as the energy and momentum conservation
laws associated therewith.  In Subsection \ref{ssec:qf_heuristic} we offer an heuristic idea of quasilocal frames before proceeding in Subsection \ref{ssec:qf_math} to present the full mathematical construction. Then in Subsection \ref{ssec:qf_by} we motivate and discuss the quasilocal stress-energy-momentum tensor used in this work, that is, the Brown-York tensor. Finally in Subsection \ref{ssec:qf_cons_laws} we review the formulation of quasilocal conservation laws using these ingredients.

\subsection{Quasilocal frames: heuristic idea\label{ssec:qf_heuristic}}

Before we enter into the technical details, we would like to offer
a heuristic picture and motivation for defining the concept of quasilocal
frames.

We would like to show how the GSF arises from general-relativistic conservation laws. For this, we require
first the embedding into our spacetime $\mathscr{M}$ of a worldtube
boundary $\mathscr{B}\simeq\mathbb{R}\times\mathbb{S}^{2}$. The worldtube interior
contains the system the dynamics of which we are
interested in describing. In principle, such a $\mathscr{B}$ can
be completely specified by choosing an appropriate \emph{radial function}
$r(x)$ on $\mathscr{M}$ and setting it equal to a non-negative constant
(such that the $r(x)={\rm const.}>0$ Lorentzian slices of $\mathscr{M}$
have topology $\mathbb{R}\times\mathbb{S}^{2}$). This would be analogous
to defining a (Riemannian, with topology $\mathbb{R}^{3}$) Cauchy
surface by the constancy of a time function $t(x)$ on $\mathscr{M}$.

However, this does not quite suffice. As we have briefly argued in
the introduction (and will shortly elaborate upon in greater technicality),
the conservation laws appropriate to GR ought to be quasilocal in
form, that is, involving stress-energy-momentum as boundary (not volume)
densities. One may readily assume that the latter are defined by a
quasilocal stress-energy-momentum tensor living on $\mathscr{B}$,
which we denote—for the moment, generally—by $\tau_{ab}$. (Later
we give an explicit definition, namely that of the Brown-York tensor,
for $\bm{\tau}$.)

To construct conservation laws, then, one would need to project this
$\bm{\tau}$ into directions on $\mathscr{B}$, giving quantities
such as energy or momenta, and then to consider their flux through
a portion of $\mathscr{B}$ (an interval of time along the worldtube
boundary). But in this case, we have to make clear what is meant by
the energy (``time-time'') and momenta (``time-space'') components
of $\bm{\tau}$ within $\mathscr{B}$, the changes in which we are
interested in studying. For this reason, additional constructions
are required.

In particular, what we need is a \emph{congruence} of observers with
respect to which projections of $\bm{\tau}$ yield stress-energy-momentum
quantities. Since $\bm{\tau}$ is only defined on $\mathscr{B}$,
this therefore needs to be a two-parameter family of (timelike) worldlines
the union of which is $\mathscr{B}$ itself. This is analogous to
how the integral curves of a time flow vector field (as in canonical
GR) altogether constitute (``fill up'') the entire spacetime $\mathscr{M}$,
except that there we are dealing with a three- (rather than
two-) parameter family of timelike worldlines. 

We refer to any set of observers, the worldlines of which form a two-parameter
family constituting $\mathscr{B}\simeq\mathbb{R}\times\mathbb{S}^{2}$,
as \emph{quasilocal} observers. A specification of such a $2$-parameter
family, equivalent to specifying the unit four-velocity $u^{a}\in T\mathscr{B}$
of these observers (the integral curves of which ``trace out'' $\mathscr{B}$),
is what is meant by a quasilocal frame.

With this, we can now meaningfully talk about projections of $\bm{\tau}$
into directions on $\mathscr{B}$ as stress-energy-momentum quantities.
For example, $\tau_{\bm{u}\bm{u}}$ may appear immediately suggestible
as a definition for the (boundary) energy density. Indeed, later we
take precisely this definition, and we will furthermore see how momenta
(the basis of the GSF problem) can be defined as well.

\subsection{Quasilocal frames: mathematical construction\label{ssec:qf_math}}

\begin{figure}
\begin{centering}
\includegraphics[scale=0.8]{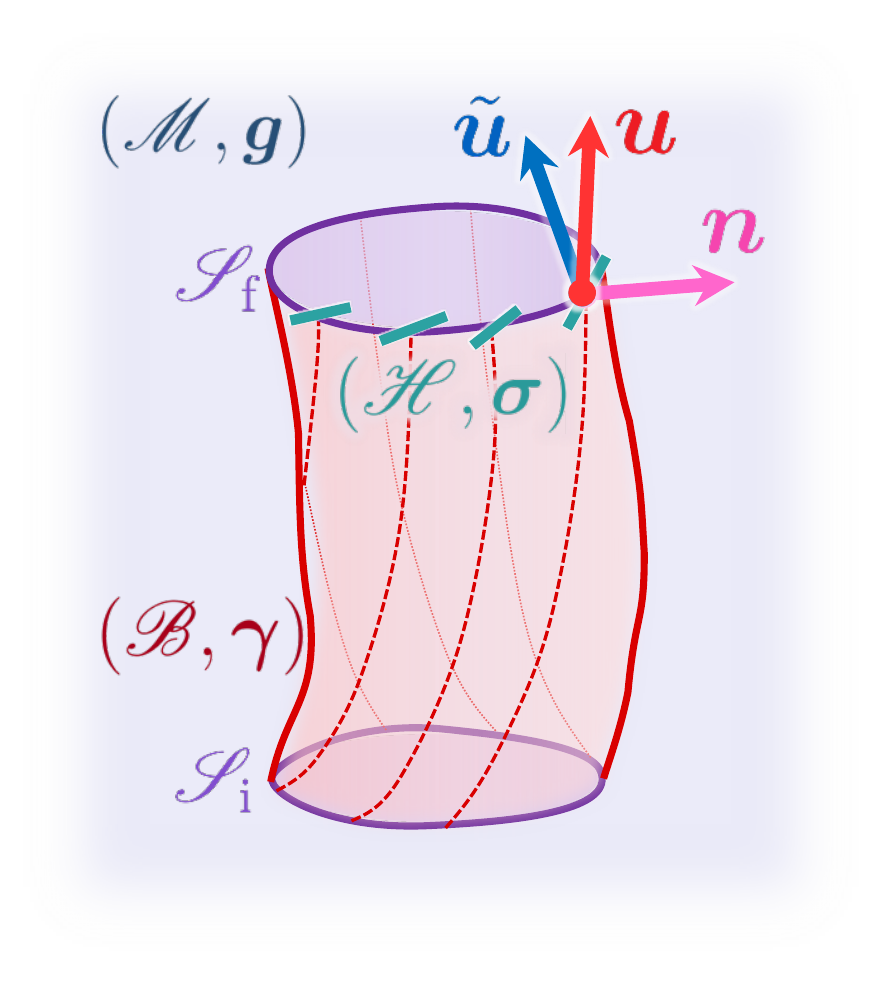}
\par\end{centering}
\caption{A portion of a quasilocal frame $(\mathscr{B};\bm{u})$ in a spacetime $\mathscr{M}$,
bounded by constant $t$ two-surfaces $\mathscr{S}_{\textrm{i}}$
and $\mathscr{S}_{\textrm{f}}$. In particular, $\mathscr{B}\simeq\mathbb{R}\times\mathbb{S}^{2}$ is the union of all integral curves (two-parameter family of timelike worldlines), depicted in the figure as dotted red lines, of the vector field $\bm{u}\in T\mathscr{B}$ which represents the unit four-velocity of quasilocal observers making up the congruence. The unit normal to $\mathscr{B}$ (in $\mathscr{M}$) is $\bm{n}$ and the normal to each constant $t$ slice $\mathscr{S}$ of $\mathscr{B}$ is $\tilde{\bm{u}}$ (not necessarily coincidental with $\bm{u}$). Finally, $\mathscr{H}$ (with induced metric $\bm{\sigma}$) is the two-dimensional subspace of $T\mathscr{B}$ consisting of the spatial vectors orthogonal to $\bm{u}$. Note that unlike $\mathscr{S}$, $\mathscr{H}$ need not be integrable (indicated in the figure by the failure of $\mathscr{H}$ to make a closed two-surface).\label{fig-qf}}
\end{figure}

Concordant with our discussion in the previous subsection, a quasilocal
frame (see Figure~\ref{fig-qf} for a graphical illustration of the construction) is defined as a two-parameter family of timelike worldlines
constituting the worldtube boundary (topologically $\mathbb{R}\times\mathbb{S}^{2}$)
of the history of a finite (closed) spatial three-volume in $\mathscr{M}$.
Let $u^{a}$ denote the timelike unit vector field tangent to these
worldlines. Such a congruence constitutes a submanifold of $\mathscr{M}$
that we call $\mathscr{B}\simeq\mathbb{R}\times\mathbb{S}^{2}$. Let
$n^{a}$ be the outward-pointing unit vector field normal to $\mathscr{B}$;
note that $\bm{n}$ is uniquely fixed once $\mathscr{B}$ is specified. There is
thus a Lorentzian metric $\bm{\gamma}$ (of signature $(-,+,+)$)
induced on $\mathscr{B}$, the components of which are given by 
\begin{equation}
\gamma_{ab}=g_{ab}-n_{a}n_{b}\,.\label{eq:gamma_ab}
\end{equation}
We denote the induced derivative operator compatible therewith by
$\bm{\mathcal{D}}$. To indicate that a topologically $\mathbb{R}\times\mathbb{S}^{2}$
submanifold $(\mathscr{B},\bm{\gamma},\bm{\mathcal{D}})$ of $\mathscr{M}$
is a quasilocal frame (that is to say, defined as a particular congruence
with four-velocity $\bm{u}$ as detailed above, and not just as an
embedded submanifold) in $\mathscr{M}$, we write $(\mathscr{B},\bm{\gamma},\bm{\mathcal{D}};\bm{u})$
or simply $(\mathscr{B};\bm{u})$.

Let $\mathscr{H}$ be the two-dimensional subspace of $T\mathscr{B}$ consisting of the ``spatial'' vectors orthogonal to $\bm{u}$. Let $\bm{\sigma}$ denote the two-dimensional (spatial) Riemannian
metric (of signature $(+,+)$) that projects tensor indices into $\mathscr{H}$,
and is induced on $\mathscr{B}$ by the choice of $\bm{u}$ (and thus
also $\bm{n}$), given by 
\begin{equation}
\sigma_{ab}=\gamma_{ab}+u_{a}u_{b}=g_{ab}-n_{a}n_{b}+u_{a}u_{b}\,.\label{eq:sigma_ab}
\end{equation}
The induced derivative operator compatible with $\bm{\sigma}$ is
denoted by $\bm{D}$. Let $\{x^{\mathfrak{i}}\}_{\mathfrak{i}=1}^{2}$
(written using Fraktur indices from the middle third of the Latin
alphabet) be spatial coordinates on $\mathscr{B}$ that label
the worldlines of the observers, and let $t$ be a time coordinate
on $\mathscr{B}$ such that surfaces of constant $t$, to which there
exists a unit normal vector that we denote by $\tilde{u}^{a}\in T\mathscr{B}$, foliate
$\mathscr{B}$ by closed spatial two-surfaces $\mathscr{S}$ (with
topology $\mathbb{S}^{2}$). Letting $N$ denote the lapse function
of $\bm{g}$, we have $\bm{u}=N^{-1}\partial/\partial t$.

Note that in general, $\mathscr{H}$ need not coincide
with the constant time slices $\mathscr{S}$. Equivalently, $\bm{u}$
need not coincide with $\tilde{\bm{u}}$. In general, there will be
a shift between them, such that 
\begin{equation}
\tilde{\bm{u}}=\tilde{\gamma}(\bm{u}+\bm{v})\,,\label{eq:u^tilde}
\end{equation}
where $v^{a}$ represents the spatial two-velocity of fiducial observers
that are at rest with respect to $\mathscr{S}$ as measured by our
congruence of quasilocal observers (the four-velocity of which is
$\bm{u}$), and $\tilde{\gamma}=1/\sqrt{1-\bm{v}\cdot\bm{v}}$ is
the Lorentz factor.

The specification of a quasilocal frame is thus equivalent to making
a particular choice of a two-parameter family of timelike worldlines comprising $\mathscr{B}$. There are,
a priori, three degrees of freedom (DoFs) available to us
for doing this. Heuristically, these can be regarded as corresponding
to the three DoFs in choosing the direction of $\bm{u}$—from which
$\bm{n}$ and all induced quantities are then computable. (Note that
$\bm{u}$ has four components, but one of the four is fixed by the
normalization requirement $\bm{u}\cdot\bm{u}=-1$, leaving three independent
direction DoFs.) Equivalently, we are in principle free to pick any
three geometrical conditions (along the congruence) to fix a quasilocal
frame. In practice, usually it is physically more natural, as well
as mathematically easier, to work with geometric quantities other
than $\bm{u}$ itself to achieve this. 

Yet, it is worth remarking that simply writing down three desired
equations (or conditions) to be satisfied by geometrical quantities
on $\mathscr{B}$ does not itself guarantee that, in general,
a submanifold $(\mathscr{B},\bm{\gamma},\bm{\mathcal{D}})$ obeying
those three particular equations will always exist—and, if
it does, that it will be the unique such submanifold—in an arbitrary
$(\mathscr{M},\bm{g},\bm{\nabla}$). Nevertheless, one choice of quasilocal
frame that is known to always exist (a claim we will qualify more
carefully in a moment) is that where the two-metric $\bm{\sigma}$
on $\mathscr{H}$ is ``rigid'' (or ``time
independent'')—these are called \emph{rigid quasilocal frames}.

Most of the past work on quasilocal frames has
in fact been done in the rigid case \cite{epp_rigid_2009,epp_existence_2012,mcgrath_quasilocal_2012,epp_momentum_2013,mcgrath_post-newtonian_2014,mcgrath_rigid_2014}.
We know however that other quasilocal frame choices are also possible,
such as \emph{geoids}---dubbed geoid quasilocal frames \cite{oltean_geoids_2016,oltean_geoids_2017}:
these are the general-relativistic generalization of ``constant gravitational
potential'' surfaces in Newtonian gravity. Regardless, the quasilocal
frame choice that we will mainly consider in this paper is the rigid
one (and we will be clear when this choice is explicitly enacted). 

Intuitively, the reason for this preference is that imposing in this
way the condition of ``spatial rigidity'' on $(\mathscr{B};\bm{u})$—a
two-dimensional (boundary) rigidity requirement, which unlike three-dimensional
rigidity, is permissible in GR—eliminates from the description of
the system any effects arising simply from the motion of the quasilocal
observers relative to each other. Thus, the physics of what is going
on inside the system (\textit{i.e.} the worldtube interior) is essentially all that affects its
dynamics. 

Technically, there is a further reason: a proof of the existence of
solutions—\textit{i.e.} the existence of a submanifold $\mathscr{B}\simeq\mathbb{R}\times\mathbb{S}^{2}$
in $\mathscr{M}$ that is also a quasilocal frame $(\mathscr{B};\bm{u})$—for
any spacetime $(\mathscr{M},\bm{g},\bm{\nabla})$ has up to now only
been fully carried out for rigid quasilocal frames\footnote{The idea of the proof is to explicitly construct the solutions order-by-order
in an expansion in the areal radius around an arbitrary worldline
in an arbitrary spacetime \cite{epp_existence_2012}.}. While, as we have commented, other quasilocal frame choices may
be generally possible in principle (and may be shown to be possible
to construct, case-by-case, in specific spacetimes—as we have done,
\textit{e.g.}, with geoid quasilocal frames \cite{oltean_geoids_2016,oltean_geoids_2017}),
they are as yet not rigorously guaranteed to exist in arbitrary spacetimes.

The quasilocal rigidity conditions can be stated in a number of ways.
Most generally, defining 
\begin{equation}
\theta_{ab}=\sigma_{ac}\sigma_{bd}\nabla^{c}u^{d}\label{eq:strain-rate}
\end{equation}
to be the \emph{strain rate tensor} of the congruence, they amount
to the requirement of vanishing expansion $\theta={\rm tr}(\bm{\theta})$
and shear $\theta_{\langle ab\rangle}=\theta_{(ab)}-\frac{1}{2}\theta\sigma_{ab}$,
\textit{i.e.} 
\begin{equation}
\theta=0=\theta_{\langle ab\rangle}\Leftrightarrow0=\theta_{(ab)}\,.\label{eq:RQF-conditions}
\end{equation}
In the adapted coordinates, these three conditions are expressible
as the vanishing of the time derivative of the two-metric on $\mathscr{H}$,
\textit{i.e.} $0=\partial_{t}\bm{\sigma}$. Both of these two equivalent mathematical
conditions, $\theta_{(ab)}=0=\partial_{t}\bm{\sigma}$, capture physically
the meaning of the quasilocal observers moving rigidly with
respect to each other (\textit{i.e.} the ``radar-ranging'' distances between
them does not change in time).

\subsection{The quasilocal stress-energy-momentum tensor\label{ssec:qf_by}}

Before we consider the formulation of conservation laws with the use
of quasilocal frames (from which our analysis of the GSF will eventually emerge),
we wish to address in a bit more detail an even more fundamental question: what are conservation
laws in GR actually supposed to be about? At the most basic level, they should
express changes (over time) in some appropriately defined notion of
energy-momentum. As we are interested in gravitational
systems (and specifically, those driven by the effect of the GSF),
this energy-momentum must include that of the gravitational field, in addition
to that of any matter fields if present.

Hence, we may assert from the outset that it does not make much sense
in GR to seek conservation laws based solely on the matter stress-energy-momentum tensor
$\bm{T}$, such as Eq.~(\ref{eq:local_cons}). It is evident that these would, by construction, account
for matter only---leaving out gravitational effects in general (which
could exist in the complete absence of matter, \textit{e.g.} gravitational
waves), and thus the GSF in particular. What is more, such conservation
laws are logically inconsistent from a general-relativistic point
of view: a non-vanishing $\bm{T}$ implies a non-trivial gravitational
field (through the Einstein equation) and thus a necessity of taking
into account that field along with the matter one(s) for a proper
accounting of energy-momentum transfer. A further technical problem is also that
the formulation of conservation laws of this sort is typically predicated
upon the existence of Killing vector fields or other types of symmetry generators in $\mathscr{M}$, which
one does not have in general---and which do not exist in spacetimes
pertinent for the GSF problem in particular. 

We are therefore led to ask: how can we meaningfully define a total---\emph{gravity
plus matter---}stress-energy-momentum tensor in GR? It turns out that the precise answer
to this question, while certainly not intractable, is unfortunately
also not unique---or at least, it lacks a clear consensus among relativists,
even today. See, \textit{e.g.}, Refs.~\cite{jaramillo_mass_2011,szabados_quasi-local_2004}
for reviews of the variety of proposals that have been put forward
towards addressing this question. Nonetheless, for reasons already touched upon and to be elaborated presently, what is clear and generally accepted is
that such a tensor cannot be local in nature (as $\bm{T}$ is), and
for this reason is referred to as \emph{quasilocal}.

Let $\tau_{ab}$ denote this quasilocal, total (matter plus gravity) stress-energy-momentum tensor that we
eventually seek to use for our conservation laws. It has long been
understood \cite{misner_gravitation_1973} that whatever the notion
of ``gravitational energy-momentum'' (defined by $\bm{\tau}$) might
mean, it is not something localizable: in other words, there is no
way of meaningfully defining an ``energy-momentum volume density''
for gravity. This is, ultimately, due to the equivalence principle:
locally, one can always find a reference frame in which all local
``gravitational fields'' (the connection coefficients), and thus
any notion of ``energy-momentum volume density'' associated therewith,
disappear. The remedy is to make $\bm{\tau}$ quasilocal: meaning that, rather
than volume densities, it should define surface densities (of energy,
momentum etc.)---a type of construction which \emph{is} mathematically realizable
and physically sensible in general.

The specific choice we make for how to define this total (matter plus
gravity), quasilocal energy-momentum tensor $\bm{\tau}$ is the so-called Brown-York
tensor, first put forward by the authors in Ref. \cite{brown_quasilocal_1993}; see also Ref. \cite{brown_action_2002}
for a detailed review. This proposal was based originally
upon a Hamilton-Jacobi analysis; here we will offer a simpler argument
for its definition, sketched out initially in Ref. \cite{epp_momentum_2013}.

Consider the standard gravitational action $S_{\textrm{G}}$ for a
spacetime volume $\mathscr{V}\subset\mathscr{M}$ such that 
$\partial \mathscr{V}=\mathscr{B}\simeq\mathbb{R}\times\mathbb{S}^{2}$ is a worldtube
boundary as in the previous subsection (possibly constituting a quasilocal
frame, but not necessarily). This action is given by the sum of two
terms, a bulk and a boundary term respectively:
\begin{equation}
S_{\textrm{G}}\left[\bm{g}\right]=S_{\textrm{EH}}\left[\bm{g}\right]+S_{\textrm{GHY}}\left[\bm{\gamma},\bm{n}\right]\,.\label{eq:action_gravitational}
\end{equation}
In particular, the first is the Einstein-Hilbert bulk term, 
\begin{equation}
S_{\textrm{EH}}\left[\bm{g}\right]=\frac{1}{2\kappa}\intop_{\mathscr{V}}\bm{\epsilon}_{\mathscr{M}}^{\,}\,R\,,
\end{equation}
and the second is the Gibbons-Hawking-York boundary term \cite{york_role_1972,gibbons_action_1977},
\begin{equation}
S_{\textrm{GHY}}\left[\bm{\gamma},\bm{n}\right]=-\frac{1}{\kappa}\intop_{\partial\mathscr{V}}\bm{\epsilon}_{\mathscr{B}}^{\,}\,\Theta\,.
\end{equation}
Here, $\bm{\epsilon}_{\mathscr{M}}^{\,}={\rm d}^{4}x\sqrt{-g}$ is
the volume form on $\mathscr{M}$ with $g={\rm \det}(\bm{g})$, $\bm{\epsilon}_{\mathscr{B}}^{\,}={\rm d}^{3}x\sqrt{-\gamma}$
is the volume form on $\mathscr{B}$ with $\gamma={\rm \det}(\bm{\gamma})$,
and $\Theta={\rm tr}(\bm{\Theta})$ is the trace of the extrinsic
curvature $\Theta_{ab}=\gamma_{ac}\nabla^{c}n_{b}$ of $\mathscr{B}$
in $\mathscr{M}$. Additionally, the matter action $S_{\textrm{M}}$
for any set of matter fields $\Psi$ described by a Lagrangian $L_{\textrm{M}}$
is 
\begin{equation}
S_{\textrm{M}}\left[\Psi\right]=\intop_{\mathscr{V}}\bm{\epsilon}_{\mathscr{M}}^{\,}\,L_{\textrm{M}}\left[\Psi\right]\,.\label{eq:action_matter}
\end{equation}

The definition of the total (quasilocal) stress-energy-momentum tensor $\bm{\tau}$
for gravity plus matter can be obtained effectively in the same way
as that of the (local) stress-energy-momentum tensor $\bm{T}$ for matter
alone—from the total action in Eq.~(\ref{eq:action_gravitational}) rather
than just, respectively, the matter action in Eq.~(\ref{eq:action_matter}).
In particular, $\bm{T}$ is defined by computing the variation $\delta$
(with respect to the spacetime metric) of the matter action: 
\begin{equation}
\delta S_{\mathrm{M}}\left[\Psi\right]=-\frac{1}{2}\intop_{\mathscr{V}}\bm{\epsilon}_{\mathscr{M}}^{\,}\,T_{ab}\delta g^{ab}\,.\label{eq:dS_matter}
\end{equation}
In other words, one defines the matter stress-energy-momentum tensor
as the functional derivative, 
\begin{equation}
T_{ab}=-\frac{2}{\sqrt{-g}}\frac{\delta S_{\textrm{M}}}{\delta g^{ab}}\,.\label{eq:T_ab_definition}
\end{equation}
The definition of the Brown-York tensor follows completely analogously,
except that now gravity is also included. That is, for the total action
of gravity (minimally) coupled to matter, 
\begin{equation}
S_{\textrm{G}+\textrm{M}}\left[\bm{g},\Psi\right]=S_{\textrm{G}}\left[\bm{g}\right]+S_{\textrm{M}}\left[\Psi\right]\,,\label{eq:GM-action}
\end{equation}
we have that the metric variation is: 
\begin{align}
\delta S_{\textrm{G}+\textrm{M}}\left[\bm{g},\Psi\right]=\, & \frac{1}{2}\Bigg\{\intop_{\mathscr{V}}\bm{\epsilon}_{\mathscr{M}}\,\left(\frac{1}{\kappa}G_{ab}-T_{ab}\right)\delta g^{ab}\nonumber \\
 & -\intop_{\partial\mathscr{V}}\bm{\epsilon}_{\mathscr{B}}^ {}\,\left(-\frac{1}{\kappa}\Pi_{ab}\right)\delta\gamma^{ab}\Bigg\}\label{eq:dS_G+M_1}\\
=\, & -\frac{1}{2}\intop_{\partial\mathscr{V}}^{\,}\bm{\epsilon}_{\mathscr{B}}^ {}\,\tau_{ab}\delta\gamma^{ab}\,.\label{eq:dS_G+M_2}
\end{align}
In the equality of Eq.~(\ref{eq:dS_G+M_1}), $\bm{\Pi}$ is the canonical
momentum of $(\mathscr{B},\bm{\gamma},\bm{\mathcal{D}})$, given by
$\bm{\Pi}=\bm{\Theta}-\Theta\bm{\gamma}$. It follows from direct
computation using Eqs.~(\ref{eq:action_gravitational}), (\ref{eq:action_matter})
and (\ref{eq:dS_matter}); for a review of this derivation carefully
accounting for the boundary term see, \textit{e.g.}, Chapter 12 of Ref.~\cite{padmanabhan_gravitation:_2010}.
In the equality of Eq.~(\ref{eq:dS_G+M_2}), the Einstein equation
$\bm{G}=\kappa\bm{T}$ has been invoked (in other words, we impose the Einstein equation to be satisfied in the bulk),
thus leading to the vanishing of the bulk term; meanwhile in the boundary
term, a gravity plus matter stress-energy-momentum tensor $\bm{\tau}$ (the
Brown-York tensor) has been defined in direct analogy with the definition
of the matter energy-momentum tensor $\bm{T}$ in Eq.~(\ref{eq:dS_matter}).
Hence just as Eq.~(\ref{eq:dS_matter}) implies Eq.~(\ref{eq:T_ab_definition}),
Eq.~(\ref{eq:dS_G+M_2}) implies 
\begin{equation}
\bm{\tau}=-\frac{1}{\kappa}\bm{\Pi}\,.\label{eq:tau_ab_definition}
\end{equation}
Henceforth, $\bm{\tau}$ refers strictly to this (Brown-York) quasilocal stress-energy-momentum tensor of Eq.~(\ref{eq:tau_ab_definition}), and not to any other definition.

It is useful to decompose  $\bm{\tau}$ in a similar way as is
ordinarily done with $\bm{T}$, so we define: 
\begin{align}
\mathcal{E}=\, & u^{a}u^{b}\tau_{ab}\,,\\
\mathcal{P}^{a}=\, & -\sigma^{ab}u^{c}\tau_{bc}\,,\\
\mathcal{S}^{ab}=\, & -\sigma^{ac}\sigma^{bd}\tau_{cd}\,,
\end{align}
as the quasilocal energy, momentum and stress, respectively, with
units of energy per unit area, momentum per unit area and force per
unit length. Equivalently, 
\begin{equation}
\tau^{ab}=u^{a}u^{b}\mathcal{E}+2u^{(a}\mathcal{P}^{b)}-\mathcal{S}^{ab}\,.
\end{equation}

\subsection{Conservation laws\label{ssec:qf_cons_laws}}

The construction of general conservation laws from $\bm{\tau}$
was first achieved in Refs. \cite{mcgrath_quasilocal_2012,epp_momentum_2013}, and proceeds along the following lines.  Let $\boldsymbol{\psi}\in T\mathscr{B}$ be an arbitrary vector field
in $\mathscr{B}$. We begin by considering a projection of $\bm{\Pi}$
in the direction of $\bm{\psi}$ (in one index), \textit{i.e.} $\Pi^{ab}\psi_{b}$, and computing its divergence in $\mathscr{B}$. By using the Leibnitz rule, we simply have
\begin{equation}
\mathcal{D}_{a}\left(\Pi^{ab}\psi_{b}\right)=\left(\mathcal{D}_{a}\Pi^{ab}\right)\psi_{b}+\Pi^{ab}\left(\mathcal{D}_{a}\psi_{b}\right)\,.
\end{equation}
Next, we integrate this equation over a portion $\Delta\mathscr{B}$
of $\mathscr{B}$ bounded by initial and final constant $t$ surfaces
$\mathscr{S}_{\textrm{i}}$ and $\mathscr{S}_{\textrm{f}}$, as depicted in Figure~\ref{fig-qf}. On the
resulting LHS we apply Stokes' theorem, and on the first term on the
RHS we use the Gauss-Codazzi identity: $\mathcal{D}_{a}\Pi^{ab}=n_{a}\gamma^{b}_{\;\;c}G^{ac}$.
Thus, using the notation for tensor projections in certain directions introduced in Section \ref{ssec:notation} for ease of readability (\textit{e.g.}, $G_{ab}n^{a}\psi^{b}=G_{\bm{n}\bm{\psi}}$ and similarly for other contractions), we obtain:
\begin{equation}
\intop_{\mathscr{S}_{\textrm{f}}-\mathscr{S}_{\textrm{i}}}\!\!\!\!\bm{\epsilon}^{\,}_{\mathscr{S}}\,\Pi_{\bm{\tilde{u}}\bm{\psi}}=-\intop_{\Delta\mathscr{B}}\!\bm{\epsilon}^{\,}_{\mathscr{B}}\,\left(G_{\bm{n}\bm{\psi}}+\Pi^{ab}\mathcal{D}_{a}\psi_{b}\right)\,,\label{eq:geometrical_identity}
\end{equation}
where $\bm{\epsilon}^{\,}_{\mathscr{S}}$ denotes the volume form on the constant time closed two-surfaces $\mathscr{S}$,
and we have used the notation: $\int_{\mathscr{S}_{\textrm{f}}-\mathscr{S}_{\textrm{i}}}(\cdot)=\int_{\mathscr{S}_{\textrm{f}}}(\cdot)-\int_{\mathscr{S}_{\textrm{i}}}(\cdot)$.
We also remind the reader that $\tilde{\bm{u}}$ represents the unit normal to each constant time closed two-surface, which in general need not coincide with the quasilocal observers' four-velocity $\bm{u}$ but is related to it by a Lorentz transformation, Eq.~(\ref{eq:u^tilde}); see also Figure~\ref{fig-qf}.

We stress that so far, Eq.~(\ref{eq:geometrical_identity}) is a purely
geometrical identity, completely general for any Lorentzian manifold
$\mathscr{M}$; in other words, thus far we have said nothing about
physics.

Now, to give this identity physical meaning, we invoke the definition
of the Brown-York tensor in Eq.~(\ref{eq:tau_ab_definition}) (giving
the boundary extrinsic geometry its meaning as stress-energy-momentum) as well as the
Einstein equation [Eq.~(\ref{eq:Einstein_eqn})], giving the spacetime
curvature its meaning as the gravitational field. With these, Eq.~(\ref{eq:geometrical_identity}) turns into:
\begin{equation}
\intop_{\mathscr{S}_{\textrm{f}}-\mathscr{S}_{\textrm{i}}}\!\!\!\!\bm{\epsilon}^{\,}_{\mathscr{S}}\,\tilde{\gamma}\left(\tau_{\bm{u}\bm{\psi}}+\tau_{\bm{v}\bm{\psi}}\right)=\intop_{\Delta\mathscr{B}}\!\bm{\epsilon}^{\,}_{\mathscr{B}}\,\left(T_{\bm{n}\bm{\psi}}-\tau^{ab}\mathcal{D}_{(a}\psi_{b)}\right)\,.\label{eq:cons_law_general}
\end{equation}
On the LHS we have inserted the relation $\tilde{\bm{u}}=\tilde{\gamma}(\bm{u}+\bm{v})$,
with $v^{a}$ representing the spatial two-velocity of fiducial observers
that are at rest with respect to $\mathscr{S}$ (the hypersurface-orthogonal
four-velocity of which is $\tilde{\bm{u}}$) as measured by our congruence
of quasilocal observers (the four-velocity of which is $\bm{u}$), and $\tilde{\gamma}=1/\sqrt{1-\bm{v}\cdot\bm{v}}$
is the Lorentz factor.

Observe that Eq.~(\ref{eq:cons_law_general}) expresses the change of
some component of the quasilocal stress-energy-momentum tensor integrated over two different $t = const.$ closed two-surfaces $\mathscr{S}$ as a flux through
the worldtube boundary $\Delta\mathscr{B}$ between them. The identification
of the different components of $\bm{\tau}$ as the various components
of the total energy-momentum of the system thus leads to the understanding of Eq.~(\ref{eq:cons_law_general}) as a general conservation law for the
system contained inside of $\Delta\mathscr{B}$. Thus, depending on
our particular choice of $\bm{\psi}\in T\mathscr{B}$, Eq.~(\ref{eq:cons_law_general})
will represent a conservation law for the total energy, momentum,
or angular momentum of this system~\cite{epp_momentum_2013}. 

Let us now assume that $(\mathscr{B};\bm{u})$ is a rigid quasilocal frame. If we choose
$\bm{\psi}=\bm{u}$, then Eq.~(\ref{eq:cons_law_general}) becomes the
energy conservation law:
\begin{equation}
\intop_{\mathscr{S}_{\textrm{f}}-\mathscr{S}_{\textrm{i}}}\!\!\!\!\bm{\epsilon}^{\,}_{\mathscr{S}}\,\tilde{\gamma}\left(\mathcal{E}-\mathcal{P}_{\bm{v}}\right)=\intop_{\Delta\mathscr{B}}\!\bm{\epsilon}^{\,}_{\mathscr{B}}\,\left(T_{\bm{n}\bm{u}}-\bm{\alpha}\cdot\bm{\mathcal{P}}\right)\,,\label{eq:cons_law_E}
\end{equation}
where $\alpha^{a}=\sigma^{ab}a_{b}$ is the $\mathscr{H}$ projection
of the acceleration of the quasilocal observers, defined by $a^{a}=\nabla_{\bm{u}}u^{a}$. 

Now suppose, on the other hand, that we instead choose $\bm{\psi}=-\bm{\phi}$
where $\bm{\phi}\in \mathscr{H}$ is orthogonal to $\bm{u}$ (with the minus sign introduced for convenience), and
represents a stationary conformal Killing vector field with
respect to $\bm{\sigma}$. This means that $\bm{\phi}$ is chosen such that it satisfies the conformal Killing equation, $\mathcal{L}_{\bm{\phi}}\bm{\sigma}=(\bm{D}\cdot\bm{\phi})\bm{\sigma}$, with $\mathcal{L}$ the Lie derivative and $\bm{D}$ the derivative on $\mathscr{H}$ (compatible with $\bm{\sigma}$). A set of six such conformal Killing vectors always exist: three for translations
and three for rotations, respectively generating the action of boosts
and rotations of the Lorentz group on the two-sphere \cite{epp_momentum_2013}. The idea, then, is that the contraction of these vectors with the quasilocal momentum integrated over a constant-time topological two-sphere boundary expresses, respectively, the total linear and angular momentum (in the three ordinary spatial directions each) at that time instant. Thus, Eq.~(\ref{eq:cons_law_general})
becomes the (respectively, linear and angular) momentum conservation
law:
\begin{multline}
\intop_{\mathscr{S}_{\textrm{f}}-\mathscr{S}_{\textrm{i}}}\!\!\!\!\bm{\epsilon}^{\,}_{\mathscr{S}}\,\tilde{\gamma}\left(\mathcal{P}_{\bm{\phi}}+\mathcal{S}_{\bm{v}\bm{\phi}}\right)\\
=-\intop_{\Delta\mathscr{B}}\!\bm{\epsilon}^{\,}_{\mathscr{B}}\,\left(T_{\bm{n}\bm{\phi}}+\mathcal{E}\alpha_{\bm{\phi}}+2\nu\epsilon_{ab}\mathcal{P}^{a}\phi^{b}+{\rm P}\bm{D}\cdot\bm{\phi}\right)\,,\label{eq:cons_law_Pa}
\end{multline}
where $\nu=\frac{1}{2}\epsilon^{ab}_{\mathscr{H}}\mathcal{D}_{a}u_{b}$ is the twist
of the congruence (with $\epsilon_{ab}^{\mathscr{H}}=\epsilon_{abcd}^{\mathscr{M}}u^{c}n^{d}$ the induced volume form on $\mathscr{H}$), and ${\rm P}=\frac{1}{2}\bm{\sigma}:\bm{\mathcal{S}}$
is the quasilocal pressure (force per unit length) between the worldlines
of $\mathscr{B}$. We remark that the latter can be shown to satisfy
the very useful general identity (which we will expediently invoke
in our later calculations):
\begin{equation}
\mathcal{E}-2{\rm P}=\frac{2}{\kappa}a_{\bm{n}}\,.\label{eq:E-P_relation}
\end{equation}

An analysis of the gravitational self-force problem should consider the conservation law in Eq.~(\ref{eq:cons_law_Pa}) for \emph{linear momentum}. Thus, we will use the fact, described in greater detail in Appendix \ref{sec:A}, that the conformal Killing vector $\bm{\phi}\in \mathscr{H}$ for linear momentum admits a multipole decomposition
of the following form:
\begin{align}
\phi^{\mathfrak{i}}=\, & \frac{1}{r}\,D^{\mathfrak{i}}\left(\Phi^{I}r_{I}+\Phi^{IJ}r_{I}r_{J}+\cdots\right)\label{eq:phi_multipole}\\
=\, & \frac{1}{r}\left(\Phi^{I}\mathfrak{B}_{I}^{\mathfrak{i}}+2\Phi^{IJ}\mathfrak{B}_{I}^{\mathfrak{i}}r_{J}+\cdots\right)\,,
\end{align}
with the dots indicating higher harmonics. Here, $r$ is the area radius of the quasilocal frame (such that $\mathscr{B}$
is a constant $r$ hypersurface in $\mathscr{M}$), $r^{I}$ denotes
the the standard direction cosines of a radial unit vector in $\mathbb{R}^{3}$
and $\mathfrak{B}_{I}^{\mathfrak{i}}=\partial^{\mathfrak{i}}r_{I}$
are the boost generators on the two-sphere. See Appendix \ref{sec:A} for a detailed
discussion regarding conformal Killing vectors and the two-sphere. In spherical coordinates
$\{\theta,\phi\}$, we have $r^{I}=(\sin\theta\cos\phi,\sin\theta\sin\phi,\cos\theta)$.
Thus Eq.~(\ref{eq:phi_multipole}) gives us a decomposition of $\bm{\phi}$
in terms of multipole moments, with the $\ell=1$ coefficients $\Phi^{I}$
simply representing vectors in $\mathbb{R}^{3}$ in the direction of which
we are considering the conservation law.

\section{General derivation of the gravitational self-force from quasilocal conservation laws\label{sec:general-analysis}}

In this section, we will show how the GSF is a general
consequence of the momentum conservation law in Eq.~(\ref{eq:cons_law_Pa})
for any system which is sufficiently localized. By that, we mean something
very simple: taking the $r\rightarrow0$ limit of a quasilocal frame around the
moving object which is treated as ``small'', \textit{i.e.} as a formal perturbation
about some background. No further assumptions are for the moment needed.
In particular, we do not even need to enter into the precise details
of how to specify the perturbation family for this problem; that will
be left to the following section, where we will carefully define and
work with the family of perturbed spacetimes typically employed for
applications of the GSF.

We first review the basic
formulation of perturbation theory in GR in Subsection \ref{ssec:general_perturbations}. While this material is well-known, we find it useful to include it here both for establishing notation as well as carefully defining the concepts that we need to work with at an adequate level of rigour. Then in Subsection \ref{ssec:general_egsf}, we show that the
first-order perturbation of the momentum conservation law in Eq.~(\ref{eq:cons_law_Pa})
always contains the GSF, and that it dominates the dynamics for localized systems.

\subsection{Perturbation theory in GR\label{ssec:general_perturbations}}

Our exposition of perturbation theory in this subsection follows closely
the treatment of Ref. \cite{bruni_perturbations_1997}. (See also Chapter 7 of Ref. \cite{wald_general_1984} for a simpler treatment of this topic but following the same philosophy.) 

\begin{figure}
\begin{centering}
\includegraphics[scale=0.7]{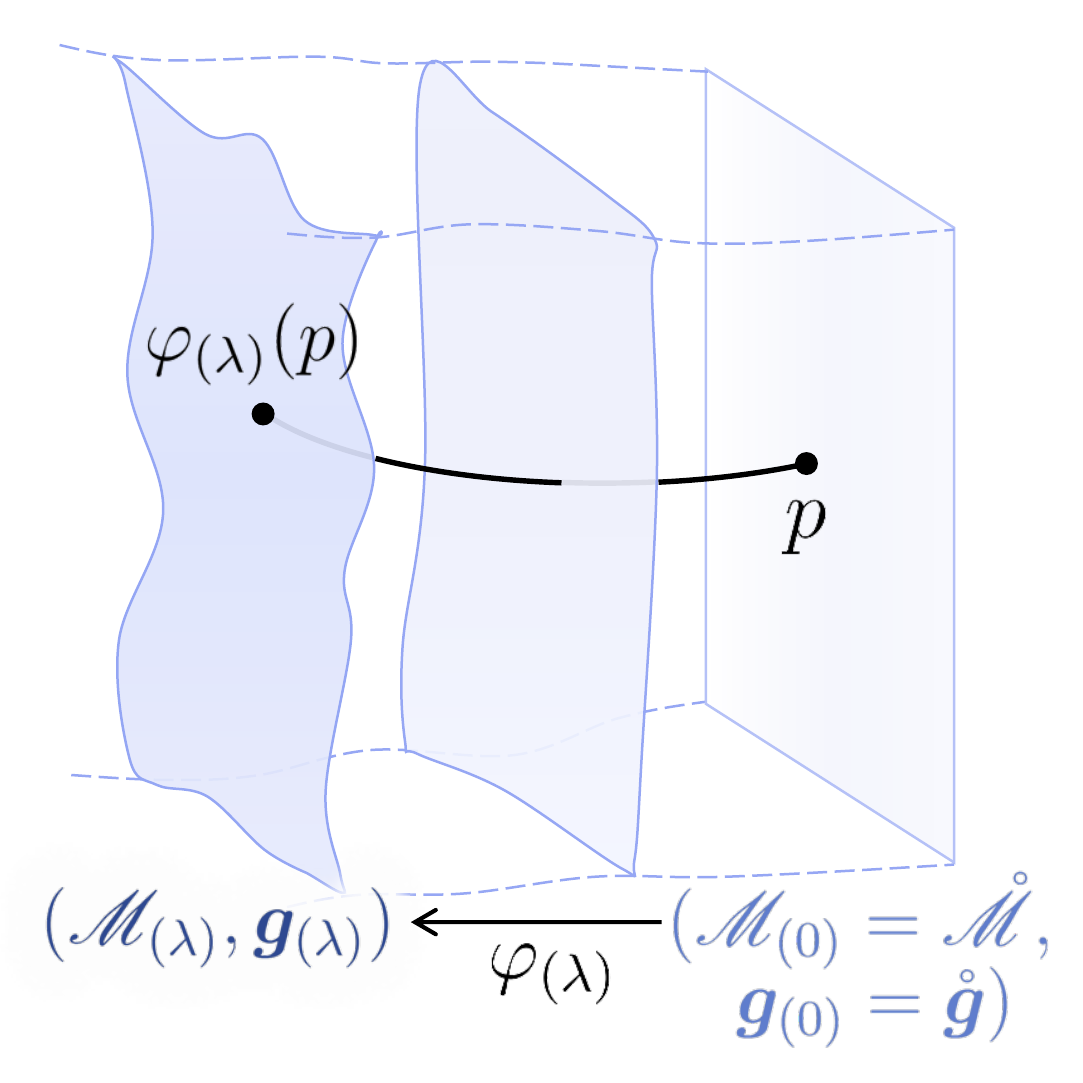}
\par\end{centering}
\caption{Representation of a one-parameter family of spacetimes  $\{\mathscr{M}_{(\lambda)}\}_{\lambda\geq0}$ used for perturbation theory. Each of the $\mathscr{M}_{(\lambda)}$ are depicted visually in $(1+1)$ dimensions,
as leaves of a (five-dimensional) product manifold $\mathscr{N}=\mathscr{M}_{(\lambda)}\times\mathbb{R}$, with the coordinate $\lambda\geq 0$ representing the perturbative expansion parameter.
A choice of a map (or gauge) $\varphi_{(\lambda)}:\mathring{\mathscr{M}}\rightarrow\mathscr{M}_{(\lambda)}$
gives us a way of identifying any point $p\in\mathring{\mathscr{M}}=\mathscr{M}_{(0)}$
on the background to one on some perturbed ($\lambda>0$)
spacetime, \textit{i.e.} $p\protect\mapsto\varphi_{(\lambda)}(p)$.}\label{fig-perturbations}
\end{figure}

Perturbation theory in GR is best made sense of from the point of
view of ``stacked'' manifolds off some known background. To
be more precise, let $\lambda\geq0$ represent our perturbation parameter.
It is a purely formal parameter, in the sense that it should be set
equal to $1$ at the end of any computation and serves only to indicate
the order of the perturbation. The idea, then, is to define a one-parameter
family of spacetimes $\{(\mathscr{M}_{(\lambda)},\bm{g}_{(\lambda)},\bm{\nabla}_{(\lambda)})\}_{\lambda\geq0}$,
where $\bm{\nabla}_{(\lambda)}$ is the connection compatible with
the metric $\bm{g}_{(\lambda)}$ in $\mathscr{M}_{(\lambda)}$, $\forall\lambda\geq0$,
such that $(\mathscr{M}_{(0)},\bm{g}_{(0)},\bm{\nabla}_{(0)})=(\mathring{\mathscr{M}},\mathring{\bm{g}},\mathring{\bm{\nabla}})$
is a known, exact spacetime---the \emph{background}. See Fig. \ref{fig-perturbations} for a visual depiction. For notational
convenience, any object with a sub-scripted ``$(0)$'' (from a one-parameter
perturbative family) is equivalently written with an overset ``$\circ$''
instead. For the GSF problem, $\mathring{\mathscr{M}}$ is usually
the Schwarzschild-Droste or Kerr spacetime. Then, one should establish a
way of smoothly relating the elements of this one-parameter family
(between each other) such that calculations on any $\mathscr{M}_{(\lambda)}$
for $\lambda>0$---which may be, in principle, intractable analytically---can
be mapped to calculations on $\mathring{\mathscr{M}}$ in the form
of infinite (Taylor) series in $\lambda$---which, provided $\mathring{\mathscr{M}}$
is chosen to be a known, exact spacetime, become tractable, order-by-order,
in $\lambda$.

Thus, one begins by defining a (five-dimensional, Lorentzian) product
manifold
\begin{equation}
\mathscr{N}=\mathscr{M}_{(\lambda)}\times\mathbb{R}^{\geq}\,,\label{eq:prodmfld}  
\end{equation}
the natural differentiable structure of which is given simply by the
direct product of those on $\mathscr{M}_{(\lambda)}$ and the non-negative real numbers (labeling the perturbation parameter), $\mathbb{R}^{\geq}=\{\lambda\in\mathbb{R}|\lambda\geq0\}$.
For any one-parameter family of $(k,l)$-tensors $\{\bm{A}_{(\lambda)}\}_{\lambda\geq0}$
such that $\bm{A}_{(\lambda)}\in\mathscr{T}^{k}\,_{l}(\mathscr{M}_{(\lambda)})$,
$\forall\lambda\geq0$, we define $\bm{\mathsf{A}}\in\mathscr{T}^{k}\,_{l}(\mathscr{N})$
by the relation $\mathsf{A}^{\alpha_{1}\cdots\alpha_{k}}\,_{\beta_{1}\cdots\beta_{l}}(p,\lambda)=A_{(\lambda)}^{\alpha_{1}\cdots\alpha_{k}}\,_{\beta_{1}\cdots\beta_{l}}(p)$,
$\forall p\in\mathscr{M}_{(\lambda)}$ and $\forall\lambda\geq0$.
Henceforth any such tensor living on the product manifold will be
denoted in serif font---instead of Roman font, which remains reserved
for tensors living on $(3+1)$-dimensional spacetimes. Furthermore,
any spacetime tensor (except for volume forms) or operator written
without a sub- or super-scripted $(\lambda)$ lives on $\mathring{\mathscr{M}}$.
Conversely, any tensor (except for volume forms) or operator living
on $\mathscr{M}_{(\lambda)}$, $\forall\lambda>0$, is indicated via
a sub- or (equivalently, if notationally more convenient) super-scripted
$(\lambda)$, \textit{e.g.} $\bm{A}_{(\lambda)}=\bm{A}^{(\lambda)}\in\mathscr{T}^{k}\,_{l}(\mathscr{M}_{(\lambda)})$
is always tensor in $\mathscr{M}_{(\lambda)}$. The volume form of
any (sub-)manifold $\mathscr{U}$ is always simply denoted by the
standard notation $\bm{\epsilon}_{\mathscr{U}}$ (and is always understood
to live on $\mathscr{U}$), as in Eq. (\ref{eq:vol_form}).

Let $\Phi_{(\lambda)}^{\bm{\mathsf{X}}}:\mathscr{N}\rightarrow\mathscr{N}$
be a one-parameter group of diffeomorphisms generated by a vector
field $\bm{\mathsf{X}}\in T\mathscr{N}$. (That is to say, the integral
curves of $\bm{\mathsf{X}}$ define a flow on $\mathscr{N}$ which
connects any two leaves of the product manifold.) For notational convenience,
we denote its restriction to maps from the background to a particular
perturbed spacetime (identified by a particular value of $\lambda>0$)
as 
\begin{align}
\varphi_{(\lambda)}^{\bm{\mathsf{X}}}=\Phi_{(\lambda)}^{\bm{\mathsf{X}}}|_{\mathring{\mathscr{M}}}:\mathring{\mathscr{M}}\, & \rightarrow\mathscr{M}_{(\lambda)}\label{eq:phi_map}\\
p\, & \mapsto\varphi_{(\lambda)}^{\bm{\mathsf{X}}}\left(p\right)\,.
\end{align}

The choice of $\bm{\mathsf{X}}$---equivalently, the choice of $\varphi_{(\lambda)}^{\bm{\mathsf{X}}}$---is
not unique; there exists freedom in choosing it, and for this reason,
$\bm{\mathsf{X}}$---equivalently, $\varphi_{(\lambda)}^{\bm{\mathsf{X}}}$---is
referred to as the \emph{perturbative gauge}. We may work with any
different gauge choice $\bm{\mathsf{Y}}$ generating a different map
$\varphi_{(\lambda)}^{\bm{\mathsf{Y}}}:\mathring{\mathscr{M}}\rightarrow\mathscr{M}_{(\lambda)}$.
If we do not need to render the issue of gauge specification explicit,
we may drop the superscript and, instead of $\varphi_{(\lambda)}^{\bm{\mathsf{X}}}$,
simply write $\varphi_{(\lambda)}$.

Consider now the transport under $\varphi_{(\lambda)}^{\bm{\mathsf{X}}}$
of any tensor $\bm{A}_{(\lambda)}\in\mathscr{T}^{k}\,_{l}(\mathscr{M}_{(\lambda)})$
from a perturbed spacetime to the background manifold. We always denote
the transport of any such tensor by simply dropping the $(\lambda)$
sub- or super-script and optionally including a superscript to indicate
the gauge---that is, $\forall\bm{A}_{(\lambda)}\in\mathscr{T}^{k}\,_{l}(\mathscr{M}_{(\lambda)})$,
\begin{equation}
(\varphi_{(\lambda)}^{\bm{\mathsf{X}}})^{*}\bm{A}_{(\lambda)}=\bm{A}^{\bm{\mathsf{X}}}=\bm{A}\in\mathscr{T}^{k}\,_{l}(\mathring{\mathscr{M}})\,,\label{eq:transport}
\end{equation}
and similarly the transport of $\bm{\nabla}_{(\lambda)}$ to $\mathring{\mathscr{M}}$
is $\bm{\nabla}$. We know, moreover, that we can express any such
$\bm{A}$ as a Taylor series around its background value, $\bm{A}_{(0)}=\mathring{\bm{A}}$, in Lie derivatives along $\bm{\mathsf{X}}$  (see~\cite{bruni_perturbations_1997}):
\begin{align}
\bm{A}=\, & \mathring{\bm{A}}+\sum_{n=1}^{\infty}\frac{\lambda^{n}}{n!}\mathcal{L}_{\bm{\mathsf{X}}}^{n}\bm{\mathsf{A}}|_{\mathring{\mathscr{M}}}\\
=\, & \mathring{\bm{A}}+\sum_{n=1}^{\infty}\lambda^{n}\delta^{n}\bm{A}\,,\label{eq:A_perturbation_series}
\end{align}
where $\mathcal{L}$ denotes the Lie derivative; in the last equality,
we have defined $\delta^{n}\bm{A}=(1/n!)(\partial_{\lambda}^{n}\bm{A})|_{\lambda=0}$
and so the (gauge-dependent) first-order perturbation is $\delta^{1}\bm{A}=\delta\bm{A}=\delta\bm{A}^{\bm{\mathsf{X}}}$.
Note that the symbol $\delta^{n}$, $\forall n$, can be thought
of as an operator $\delta^{n}=(1/n!)\partial_{\lambda}^{n}|_{\lambda=0}$
that acts upon and extracts the $\mathcal{O}(\lambda^{n})$ part of
any tensor in $\mathring{\mathscr{M}}$. So now, in particular, we have
that the background value of $\bm{g}=(\varphi_{(\lambda)}^{\bm{\mathsf{X}}})^{*}\bm{g}_{(\lambda)}$
is $\mathring{\bm{g}}$ and we denote its first-order perturbation
for convenience and according to convention as $\bm{h}=\delta\bm{g}$. Thus we have 
\begin{equation}
\bm{g}=\mathring{\bm{g}}+\lambda\bm{h}+\mathcal{O}(\lambda^{2})\,,
\end{equation}
where we have omitted explicitly specifying the gauge ($\bm{\mathsf{X}}$)
dependence for now.

Let us define one further piece of notation that we shall need to
use: let  $\mathring{\bm{\Gamma}}$ and $\bm{\Gamma}=(\varphi_{(\lambda)}^{\bm{\mathsf{X}}})^{*}\bm{\Gamma}_{(\lambda)}$
denote the Christoffel symbols (living on $\mathring{\mathscr{M}}$)
associated respectively with $\mathring{\bm{g}}$ and $\bm{g}$, defined
in the usual way (as the connection coefficients between their respective
covariant derivatives and the partial derivative). Then their difference
$\bm{C}=\bm{\Gamma}-\mathring{\bm{\Gamma}}$ is the connection coefficient
relating $\bm{\nabla}$ and $\mathring{\bm{\nabla}}$ on $\mathring{\mathscr{M}}$, which is in fact a tensor.
Note that $\mathring{\bm{C}}=0$, \textit{i.e.} $\bm{C}=\lambda\delta\bm{C}+\mathcal{O}(\lambda^{2})$.
In particular, it is given by
\begin{equation}
C^{a}\,_{bc}=\frac{\lambda}{2}\mathring{g}^{ad}\left(\mathring{\nabla}_{b}h_{cd}+\mathring{\nabla}_{c}h_{bd}-\mathring{\nabla}_{d}h_{bc}\right)+\mathcal{O}\left(\lambda^{2}\right)\,.\label{eq:connection_coeff}
\end{equation}

\subsection{Gravitational self-force from the general momentum conservation law\label{ssec:general_egsf}}

\begin{figure}
\begin{centering}
\includegraphics[scale=0.7]{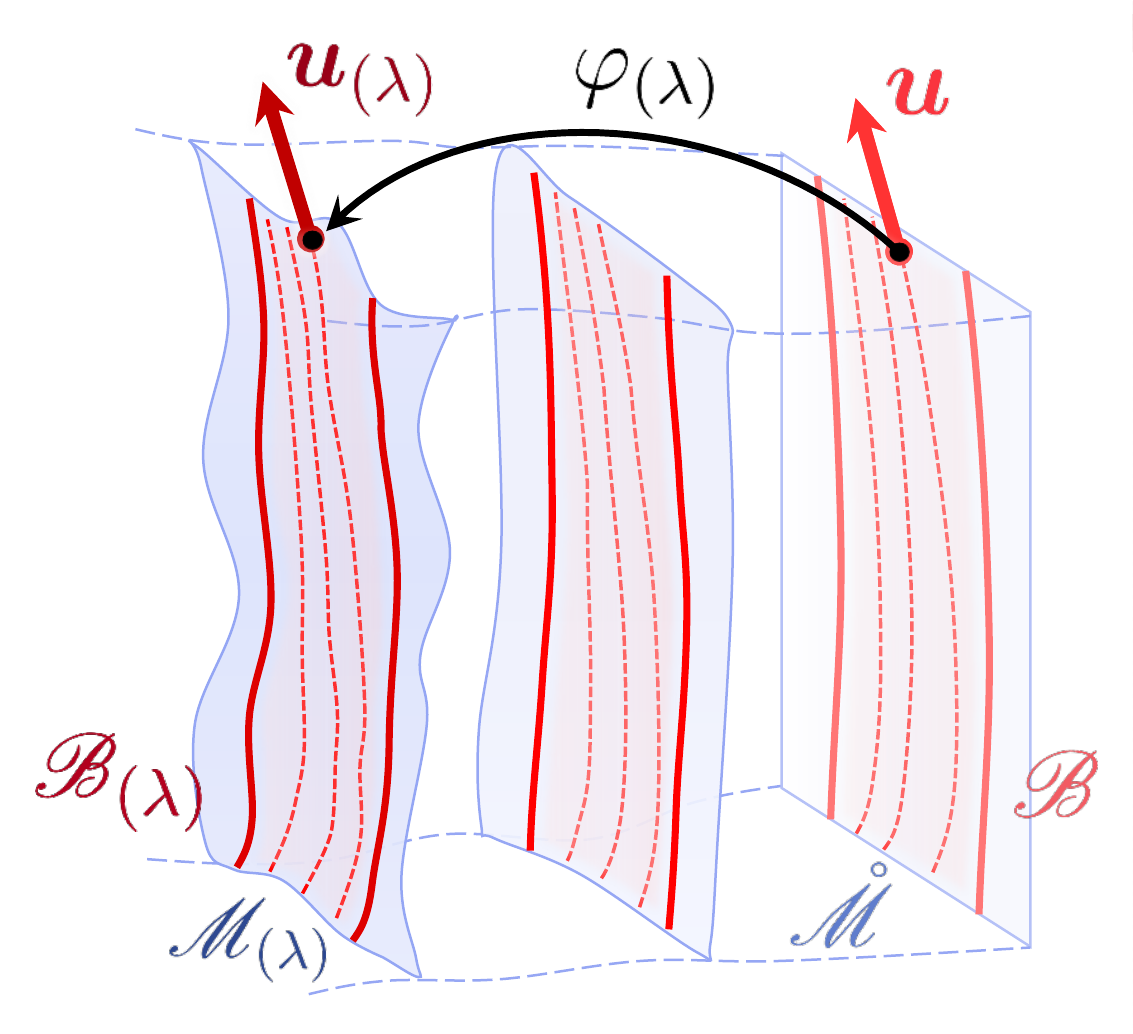}
\par\end{centering}
\caption{Representation of a one-parameter family of quasilocal frames $\{(\mathscr{B}_{(\lambda)};\bm{u}_{(\lambda)})\}_{\lambda\geq0}$
embedded correspondingly in a family of spacetimes $\{\mathscr{M}_{(\lambda)}\}_{\lambda\geq0}$.}\label{fig-pert-qf}
\end{figure}

Let $\{(\mathscr{B}_{(\lambda)};\bm{u}_{(\lambda)})\}_{\lambda\geq0}$
be an arbitrary one-parameter family of quasilocal frames (defined as in Section
\ref{sec:qf}) each of which is embedded, respectively, in the corresponding
element of the family of perturbed spacetimes $\{(\mathscr{M}_{(\lambda)},\bm{g}_{(\lambda)},\bm{\nabla}_{(\lambda)})\}_{\lambda\geq0}$
described in the previous subsection. See Fig. \ref{fig-pert-qf}. Consider the general geometrical
identity (\ref{eq:geometrical_identity}) in $\mathscr{M}_{(\lambda)}$,
$\forall\lambda\geq0$:
\begin{multline}
\intop_{\mathscr{S}_{\textrm{f}}^{(\lambda)}-\mathscr{S}_{\textrm{i}}^{(\lambda)}}\!\!\!\!\!\!\!\!\bm{\epsilon}_{\mathscr{S}_{(\lambda)}}\,\Pi_{\bm{\tilde{u}}_{(\lambda)}\bm{\psi}_{(\lambda)}}^{(\lambda)}\\
=-\!\!\intop_{\Delta\mathscr{B}_{(\lambda)}}\!\!\!\bm{\epsilon}_{\mathscr{B}_{(\lambda)}}\,\left(G_{\bm{n}_{(\lambda)}\bm{\psi}_{(\lambda)}}^{(\lambda)}+\Pi_{(\lambda)}^{ab}\mathcal{D}_{a}^{(\lambda)}\psi_{b}^{(\lambda)}\right)\,.\label{eq:general_cons_law_lambda}
\end{multline}
This gives us our conservation laws in the background for $\lambda=0$, and in a perturbed spacetime for $\lambda>0$. It is the
latter that we are interested in, but since we do not know how to
do calculations in $\mathscr{M}_{(\lambda)}$ $\forall\lambda>0$,
we have to work with Eq. (\ref{eq:general_cons_law_lambda}) transported
to $\mathring{\mathscr{M}}$. This is easily achieved by using the
fact that for any diffeomorphism $f:\mathscr{U}\rightarrow\mathscr{V}$
between two (oriented) smooth $n$-dimensional manifolds $\mathscr{U}$ and $\mathscr{V}$
and any (compactly supported) $n$-form $\bm{\omega}$ in $\mathscr{V}$,
we have that $\int_{\mathscr{V}}\bm{\omega}=\int_{\mathscr{U}}f^{*}\bm{\omega}$.
Applying this to the LHS and RHS of Eq. (\ref{eq:general_cons_law_lambda})
respectively, we simply get
\begin{multline}
\nn\nn\nn\nn\nn\nn\nn\nn\intop_{\quad\quad\quad\quad\quad\quad\quad\varphi_{(\lambda)}^{-1}(\mathscr{S}_{\textrm{f}}^{(\lambda)})-\varphi_{(\lambda)}^{-1}(\mathscr{S}_{\textrm{i}}^{(\lambda)})}\nn\nn\nn\nn\nn\nn\nn\nn\left(\varphi_{(\lambda)}^{*}\bm{\epsilon}_{\mathscr{S}_{(\lambda)}}\right)\,\varphi_{(\lambda)}^{*}\Pi_{\bm{\tilde{u}}_{(\lambda)}\bm{\psi}_{(\lambda)}}^{(\lambda)}\\
=\nn\nn\nn\nn\intop_{\quad\quad\quad\varphi_{(\lambda)}^{-1}(\Delta\mathscr{B}_{(\lambda)})}\nn\nn\nn\nn\left(\varphi_{(\lambda)}^{*}\bm{\epsilon}_{\mathscr{B}_{(\lambda)}}\right)\,\varphi_{(\lambda)}^{*}\left(G_{\bm{n}_{(\lambda)}\bm{\psi}_{(\lambda)}}^{(\lambda)}+\Pi_{(\lambda)}^{ab}\mathcal{D}_{a}^{(\lambda)}\psi_{b}^{(\lambda)}\right)\,.
\end{multline}
Denoting $\mathscr{S}=\varphi_{(\lambda)}^{-1}(\mathscr{S}_{(\lambda)})\subset\mathring{\mathscr{M}}$
as the inverse image of a constant time two-surface and similarly
$\mathscr{B}=\varphi_{(\lambda)}^{-1}(\mathscr{B}_{(\lambda)})\subset\mathring{\mathscr{M}}$
as the inverse image of the worldtube boundary (quasilocal frame) in the background
manifold, and using the fact that the tensor transport commutes with
contractions, the above can simply be written in the notation we have
established as
\begin{multline}
\intop_{\mathscr{S}_{\textrm{f}}-\mathscr{S}_{\textrm{i}}}\left(\varphi_{(\lambda)}^{*}\bm{\epsilon}_{\mathscr{S}_{(\lambda)}}\right)\Pi_{\bm{\tilde{u}}\bm{\psi}}\\
=\intop_{\Delta\mathscr{B}}\left(\varphi_{(\lambda)}^{*}\bm{\epsilon}_{\mathscr{B}_{(\lambda)}}\right)\,\left(G_{\bm{n}\bm{\psi}}+\Pi^{ab}\mathcal{D}_{a}\psi_{b}\right)\,.\label{eq:general_cons_law_background}
\end{multline}

So far we have been completely general. Now, let us restrict our attention to the momentum conservation law ($\bm{\psi}=-\bm{\phi}\in \mathscr{H}$)
given by Eq. (\ref{eq:general_cons_law_background}), and let us assume
that we do not have any matter on $\Delta\mathscr{B}$ (hence, by
the Einstein equation, $G_{\bm{n}\bm{\phi}}|_{\Delta\mathscr{B}}=\kappa T_{\bm{n}\bm{\phi}}|_{\Delta\mathscr{B}}=0$),
or even simply that any matter if present there is subdominant to
the linear perturbation, \textit{i.e.} $\bm{T}|_{\Delta\mathscr{B}}=\mathcal{O}(\lambda^{2})$.
The LHS then expresses the change in momentum of the system (inside
the worldtube interval in the perturbed spacetime) between
some initial and final time slices; for notational ease, we will simply
denote this by $\Delta\mathtt{p}^{(\bm{\phi})}$. (Note that we prefer to use typewriter font for the total quasilocal momentum, so as to avoid any confusion with matter four-momentum defined in the typical way from $T_{ab}$ and traditionally labelled by $P^{a}$, as \textit{e.g.} in Eq. \ref{eq:local_cons}.) Then, inserting
also the definition of the Brown-York tensor [Eq. (\ref{eq:tau_ab_definition})]
on the RHS and replacing $\bm{\mathcal{D}}$ with $\bm{\nabla}$ since
it does not affect the contractions, Eq. (\ref{eq:general_cons_law_background})
becomes:
\begin{equation}
\Delta\mathtt{p}^{(\bm{\phi})}=\intop_{\Delta\mathscr{B}}\left(\varphi_{(\lambda)}^{*}\bm{\epsilon}_{\mathscr{B}_{(\lambda)}}\right)\,\tau^{ab}\nabla_{a}\phi_{b}\,.\label{eq:Delta_p_general}
\end{equation}
We claim, and will now demonstrate, that the $\mathcal{O}(\lambda)$
part of this always contains the GSF.

Let us consider Eq. (\ref{eq:Delta_p_general}) term by term. First
we have the transport---in this case, the pullback---under $\varphi_{(\lambda)}$
of the volume form of $\mathscr{B}_{(\lambda)}$. Now, we know that
the pullback under a diffeomorphism of the volume form of a manifold is,
in general, not simply the volume form of the inverse
image of that manifold under the diffeomorphism. However, it is always true (see, \textit{e.g.},
Chapter 7 of Ref. \cite{abraham_manifolds_2001}) that they are proportional,
with the proportionality given by a smooth function called the Jacobian
determinant and usually denoted by $J$. That is, in our case we have
$\varphi_{(\lambda)}^{*}\bm{\epsilon}_{\mathscr{B}_{(\lambda)}}=J\bm{\text{\ensuremath{\epsilon}}}_{\mathscr{B}}$,
with $J\in C^{\infty}(\mathscr{B})$. In particular, this function
is given by $J(p)=\det(T_{p}\varphi_{(\lambda)})$, $\forall p\in\mathscr{B}$,
where $T_{p}\varphi_{(\lambda)}=(\varphi_{(\lambda)})_{*}:T_{p}\mathscr{B}\rightarrow T_{\varphi_{(\lambda)}}\mathscr{B}_{(\lambda)}$
is the pushforward, and the determinant is computed with respect to
the volume forms $\bm{\text{\ensuremath{\epsilon}}}_{\mathscr{B}}(p)$
on $T_{p}\mathscr{B}$ and $\bm{\epsilon}_{\mathscr{B}_{(\lambda)}}(\varphi_{(\lambda)}(p))$
on $T_{\varphi_{(\lambda)}(p)}\mathscr{B}_{(\lambda)}$. 
Now, it is clear that we have $J=1+\mathcal{O}(\lambda)$, as $\varphi_{(0)}$
is simply the identity map. Therefore, we have
\begin{equation}
\varphi_{(\lambda)}^{*}\bm{\epsilon}_{\mathscr{B}_{(\lambda)}}=\left(1+\mathcal{O}\left(\lambda\right)\right)\bm{\text{\ensuremath{\epsilon}}}_{\mathscr{B}}\,.\label{eq:volume_form_pullback}
\end{equation}
As for the other terms in the integrand of Eq.  (\ref{eq:Delta_p_general}),
we simply have
\begin{align}
\tau^{ab}=\, & \mathring{\tau}^{ab}+\lambda\delta\tau^{ab}+\mathcal{O}(\lambda^{2})\,,\label{eq:tau_series}\\
\nabla_{a}\phi_{b}= & \mathring{\nabla}_{a}\phi_{b}+\lambda\delta\left(\nabla_{a}\phi_{b}\right)+\mathcal{O}(\lambda^{2})\,.\label{eq:Dphi_series}
\end{align}

Hence we can see that there will be three contributions to the $\mathcal{O}(\lambda)$
RHS of Eq. (\ref{eq:Delta_p_general}). Respectively, from Eqs.
(\ref{eq:volume_form_pullback})-(\ref{eq:Dphi_series}), these are
the $\mathcal{O}(\lambda)$ parts of: the volume form pullback, which
may not be easy to compute in practice; the Brown-York tensor $\bm{\tau}$,
which may be computed from its definition [Eq. (\ref{eq:tau_ab_definition})];
and the derivative of the conformal Killing vector $\bm{\phi}$, which may be readily carried
out and, as we will presently show, always contains the GSF. Thus
we denote this contribution to the $\mathcal{O}(\lambda)$ part of
$\Delta\mathtt{p}^{(\bm{\phi})}$ as $\Delta\mathtt{p}_{\textrm{self}}^{(\bm{\phi})}$,
\begin{equation}
\Delta\mathtt{p}_{\textrm{self}}^{(\bm{\phi})}=\lambda\intop_{\Delta\mathscr{B}}\bm{\text{\ensuremath{\epsilon}}}_{\mathscr{B}}\,\mathring{\tau}^{ab}\delta\left(\nabla_{a}\phi_{b}\right)\,.\label{eq:Delta_p_GSF}
\end{equation}

Now we proceed with the computation of Eq. (\ref{eq:Delta_p_GSF}). In
particular, let us consider the series expansion of Eq. (\ref{eq:Delta_p_GSF})
in the areal radius $r$ of $\mathscr{B}$. This can be defined for
any time slice by $r=(\frac{1}{4\pi}\int_{\mathscr{S}}\bm{\epsilon}^{\,}_{\mathscr{S}})^{1/2}$,
such that a constant $r$ slice of $\mathring{\mathscr{M}}$ defines
$\mathscr{B}$ (and $\bm{n}=M\mathring{\bm{\nabla}}r$ for some positive
function $M$ on $\mathscr{B}$). It has been shown \cite{epp_existence_2012} that the Brown-York
tensor has, in general, the following expansion in $r$:
\begin{equation}
\mathring{\tau}^{ab}=\mathring{u}^{a}\mathring{u}^{b}\mathcal{E}_{\textrm{vac}}-\mathring{\sigma}^{ab}{\rm P}_{\textrm{vac}}+\mathcal{O}(r)\,,\label{eq:tau_dominant}
\end{equation}
where 
\begin{align}
\mathcal{E}_{\textrm{vac}}=\, & -\frac{2}{\kappa r}\,,\label{eq:E_vac}\\
{\rm P}_{\textrm{vac}}=\, & -\frac{1}{\kappa r}\,,\label{eq:P_vac}
\end{align}
are called the \emph{vacuum energy} and \emph{vacuum pressure} respectively. Some
remarks regarding these are warranted before we move on. In particular,
these are terms which have sometimes been argued to play the role
of ``subtraction terms'' (to be removed from the quasilocal energy-momentum
tensor); see \textit{e.g.} Ref. \cite{brown_action_2002}. From this point of view, the definition of the Brown-York tensor [Eq. (\ref{eq:tau_ab_definition})]
may be regarded as carrying a certain amount of freedom, inasmuch as any
freedom may be assumed to exist to define a ``reference'' action
$S_{0}$ to be subtracted from the total (gravitational plus matter)
action $S_{\textrm{G+M}}$ in the variational principle discussed
in Subsection \ref{ssec:qf_by}. Such a subtraction of a ``reference'' action, while
common practice in gravitational physics, has the sole function of
shifting the numerical value of the action such that, ultimately,
the numerical value of the Hamiltonian constructed from the modified
action $S_{\textrm{G+M}}-S_{0}$ may be interpreted as the ADM energy.
However, this essentially amounts to a presumption that we are free
to pick the zero of the energy---in other words, that the vacuum
energy may be freely subtracted away without affecting the physics.
Though we refrain from entering into much further detail here, it
has been shown \cite{epp_momentum_2013} that these vacuum terms, Eqs. (\ref{eq:E_vac})-(\ref{eq:P_vac}),
are in fact crucial for our conservation laws to yield physically
reasonable answers and to make mathematical sense---evidencing that
the vacuum energy/pressure should be taken seriously as having physically
real significance. We will now lend further credibility to this by
showing that they are precisely the energy (and pressure) associated
with the momentum flux that are typically interpreted as the GSF. Actually, we argue in this paper that
the term implicating the vacuum energy yields the standard form
of the GSF, and the vacuum pressure term is novel in our analysis.

Now that we have an expansion [Eq. (\ref{eq:tau_dominant})] of $\mathring{\bm{\tau}}$
in $r$, let us consider the $\delta(\bm{\nabla}\bm{\phi})$ term.
We see that
\begin{equation}
\delta\left(\nabla_{a}\phi_{b}\right)=\delta\left(\mathring{\nabla}_{a}\phi_{b}-C^{d}\,_{ab}\phi_{d}\right)=-\delta C^{c}\,_{ab}\phi_{c}\,.\label{eq:delta_nabla_phi}
\end{equation}
Collecting all of our results so far---inserting Eqs. (\ref{eq:tau_dominant})-(\ref{eq:delta_nabla_phi})
into Eq. (\ref{eq:Delta_p_GSF})---we thus get:
\begin{equation}
\Delta\mathtt{p}_{\textrm{self}}^{(\bm{\phi})}=\lambda\frac{2}{\kappa}\intop_{\Delta\mathscr{B}}\bm{\text{\ensuremath{\epsilon}}}_{\mathscr{B}}\,\frac{1}{r}\left(\mathring{u}^{a}\mathring{u}^{b}-2\mathring{\sigma}^{ab}\right)\delta C^{c}\,_{ab}\phi_{c}+\mathcal{O}\left(r\right)\,.\label{eq:Delta_p_GSF_2}
\end{equation}

Let us now look at the contractions in the integrand. For the first
(energy) term, inserting the connection coefficient (\ref{eq:connection_coeff}),
we have by direct computation:
\begin{align}
\mathring{u}^{a}\mathring{u}^{b}\delta C^{c}\,_{ab}\phi_{c}=\, & \mathring{g}^{cd}\left(\mathring{\nabla}_{a}h_{bd}-\frac{1}{2}\mathring{\nabla}_{d}h_{ab}\right)\mathring{u}^{a}\mathring{u}^{b}\phi_{c}\\
=\, & -F^{c}[\bm{h};\mathring{\bm{u}}]\phi_{c}\,,
\end{align}
where the functional $\bm{F}$ is precisely the GSF four-vector functional defined
in the introduction [Eq. \ref{eq:intro_GSF_functional}], and to write the final equality we have used the orthogonality
property $\phi_{\mathring{\bm{u}}}=0$. Thus we see that this is indeed
the term that yields the GSF. For the second (pressure) term in Eq. (\ref{eq:Delta_p_GSF_2}),
we similarly obtain by direct computation:
\begin{equation}
\mathring{\sigma}^{ab}\delta C^{c}\,_{ab}\phi_{c}=2\wp^{c}[\bm{h};\mathring{\bm{\sigma}}]\phi_{c}\,,
\end{equation}
where in expressing the RHS, it is convenient to define a general
functional of two $(0,2)$-tensors similar to the GSF functional:
\begin{equation}
\wp^{c}[\bm{H};\bm{S}]=\frac{1}{2}\mathring{g}^{cd}\left(\mathring{\nabla}_{a}H_{bd}-\frac{1}{2}\mathring{\nabla}_{d}H_{ab}\right)S^{ab}\,.\label{eq:pressure_functional}
\end{equation}
We call this novel term the \emph{gravitational self-pressure force}.

Now we can collect all of the above and insert them into (\ref{eq:Delta_p_GSF_2}).
Before writing down the result, it is convenient to define a total
functional $\bm{\mathcal{F}}$ as the sum of $\bm{F}$ and $\bm{\wp}$,
\begin{equation}
\mathcal{F}^{a}[\bm{h};\mathring{\bm{u}}]=F^{a}[\bm{h};\mathring{\bm{u}}]+\wp^{a}[\bm{h};\mathring{\bm{\sigma}}]\,.\label{eq:generalized_GSF}
\end{equation}
We refer to this as the \emph{extended GSF functional}. Note that
for $\bm{\mathcal{F}}$ we write only the functional dependence on
$\bm{h}$ and $\mathring{\bm{u}}$ since the two-metric $\mathring{\bm{\sigma}}$
is determined uniquely by $\mathring{\bm{u}}$. With this, and setting the perturbation
parameter to unity, Eq. (\ref{eq:Delta_p_GSF_2})
becomes:
\begin{equation}
\boxed{\Delta\mathtt{p}_{\textrm{self}}^{(\bm{\phi})}=-\frac{1}{4\pi}\intop_{\Delta\mathscr{B}}\!\bm{\text{\ensuremath{\epsilon}}}_{\mathscr{B}}\,\frac{1}{r}\bm{\phi}\cdot\boldsymbol{\mathcal{F}}[\bm{h};\mathring{\bm{u}}]+\mathcal{O}\left(r\right)\,.}\label{eq:Delta_p_GSF_general}
\end{equation}

This is to be compared with Gralla's formula \cite{gralla_gauge_2011} discussed in the introduction, Eq. (\ref{eq:intro_Gralla}).
While the equivalence thereto is immediately suggestive based on
the general form of our result, we have to do a bit more work to show
that indeed Eqn. (\ref{eq:Delta_p_GSF_general}), both on the LHS
and the RHS, recovers---though in general will, evidently
at least from our novel gravitational self-pressure force, also have extra terms added to---Eq.  (\ref{eq:intro_Gralla}).
We leave this task to the following section, the purpose of which is to
consider in detail the application of our conservation law formulation to a concrete
example of a perturbative family of spacetimes defined for a self-force
analysis, namely the Gralla-Wald family.

Concordantly, we emphasize that the result above  [Eq. (\ref{eq:Delta_p_GSF_general})]
holds for any family of perturbed manifolds $\{\mathscr{M}_{(\lambda)}\}_{\lambda\geq0}$
and is completely independent of the internal description of our system,
\textit{i.e.} the worldtube interior.
In other words, what we have just demonstrated---provided only that
one accepts a quasilocal notion of energy-momentum---is that the (generalized)
GSF is a completely generic perturbative effect in GR for localized
systems: it arises as a linear order contribution of any spacetime
perturbation to the momentum flux of a system in the limit where its
areal radius is small. 

This view of the self-force may cast fresh conceptual light on the
old and seemingly arcane problem of deciphering its physical origin and meaning.
In particular, recall the common view that the GSF is caused by the
backreaction of the ``mass'' of a small object upon its own motion. Yet
what we have seen here is that it is actually the vacuum ``mass'',
or vacuum energy that is responsible for the GSF. We may still regard
the effect as a ``backreaction,'' in the sense that it is the boundary metric perturbations of the system---the $\bm{h}$ on $\mathscr{B}$---which
determine its momentum flux, but the point is that this flux is inexorably
present and given by Eq. (\ref{eq:Delta_p_GSF_general}) regardless of
where exactly this $\bm{h}$ is coming from. Presumably, the dominant
part of $\bm{h}$ would arise from the system itself---if we further
assume that the system itself is indeed what is being treated perturbatively
by the family $\{\mathscr{M}_{(\lambda)}\}_{\lambda\geq0}$, as is
the case with typical self-force analyses---but in principle $\bm{h}$
can comprise absolutely any perturbations, \textit{i.e.} its physical origin
doesn't even have to be from inside the system.

In this way, we may regard the GSF as a completely geometrical, purely
general-relativistic backreaction of the mass (and pressure) of the
spacetime vacuum---\emph{not} of the object inside---upon the motion of a localized
system (\textit{i.e.} its momentum flux). This point of view frees
us from having to invoke such potentially ambiguous notions as ``mass
ratios'' (in a two-body system for example), let alone ``Coulombian
$m/r$ fields,'' to make basic sense of self-force effects. They
simply---and always---happen from the interaction of the vacuum
with any boundary perturbation, and are  dominant if that boundary
is not too far out.

\section{Application to the Gralla-Wald approach to the gravitational self-force\label{sec:gralla-wald-analysis}}

In this section we will consider in detail the application of our
ideas to a particular approach to the self-force: that is to say, a particular
specification of $\{(\mathscr{M}_{(\lambda)},\bm{g}_{(\lambda)})\}$ via a few additional assumptions aimed
at encoding the notion of a ``small'' object being ``scaled down''
to zero ``size'' and ``mass'' as $\lambda\rightarrow0$. In other
words, we now identify the perturbation (which has up to this point
been treated completely abstractly) defined by $\{(\mathscr{M}_{(\lambda)},\bm{g}_{(\lambda)})\}$
as actually being that caused by the presence of the ``small'' object:
that could mean regular matter (in particular, a compact object such
as a neutron star) or a black hole. 

The assumptions (on $\{\bm{g}_{(\lambda)}\}$) that we choose to work
with here are those of the approach of Gralla and Wald \cite{gralla_rigorous_2008}. Certainly,
the application of our perturbed quasilocal conservation laws could
just as well be carried out in the context of any other self-force
analysis---such as, \textit{e.g.}, the self-consistent approximation of Pound \cite{pound_self-consistent_2010} (the
mathematical correspondence of which to the Gralla-Wald approach has,
in any case, been shown in Ref. \cite{pound_gauge_2015}). 

Our motivation for starting with the Gralla-Wald
approach in particular is two-fold. On the one hand, it furnishes
a mathematically rigorous and physically clear picture (which we show in Fig. \ref{fig-gralla-wald-family})---arguably
more so than any other available GSF treatment---of what it means
to ``scale down'' a small object to zero ``size'' and ``mass''
(or, equivalently, of perturbing any spacetime by the presence of
an object with small ``size'' and ``mass''---we will be more
precise momentarily). On the other hand, it is within this approach
that the formula for the GSF has been obtained (in Ref. \cite{gralla_gauge_2011}) as a closed two-surface
(small two-sphere) integral around the object (in lieu of evaluating
the GSF at a spacetime point identified as the location of the object),
in the form of the Gralla ``angle averaging'' formula [Eq. (\ref{eq:intro_Gralla})]---with which our
extended GSF formula (\ref{eq:Delta_p_GSF_general}) is to be compared.

In Subsection \ref{ssec:gralla-wald_review}, we provide an overview of the assumptions and consequences
of the Gralla-Wald approach to the GSF. Afterwards, in Subsection \ref{ssec:gralla-wald_rqf_general}, we describe the general embedding of rigid quasilocal frames in the Gralla-Wald family of spacetimes, and then in Subsection \ref{ssec:gralla-wald_rqf_background} we describe their detailed construction in the background spacetime in this family. Having established this, we then proceed to derive equations of motion in two ways. In particular, we carry out the analysis with two separate choices of rigid quasilocal frames (``frames of reference''): first, inertially with the point particle approximation of the moving object in the background in Subsection \ref{ssec:gralla-wald_pp-inertial}, and second, inertially with the object itself in the perturbed spacetime in Subsection \ref{ssec:gralla-wald_sco-inertial}.

\subsection{The Gralla-Wald approach to the GSF\label{ssec:gralla-wald_review}}

The basic idea of Gralla and Wald \cite{gralla_rigorous_2008} for defining a family $\{(\mathscr{M}_{(\lambda)},\bm{g}_{(\lambda)})\}_{\lambda\geq0}$
such that $\lambda>0$ represents the inclusion of perturbations generated
by a ``small'' object is the following one. One begins by imposing
certain smoothness conditions on $\{\bm{g}_{(\lambda)}\}_{\lambda\geq0}$
corresponding to the existence of certain limits of each $\bm{g}_{(\lambda)}$. In particular, two limits are sought corresponding intuitively to two limiting views of the
system: first, a view from ``far away'' from which the ``motion''
of the (extended but localized) object reduces to a worldline; second,
a view from ``close by'' the object from which the rest of the universe
(and in particular, the MBH it might be orbiting as in an EMRI) looks
``pushed away'' to infinity. A third requirement must be added to this, namely that both
of these limiting pictures nonetheless coexist in the same spacetime,
\textit{i.e.} the two limits are smoothly related (or, in other words, there is no pathological behaviour when taking these limits along different directions). While in principle this
may sound rather technical, one can actually motivate each of these
conditions with very sensible physical arguments as we shall momentarily
elaborate further upon. From them, Gralla and Wald have shown \cite{gralla_rigorous_2008} that
it is possible to derive a number of consequences, including geodesic
motion in the background at zeroth order and the MiSaTaQuWa equation \cite{mino_gravitational_1997,quinn_axiomatic_1997}
for the GSF at first order in $\lambda$.

\begin{figure}
\begin{centering}
\includegraphics[scale=0.6]{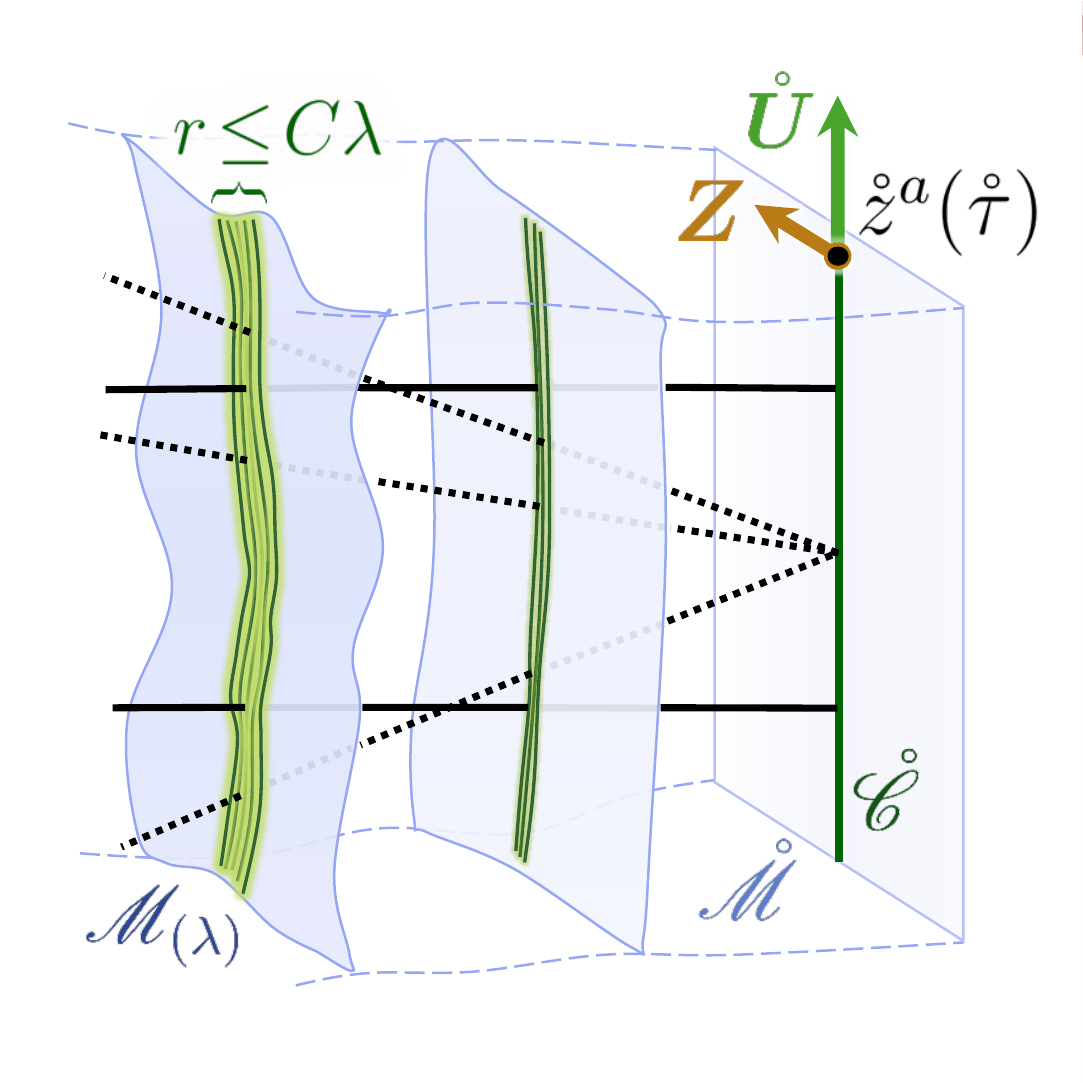}
\par\end{centering}
\caption{Representation of the Gralla-Wald family of spacetimes $\{\mathscr{M}_{(\lambda)}\}_{\lambda\geq0}$. (This is an adaptation of Fig. 1 of Ref. \cite{gralla_rigorous_2008}.)
The lined green region that ``fills in'' $\mathscr{M}_{(\lambda)}$
for $r\leq C\lambda$ is the ``small'' object which ``scales down''
to zero ``size'' and ``mass'' in the background $\mathring{\mathscr{M}}$.
The solid black lines represent taking the ``ordinary limit'' (the
``far away'' view where the motion appears reduced to a worldline)
and the dashed black lines the ``scaled limit'' (the ``close by''
view where the rest of the universe appears ``pushed away'' to infinity).
The worldline $\mathring{\mathscr{C}}$, which can be proven to be
a geodesic, is parametrized by $\mathring{z}^{a}(\mathring{\tau})$
and has four-velocity $\mathring{\bm{U}}$. The deviation vector
$\bm{Z}$ on $\mathring{\mathscr{C}}$ is used for formulating the
first-order correction to the motion.}\label{fig-gralla-wald-family}
\end{figure}

Let us now be more precise. Let $\{(\mathscr{M}_{(\lambda)},\bm{g}_{(\lambda)})\}_{\lambda\geq0}$
be a perturbative one-parameter family of spacetimes as in the previous
section. We assume that $\{\bm{g}_{(\lambda)}\}_{\lambda\geq0}$ satisfies
the following conditions, depicted visually in Fig. \ref{fig-gralla-wald-family}:

\bigskip{}

\noindent \emph{(i) Existence of an ``ordinary limit'':} There exist
coordinates $\{x^{\alpha}\}$ in $\mathscr{M}_{(\lambda)}$ such that
$g_{\beta\gamma}^{(\lambda)}(x^{\alpha})$ is jointly smooth in $(\lambda,x^{\alpha})$
for $r>C\lambda$ where $C>0$ is a constant and $r=(x_{i}x^{i})^{1/2}$.
For all $\lambda\geq0$ and $r>C\lambda$, $\bm{g}_{(\lambda)}$ is
a vacuum solution of the Einstein equation. Furthermore, $\mathring{g}_{\beta\gamma}(x^{\alpha})$
is smooth in $x^{\alpha}$ including at $r=0$, and the curve $\mathring{\mathscr{C}}=\{r=0\}\subset\mathring{\mathscr{M}}$
is timelike.

\bigskip{}

\noindent \emph{(ii) Existence of a ``scaled limit'':} For all $t_{0}$,
define the ``scaled coordinates'' $\{\bar{x}^{\alpha}\}=\{\bar{t},\bar{x}^{i}\}$
by $\bar{t}=(t-t_{0})/\lambda$ and $\bar{x}^{i}=x^{i}/\lambda$.
Then the ``scaled metric'' $\bar{g}_{\bar{\beta}\bar{\gamma}}^{(\lambda)}(t_{0};\bar{x}^{\alpha})=\lambda^{-2}g_{\bar{\beta}\bar{\gamma}}^{(\lambda)}(t_{0};\bar{x}^{\alpha})$
is jointly smooth in $(\lambda,t_{0};\bar{x}^{\alpha})$ for $\bar{r}=r/\lambda>C$.

\bigskip{}

\noindent \emph{(iii) Uniformity condition:} Define $A=r$, $B=\lambda/r$
and $n^{i}=x^{i}/r$. Then each $g_{\beta\gamma}^{(\lambda)}(x^{\alpha})$
is jointly smooth in $(A,B,n^{i},t)$.

\bigskip{}

Mathematically, the first two conditions respectively ensure the existence
of an appropriate Taylor expansion (in $r$ and $\lambda$) of the
metric in a ``far zone'' (on length scales comparable with the mass
of the MBH in an EMRI, $r\sim M$) and a ``near zone'' (on length
scales comparable with the mass of the object, $r\sim m$) . Meanwhile,
the third is simply a consistency requirement ensuring the existence
of a ``buffer zone'' ($m\ll r\ll M$) where both expansions are
valid. (This idea is in many ways similar to the method of ``matched asymptotic expansions'' \cite{mino_gravitational_1997}). 

From a physical point of view, what is happening in the first (``ordinary'')
limit is that the body is shrinking down to a worldline $\mathring{\mathscr{C}}$
with its ``mass'' (understood as defining the perturbation) going
to zero at least as fast as its radius. (As we increase the perturbative
parameter $\lambda$ from zero, the radius is not allowed to grow
faster than linearly with $\lambda$; viewed conversely, this condition
ensures that the object does not collapse to a black hole if it was
not one already before reaching the point particle limit.) In the
second (``scaled'') limit, the object is shrinking down to zero
size in an asymptotically self-similar manner (its mass is proportional
to its size, and its ``shape'' is not changing). Finally, the uniformity
condition ensures that there are no ``bumps of curvature'' in the
one-parameter family. (Essentially, this guarantees that there are
no inconsistencies in evaluating the limits along different directions.)

From these assumptions alone, Gralla and Wald \cite{gralla_rigorous_2008}  are able to derive the
following consequences: 

\bigskip{}

\noindent \emph{(a) Background motion:} The worldline $\mathring{\mathscr{C}}$
is a geodesic in $\mathring{\mathscr{M}}$; writing its parametrization
in terms of proper time $\mathring{\tau}$ as $\mathring{\mathscr{C}}=\{\mathring{z}^{a}(\mathring{\tau})\}_{\mathring{\tau}\in\mathbb{R}}$
and denoting its four-velocity by $\mathring{U}^{a}={\rm d}\mathring{z}^{a}(\mathring{\tau})/{\rm d}\mathring{\tau}$,
this means that
\begin{equation}
\mathring{\nabla}_{\mathring{\bm{U}}}\mathring{\bm{U}}=0\,.\label{eq:background_geodesic}
\end{equation}

\bigskip{}

\noindent \emph{(b) Background ``scaled'' metric:} $\mathring{\bar{\bm{g}}}$
is stationary and asymptotically flat.

\bigskip{}

\noindent \emph{(c) First-order field equation:} At $\mathcal{O}(\lambda)$,
the Einstein equation is sourced by the matter energy-momentum tensor of a point
particle $\bm{T}^{\textrm{PP}}$ supported on $\mathring{\mathscr{C}}$,
\textit{i.e.} the field equation is
\begin{equation}
\delta G_{ab}\left[\bm{h}\right]=\kappa T_{ab}^{\textrm{PP}}\label{eq:first-order_Einstein_equation}
\end{equation}
where
\begin{equation}
T_{ab}^{\textrm{PP}}=m\intop_{\mathring{\mathscr{C}}}{\rm d}\mathring{\tau}\,\mathring{U}_{a}\left(\mathring{\tau}\right)\mathring{U}_{b}\left(\mathring{\tau}\right)\delta_{4}\left(x^{c}-z^{c}\left(\mathring{\tau}\right)\right)\,.\label{eq:T_ab^PP}
\end{equation}
Here, $m$ is a constant along $\mathring{\mathscr{C}}$ and is interpreted
as representing the mass of the object---or, more precisely,
the mass of the point particle which approximates the object in the
background. (This is a subtle point that should be kept in mind, and which will be better elucidated in our analysis further on.)

\bigskip{}

\noindent \emph{(d) First-order equation of motion:} At $\mathcal{O}(\lambda)$,
the correction to the motion in the Lorenz gauge---corresponding
to the choice of a certain gauge vector $\bm{\mathsf{L}}\in T\mathscr{N}$
defined by the condition
\begin{equation}
\mathring{\nabla}^{b}(h_{ab}^{\bm{\mathsf{L}}}-\frac{1}{2}h^{\bm{\mathsf{L}}}\mathring{g}_{ab})=0\,,
\end{equation}
where $h={\rm tr}(\bm{h})$---is given by the MiSaTaQuWa equation \cite{mino_gravitational_1997,quinn_axiomatic_1997},
\begin{equation}
\mathring{\nabla}_{\mathring{\bm{U}}}\mathring{\nabla}_{\mathring{\bm{U}}}Z^{a}=-\mathring{E}_{b}\,^{a}Z^{b}+F^{a}[\bm{h}^{\textrm{tail}};\mathring{\bm{U}}]\,,\label{eq:MiSaTaQuWa}
\end{equation}
where $\mathring{E}_{b}\,^{a}=\mathring{R}_{cbd}\,^{a}\mathring{U}^{c}\mathring{U}^{d}$ is the electric part of the Weyl tensor
and $\bm{h}^{\textrm{tail}}$ is a ``tail'' integral of the retarded Green's
functions of $\bm{h}$. The above is an equation for a four-vector
$\bm{Z}$ called the ``deviation'' vector; the LHS is the acceleration
associated therewith and the RHS is a geodesic deviation term plus
the GSF. This deviation vector is defined on $\mathring{\mathscr{C}}$
and represents the first-order correction needed to move off $\mathring{\mathscr{C}}$
and onto the worldline representing the ``center of mass'' of the
perturbed spacetime, defined as in the Hamiltonian analysis of Regge and Teitelboim \cite{regge_role_1974}.\bigskip{}

Let us make a few comments on these results, specifically concerning
\emph{(a)} and \emph{(c)}. On the one hand, it is quite remarkable
that geodesic motion can be recovered as a consequence\footnote{See again footnote 2 and the references mentioned therein for more on this topic.} of this analysis---\textit{i.e.}
without having to posit it as an assumption---just from smoothness
properties (existence of appropriate limits) of our family of metrics
$\{\bm{g}_{(\lambda)}\}$; and on the other, this analysis offers
sensible meaning to the usual ``delta function cartoon'' (ubiquitous
in essentially all self-force analyses) of the matter stress-energy-momentum tensor describing
the object in the background spacetime. The point is that the description
of the object is completely arbitrary inside the region that is not
covered by the smoothness conditions of the family $\{\bm{g}_{(\lambda)}\}$,
\textit{i.e.} for $r\leq C\lambda$ when $\lambda>0$. (Indeed, this region
can be ``filled in'' even with exotic matter, \textit{e.g.} failing to satisfy
the dominant energy condition, or a naked singularity, as long as
a well-posed initial value formulation exists.) Regardless of what
this description is, the smoothness conditions essentially ensure
that its ``reduction'' to $\mathring{\mathscr{M}}$ (or, more precisely,
the transport of any effect thereof with respect to the family $\{\bm{g}_{(\lambda)}\}$)
simply becomes that of a point particle sourcing the field equation
at $\mathcal{O}(\lambda)$. In this way, the background ``point particle
cartoon'' is justified as the simplest possible idealization of a
``small'' object.

What we are going to do, essentially, is to accept consequences \emph{(a)}-\emph{(c)}
(in fact, we will not even explicitly need \emph{(b)}), the proofs
of which do not rely upon any further limiting conditions such as
a restriction of the perturbative gauge, and to obtain, using our
perturbed momentum conservation law, a more general version of the
EoM, \textit{i.e.} consequence \emph{(d)}. For the latter, Gralla and Wald \cite{gralla_rigorous_2008}
instead rely on the typical but laborious Hadamard expansion techniques of DeWitt and Brehme \cite{dewitt_radiation_1960},
wherein the ``mass dipole moment'' of the object is set to zero.
It is possible \cite{regge_role_1974} to have such a notion in a well-defined Hamiltonian
sense by virtue of \emph{(b)}. While mathematically rigorous and conducive
to obtaining the correct known form of the MiSaTaQuWa equation, their
derivation and final result suffer not only from the limitation of
having to fix the perturbative gauge, but also from the (as we shall
see, potentially avoidable) technical complexity of arriving at the
final answer---including the evaluation of $\boldsymbol{h}^{\textrm{tail}}$
(or otherwise taking recourse to a regularization procedure).

The link between this approach and our conservation law derivation
of the EoM which we are about to carry out is established by the work
of Gralla \cite{gralla_gauge_2011}, who discovered that Eq. (\ref{eq:MiSaTaQuWa}) can be equivalently
written as:
\begin{equation}
\mathring{\nabla}_{\mathring{\bm{U}}}\mathring{\nabla}_{\mathring{\bm{U}}}Z^{a}=-\mathring{E}_{b}\,^{a}Z^{b}+\frac{1}{4\pi}\lim_{r\rightarrow0}\intop_{\mathbb{S}_{r}^{2}}\bm{\epsilon}^{}_{\mathbb{S}^{2}}\,F^{a}[\mathring{\bm{U}},\bm{h}]\,.\label{eq:Gralla_ange-average}
\end{equation}
Here, the GSF term $\bm{F}[\bm{h}^{\textrm{tail}};\mathring{\bm{U}}]$
in the MiSaTaQuWa equation [Eq. (\ref{eq:MiSaTaQuWa})] is substituted by
an integral expression---an average over the angles---of $\bm{F}$.
In particular (as, strictly speaking, one cannot define integrals of vectors as such), this is evaluated by using the exponential map based on
$\mathring{\mathscr{C}}$ to associate a flat metric, in terms of
which the integration is performed over a two-sphere of radius $r$,
$\mathbb{S}_{r}^{2}$, with $\bm{\epsilon}^{}_{\mathbb{S}^{2}}$ denoting the volume
form of $\mathbb{S}^{2}$. 

Observe that, here, the functional dependence of $\bm{F}$ is on $\bm{h}$
itself (and not on $\bm{h}^{\textrm{tail}}$ or any sort of regularized
$\bm{h}$) and for this reason is referred to as the ``bare'' GSF.
Moreover, this formula is actually valid in a wider class of
gauges than just the Lorenz gauge: in particular, it holds in what
are referred to as ``parity-regular'' gauges \cite{gralla_gauge_2011}. We refrain from entering
here into the technical details of exactly how such gauges are defined,
except to say that the eponymous ``parity condition'' that they
need to satisfy has its ultimate origin in the Hamiltonian analysis
of Regge and Teitleboim \cite{regge_role_1974} and is imposed so as to make certain Hamiltonian
definitions---and in particular for Gralla's analysis \cite{gralla_gauge_2011}, the Hamiltonian
``center of mass''---well defined. These, however, are \emph{not} limitations
of our quasilocal formalism, where we know how to define energy-momentum notions
more generally than any Hamiltonian approach. Thus, in our result,
there will be \emph{no restriction} on the perturbative gauge. This may constitute
a great advantage, as the ``parity-regular'' gauge class---though
an improvement from being limited to the Lorenz gauge in formulating
the EoM---still excludes entire classes of perturbative gauges convenient for formulating black hole perturbation theory (\textit{e.g.} the Regge-Wheeler gauge in Schwarzschild-Droste) and hence for carrying out practical EMRI
calculations.

We proceed to apply our quasilocal analysis to the Gralla-Wald family of spacetimes, beginning with a general setup in this family of rigid quasilocal frames.

\subsection{General setup of rigid quasilocal frames in the Gralla-Wald family\label{ssec:gralla-wald_rqf_general}}

Let $(\mathscr{B}_{(\lambda)};\bm{u}_{(\lambda)})$ be a quasilocal frame in $(\mathscr{M}_{(\lambda)},\bm{g}_{(\lambda)},\bm{\nabla}_{(\lambda)})$,
for any $\lambda>0$, constructed just as described in Section \ref{sec:qf}: with unit four-velocity $\bm{u}_{(\lambda)}$, unit normal
$\bm{n}_{(\lambda)}$, induced metric $\bm{\gamma}_{(\lambda)}$ and
so on. Using the fact that the tensor transport is linear and commutes
with tensor products, we can compute the transport (in the five-dimensional ``stacked'' manifold $\mathscr{N}=\mathscr{M}_{(\lambda)}\times\mathbb{R}^{\geq}$ used in our perturbative setup, as in Subsection \ref{ssec:general_perturbations}) of any geometrical
quantity of interest to the background. For example,
\begin{align}
\gamma_{ab}=\, & \varphi_{(\lambda)}^{*}\gamma_{ab}^{(\lambda)}\\
=\, & \varphi_{(\lambda)}^{*}(g_{ab}^{(\lambda)}-n_{a}^{(\lambda)}n_{b}^{(\lambda)})\\
=\, & g_{ab}-n_{a}n_{b}\\
=\, & \mathring{\gamma}_{ab}+\lambda\delta\gamma_{ab}+\mathcal{O}\left(\lambda^{2}\right)\,,
\end{align}
where
\begin{align}
\mathring{\gamma}_{ab}=\, & \mathring{g}_{ab}-\mathring{n}_{a}\mathring{n}_{b}\,,\\
\delta\gamma_{ab}=\, & h_{ab}-2\mathring{n}_{(a}\delta n_{b)}\,.
\end{align}
Similarly, 
\begin{equation}
\sigma_{ab}=\mathring{\sigma}_{ab}+\lambda\delta\sigma_{ab}+\mathcal{O}(\lambda^{2})\,,
\end{equation}
 where
\begin{align}
\mathring{\sigma}_{ab}=\, & \mathring{\gamma}_{ab}+\mathring{u}_{a}\mathring{u}_{b}\,,\\
\delta\sigma_{ab}=\, & \delta\gamma_{ab}+2\mathring{u}_{(a}\delta u_{b)}\,.
\end{align}

Now let us assume that $(\mathscr{B}_{(\lambda)};\bm{u}_{(\lambda)})$
is a rigid quasilocal frame, meaning that the congruence defining it has a vanishing
symmetrized strain rate tensor in $\mathscr{M}_{(\lambda)}$,
\begin{equation}
\theta_{(ab)}^{(\lambda)}=0\,.\label{eq:RQF_lambda}
\end{equation}

Let $\mathscr{B}=\varphi_{(\lambda)}^{-1}(\mathscr{B}_{(\lambda)})$
be the inverse image of $\mathscr{B}_{(\lambda)}$ in the background
$\mathring{\mathscr{M}}$, with $\boldsymbol{u}=\varphi_{(\lambda)}^{*}\boldsymbol{u}_{(\lambda)}=\mathring{\bm{u}}+\lambda\delta\bm{u}+\mathcal{O}(\lambda^{2})$
giving the transport of the quasilocal observers' four-velocity, $\bm{n}=\mathring{\bm{n}}+\lambda\delta\bm{n}+\mathcal{O}(\lambda^{2})$
the unit normal and so on. In other words, $(\mathscr{B};\bm{u})$
is the background mapping of the perturbed congruence $(\mathscr{B}_{(\lambda)};\bm{u}_{(\lambda)})$,
and so  will itself constitute a congruence (in the background),
\textit{i.e.} a quasilocal frame defined by a two-parameter family of worldlines with unit
four-velocity $\bm{u}$ in $\mathring{\mathscr{M}}$. 

However, although $(\mathscr{B}_{(\lambda)};\bm{u}_{(\lambda)})$
is a rigid quasilocal frame in $\mathscr{M}_{(\lambda)}$, $(\mathscr{B};\bm{u})$ is
\emph{not} in general a rigid quasilocal frame in $\mathring{\mathscr{M}}$ (with
respect to the background metric $\mathring{\bm{g}}$). One can see
this easily as follows. Let $\bm{\vartheta}\in\mathscr{T}^{0}\,_{2}(\mathring{\mathscr{M}})$
be the strain rate tensor of $(\mathscr{B};\bm{u})$, so that it is
given by 
\begin{equation}
\vartheta_{ab}=\sigma_{ca}\sigma_{bd}\mathring{\nabla}^{c}u^{d}\,.
\end{equation}
The RHS is an series in $\lambda$, owing to the fact that $\bm{u}$
(and therefore $\bm{\sigma}$, the two-metric on the space $\mathscr{H}$ orthogonal to $\bm{u}$ in $\mathscr{B}$) are transported from
a perturbed congruence in $\mathscr{M}_{(\lambda)}$. Upon expansion
we obtain
\begin{equation}
\vartheta_{ab}=\mathring{\vartheta}_{ab}+\lambda\delta\vartheta_{ab}+\mathcal{O}\left(\lambda^{2}\right)\,,\label{eq:vartheta_espansion}
\end{equation}
where
\begin{equation}
\mathring{\vartheta}_{ab}=\mathring{\sigma}_{c(a}\mathring{\sigma}_{b)d}\mathring{\nabla}^{c}\mathring{u}^{d}
\end{equation}
is just the strain rate tensor of the background congruence---\textit{i.e.}
the congruence defined by $\mathring{\bm{u}}$---and 
\begin{equation}
\delta\vartheta_{ab}=2\mathring{\sigma}^{(c}\,_{(a}\delta\sigma^{d)}\,_{b)}\mathring{\nabla}_{c}\mathring{u}_{d}+\mathring{\sigma}^{c}\,_{(a}\mathring{\sigma}_{b)d}\mathring{\nabla}_{c}\delta u^{d}
\end{equation}
is the first-order term in $\lambda$. Note that we are abusing our
established notation slightly in writing Eq. (\ref{eq:vartheta_espansion}),
as there exists no $\bm{\vartheta}_{(\lambda)}$ in $\mathscr{M}_{(\lambda)}$
the transport (to $\mathring{\mathscr{M}}$) of which yields such
a series expansion; instead $\bm{\vartheta}$ is defined directly
on $\mathring{\mathscr{M}}$ (relative to the metric $\mathring{\bm{g}}$)
as the strain rate tensor of a conguence with four-velocity $\bm{u}$---which
itself contains the expansion in $\lambda$.

Now let us compute the transport of the rigidity condition on $(\mathscr{B}_{(\lambda)};\bm{u}_{(\lambda)})$
[Eq. (\ref{eq:RQF_lambda})] to $\mathring{\mathscr{M}}$: we have
\begin{align}
0=\, & \varphi_{(\lambda)}^{*}\theta_{(ab)}^{(\lambda)}\\
=\, & \varphi_{(\lambda)}^{*}(\sigma_{c(a}^{(\lambda)}\sigma_{b)d}^{(\lambda)}\nabla_{(\lambda)}^{c}u_{(\lambda)}^{d})\\
=\, & \sigma_{c(a}\sigma_{b)d}\nabla^{c}u^{d}\\
=\, & \mathring{\theta}_{(ab)}+\lambda\delta\theta_{(ab)}+\mathcal{O}(\lambda^{2})\,,\label{eq:theta_expansion}
\end{align}
where
\begin{align}
\mathring{\theta}_{(ab)}=\, & \mathring{\vartheta}_{(ab)}\,,\\
\delta\theta_{(ab)}=\, & \delta\vartheta_{(ab)}+\mathring{\sigma}^{c}\,_{(a}\mathring{\sigma}_{b)d}\delta C^{d}\,_{ce}\mathring{u}^{e}\,.
\end{align}
Since $0=\theta_{(ab)}^{(\lambda)}$ identically in $\mathscr{M}_{(\lambda)}$
(as we demand that $(\mathscr{B}_{(\lambda)};\bm{u}_{(\lambda)})$
is a rigid quasilocal frame), Eq. (\ref{eq:theta_expansion}) must vanish order by order
in $\lambda$. That implies, in particular, that the zeroth-order
congruence (defined by $\mathring{\bm{u}}$) is a rigid quasilocal frame, and that the
symmetrized strain rate tensor of the background-mapped perturbed
congruence (defined by $\bm{u}$) is given by
\begin{equation}
\vartheta_{(ab)}=-\lambda\mathring{\sigma}^{c}\,_{(a}\mathring{\sigma}_{b)d}\delta C^{d}\,_{ce}\mathring{u}^{e}+\mathcal{O}\left(\lambda^{2}\right)\,.
\end{equation}

This tells us that the deviation from rigidity of $(\mathscr{B};\bm{u})$
in $\mathring{\mathscr{M}}$ occurs only at $\mathcal{O}(\lambda)$
(and, in particular, is caused by the same perturbed connection coefficient
term that is responsible for the GSF). In other words, we can treat
$(\mathscr{B};\bm{u})$ as a rigid quasilocal frame at zeroth order. This
zeroth order congruence actually makes up a different worldtube boundary
$\mathring{\mathscr{B}}\neq \mathscr{B}$ in $\mathring{\mathscr{M}}$, \textit{i.e.} one defined
by a congruence with four-velocity $\mathring{\bm{u}}\neq\bm{u}$ in general. Clearly,
for a rigid quasilocal frame with a small areal radius $r$ constructed around a worldline $\mathscr{G}$
in $\mathring{\mathscr{M}}$ with four-velocity $\bm{U}_{\mathscr{G}}$,
we would simply have $\mathring{\bm{u}}=\bm{U}_{\mathscr{G}}$ (where
the RHS is understood to be transported off $\mathscr{G}$ and onto
$\mathring{\mathscr{B}}$ via the exponential map), and $\mathring{\bm{\sigma}}=r^{2}\bm{\mathfrak{S}}$,
\textit{i.e.} it is the metric of $\mathbb{S}_{r}^{2}$. This is the most trivial
possible rigid quasilocal frame: at any instant of time, a two-sphere worth of quasilocal
observers moving with the same four-velocity as is the point at its
center (parametrizing the given worldline).

At first order, the equation $0=\delta\theta_{(ab)}$ can be regarded
as the constraint on the linear perturbations ($\delta\bm{u}$) in
the motion of the quasilocal observers in terms of the metric perturbations
guaranteeing that the perturbed congruence is rigid in the perturbed
spacetime. (So presumably, going to $n$-th order in $\lambda$ would
yield equations for every term up to the $n$-th order piece of the
motion of the quasilocal observers, $\delta^{n}\bm{u}$.)

Now recall the momentum conservation
law for rigid quasilocal frames, Eq. (\ref{eq:cons_law_Pa}). This holds for $(\mathscr{B}_{(\lambda)};\bm{u}_{(\lambda)})$
in $\mathscr{M}_{(\lambda)}$. Just as we did in the previous section
with the general conservation law, we can use $\varphi_{(\lambda)}$
to turn this into an equation in $\mathring{\mathscr{M}}$:
\begin{equation}
\Delta\mathtt{p}^{(\bm{\phi})}\!=\!-\!\!\intop_{\Delta\mathscr{B}}\!\!\varphi_{(\lambda)}^{*}\bm{\epsilon}_{\mathscr{B}_{(\lambda)}}\left(\mathcal{E}\alpha_{\bm{\phi}}+2\nu\epsilon_{ab}\mathcal{P}^{a}\phi^{b}+{\rm P}\bm{D}\!\cdot\!\bm{\phi}\right)\,.
\end{equation}

Let us now further assume that we can ignore the Jacobian determinant
(discussed in the previous section) as well as the shift $\bm{v}$
of the quasilocal observers (relative to constant time surfaces).
Then, dividing the above equation by $\Delta t$, where $t$ represents
the adapted time coordinate on $\mathscr{B}$, and taking the
$\Delta t\rightarrow0$ limit, we get the time rate of change of the
momentum,
\begin{equation}
\dot{\mathtt{p}}^{(\bm{\phi})}=-\intop_{\mathscr{S}}\bm{\epsilon}^{\,}_{\mathscr{S}}N\tilde{\gamma}\left(\mathcal{E}\alpha_{\bm{\phi}}+2\nu\epsilon_{ab}\mathcal{P}^{a}\phi^{b}+{\rm P}\bm{D}\cdot\bm{\phi}\right)\,.\label{eq:dpdt_RQF}
\end{equation}
where $\dot{\mathtt{p}}^{(\bm{\phi})}={\rm d}\mathtt{p}^{(\bm{\phi})}/{\rm d}t$,
and we must keep in mind that the derivative is with respect to the
adapted time on (the inverse image on the background of) our congruence.

\subsection{Detailed construction of background rigid quasilocal frames\label{ssec:gralla-wald_rqf_background}}

Let $\mathscr{G}$ be any timelike worldline in $\mathring{\mathscr{M}}$.
Any background metric $\mathring{\bm{g}}$ on $\mathring{\mathscr{M}}$
in a neighborhood of $\mathscr{G}$ admits an expression in Fermi
normal coordinates \cite{misner_gravitation_1973,poisson_motion_2011}, which we label by $\{X^{\alpha}\}=\{T=X^{0},X^{I}\}_{I=1}^{3}$,
as a power series in the areal radius. Denoting by $A_{K}(T)$ and
$W_{K}(T)$ the proper acceleration and proper rate of rotation of
the spatial axes (triad) along $\mathscr{G}$ (as functions of the
proper time $T$ along $\mathscr{G}$), respectively, this is given
by:
\begin{align}
\mathring{g}_{00}=\, & -\left(1+A_{K}X^{K}\right)^{2}+R^{2}W_{K}W_{L}P^{KL}\nonumber \\
 & -\mathring{R}_{0K0L}X^{K}X^{L}+\mathcal{O}\left(R^{3}\right)\,,\label{eq:FNC_g00}\\
\mathring{g}_{0J}=\, & \epsilon_{JKL}W^{K}X^{L}-\frac{2}{3}\mathring{R}_{0KJL}X^{K}X^{L}+\mathcal{O}\left(R^{3}\right)\,,\label{eq:FNC_g0J}\\
\mathring{g}_{IJ}=\, & \delta_{IJ}-\tfrac{1}{3}\mathring{R}_{IKJL}X^{K}X^{L}+\mathcal{O}\left(R^{3}\right)\,,\label{eq:FNC_gIJ}
\end{align}
where $R^{2}=\delta_{IJ}X^{I}X^{J}$ is the square of the radius in
these coordinates (not the square of the Ricci scalar) and $P^{KL}=\delta^{KL}-X^{K}X^{L}/R^{2}$
projects vectors perpendicular to the radial direction $X^{I}/R$. Here we have
to remember that the Riemann tensor $\mathring{R}_{IJKL}$ (along with
$\bm{A}$ and $\bm{W}$) are understood to be evaluated on $\mathscr{G}$.

For all cases that we will be interested in, we will ignore the possibility
of rotation so we set $W_{I}=0$ from now on.

Let us now assume that our background rigid quasilocal frame $(\mathring{\mathscr{B}};\mathring{\bm{u}})$
is constructed around $\mathscr{G}$: that is to say, into this coordinate
system there is embedded a two-parameter family of worldlines representing
a topological two-sphere worth of observers, \textit{i.e.} a fibrated timelike worldtube
$\mathring{\mathscr{B}}$ surrounding $\mathscr{G}$. This may be
conveniently described, as detailed in Subsection \ref{ssec:qf_math}, by defining a new set
of coordinates $\{x^{\alpha}\}=\{t,r,x^{\mathfrak{i}}\}_{\mathfrak{i}=1}^{2}$
given simply by the adapted coordinates $\{t,x^{\mathfrak{i}}\}_{\mathfrak{i}=1}^{2}$
on $\mathring{\mathscr{B}}$ supplemented with a radial coordinate
$r$. Then denoting $\{x^{\mathfrak{i}}\}=\{\theta,\phi\}$ we introduce,
as done in previous calculations with rigid quasilocal frames in Fermi normal coordinates \cite{epp_existence_2012}, the following coordinate
transformation: 
\begin{align}
T\left(t,r,\theta,\phi\right)=\, & t+\mathcal{O}\left(r^{2},\mathcal{R}\right),\,\label{eq:FNC_T}\\
X^{I}\left(t,r,\theta,\phi\right)=\, & rr^{I}\left(\theta,\phi\right)+\mathcal{O}\left(r^{2},\mathcal{R}\right),\,\label{eq:FNC_XI}
\end{align}
where 
\begin{equation}
r^{I}(\theta,\phi)=(\sin\theta\cos\phi,\sin\theta\sin\phi,\cos\theta)
\end{equation}
are the standard direction cosines of a radial unit vector in spherical
coordinates in $\mathbb{R}^{3}$, and $\mathcal{R}$ here represents the
order of the perturbations of the quasilocal frame away from the round two-sphere
due to the background curvature effects. In particular, for rigid quasilocal frames, we know that this is in fact simply the order of the Riemann
tensor on $\mathscr{G}$, \textit{i.e.} $\mathring{R}_{IJKL}=\mathcal{O}(\mathcal{R})$.
Thus, one may ultimately desire to take $\mathcal{O}(\mathcal{R})$
effects into account for a full calculation, but for the moment---since,
in principle, this $\mathcal{R}$ is unrelated to $\lambda$ and we
can assume it to be subdominant thereto---we simply omit them. Thus
we can simply take $\mathring{\mathscr{S}}=\mathbb{S}^{2}_{r}$, and we
can assume that there is no shift, so that $\tilde{\gamma}=1$.

Applying the coordinate transformation in Eqs. (\ref{eq:FNC_T})-(\ref{eq:FNC_XI})
to the background metric given by Eqs. (\ref{eq:FNC_g00})-(\ref{eq:FNC_gIJ}) with $\bm{W}=0$, and then
using all of the definitions that we have established so far, it is
possible to obtain by direct computation all of the quantities appearing in the integrand of the conservation law [Eq. (\ref{eq:dpdt_RQF})] as series in $r$. We display the results
only up to leading order in $r$, including the possibility of setting
$\bm{A}=0$:
\begin{align}
\mathring{N}=\, & 1+rA_{I}r^{I}+\frac{1}{2}r^{2}\mathring{E}_{IJ}r^{I}r^{J}+\mathcal{O}\left(r^{3}\right)\,,\label{eq:N0}\\
\mathring{\mathcal{E}}=\, & \mathcal{E}_{\textrm{vac}}+\mathcal{O}\left(r\right)\nonumber \\
=\, & -\frac{2}{\kappa r}+\mathcal{O}\left(r\right)\,,\label{eq:E0}\\
\mathring{\alpha}_{\mathfrak{i}}=\, & rA_{I}\mathfrak{B}_{\mathfrak{i}}^{I}+r^{2}\left(\mathring{E}_{IJ}-A_{I}A_{J}\right)\mathfrak{B}_{\mathfrak{i}}^{I}r^{J}+\mathcal{O}\left(r^{3}\right)\,,\label{eq:alpha0}\\
\mathring{\nu}=\, & -r\mathring{B}_{IJ}r^{I}r^{J}+\mathcal{O}\left(r^{2}\right)\,,\label{eq:nu0}\\
\mathring{\mathcal{P}}_{\mathfrak{i}}=\, & -\frac{1}{\kappa}r^{2}\mathring{B}_{IJ}\mathfrak{R}_{\mathfrak{i}}^{I}r^{J}+\mathcal{O}\left(r^{3}\right)\,,\label{eq:P0_i}\\
\mathring{{\rm P}}=\, & {\rm P}_{\textrm{vac}}-\frac{1}{\kappa}A_{I}r^{I}+\mathcal{O}\left(r\right)\nonumber \\
=\, & -\frac{1}{\kappa r}-\frac{1}{\kappa}A_{I}r^{I}+\mathcal{O}\left(r\right)\,.\label{eq:P0}
\end{align}
Here, $\mathring{E}_{IJ}=\mathring{C}_{0I0J}|_{\mathscr{G}}$ and
$\mathring{B}_{IJ}=\frac{1}{2}\epsilon_{I}\,^{KL}\mathring{C}_{0JKL}|_{\mathscr{G}}$
are respectively the electric and magnetic parts of the Weyl tensor evaluated on
on $\mathscr{G}$. Also, $\mathfrak{B}_{\mathfrak{i}}^{I}=\partial_{\mathfrak{i}}r^{I}$
and $\mathfrak{R}_{\mathfrak{i}}^{I}=\epsilon_{\,\,\mathfrak{i}}^{\mathbb{S}^2}\,^{\mathfrak{j}}\mathfrak{B}_{\mathfrak{j}}^{I}$
are respectively the boost and rotation generators of $\mathbb{S}^{2}$. See Appendix \ref{sec:A} for more technical details on this.
We remind the reader that $\mathcal{E}_{\textrm{vac}}$
and ${\rm P}_{\textrm{vac}}$ are respectively
the vacuum energy and pressure, Eqs. (\ref{eq:E_vac})-(\ref{eq:P_vac}) respectively.

The way to proceed is now clear: we expand Eq. (\ref{eq:dpdt_RQF})
as a series in $\lambda$,
\begin{equation}
\dot{\mathtt{p}}^{(\bm{\phi})}=(\dot{\mathtt{p}}^{(\bm{\phi})})_{(0)}+\lambda\delta\dot{\mathtt{p}}^{(\bm{\phi})}+\mathcal{O}\left(\lambda^{2}\right)\,,
\end{equation}
using the zeroth-order parts of the various terms written above. We
need only to specify the worldline $\mathscr{G}$ in $\mathring{\mathscr{M}}$
about which we are carrying out the Fermi normal coordinate expansion (in $r$). We will consider
two cases: $\mathscr{G}=\mathring{\mathscr{C}}$ (the geodesic, such
that $\mathscr{B}$ is inertial with the point particle in $\mathring{\mathscr{M}}$)
and $\mathscr{G}=\mathscr{C}$ (an accelerated worldline such that
$\mathscr{B}_{(\lambda)}$ is inertial with the object in $\mathscr{M}_{(\lambda)}$,
\textit{i.e.} it is defined by a constant $r>C\lambda$ in $\mathscr{M}_{(\lambda)}$).
These will give us equivalent descriptions of the dynamics of the
system, from two different ``points of view'', or (quasilocal) frames of reference.

Before entering into the calculations, we can simplify things further
by remarking that the zeroth order expansions in Eqs. (\ref{eq:N0})-(\ref{eq:P0})
will always make the twist ($\nu$) term in the conservation law [Eq. (\ref{eq:dpdt_RQF})] appear at
$\mathcal{O}(r)$ or higher, both in $(\dot{\mathtt{p}}^{(\bm{\phi})})_{(0)}$
and $\delta\dot{\mathtt{p}}^{(\bm{\phi})}$, regardless of our choice
of $\mathscr{G}$. Hence we can safely ignore it, as we are interested (at least for this work)
only in the part of the conservation law which is zeroth-order in $r$. Thus we simply work
with
\begin{equation}
\dot{\mathtt{p}}^{(\bm{\phi})}=-\intop_{\mathbb{S}_{r}^{2}}\bm{\epsilon}^{}_{\mathbb{S}^{2}}\,r^{2}N\left(\mathcal{E}\alpha_{\bm{\phi}}+{\rm P}\bm{D}\cdot\bm{\phi}\right)\,.\label{eq:dpdt_RQF_simplified}
\end{equation}

Into this, we furthermore have to insert the multipole expansion of
the conformal Killing vector $\bm{\phi}$ given by Eq. (\ref{eq:phi_multipole}). We correspondingly write
\begin{equation}
\dot{\mathtt{p}}^{(\bm{\phi})}=\sum_{\ell\in\mathbb{N}}\dot{\mathtt{p}}^{(\bm{\phi}_{\ell})}\,,
\end{equation}
such that for any $\ell\in\mathbb{N}$, we have
\begin{equation}
\dot{\mathtt{p}}^{(\bm{\phi}_{\ell})}=-\Phi^{I_{1}\cdots I_{\ell}}\intop_{\mathbb{S}_{r}^{2}}\bm{\epsilon}^{}_{\mathbb{S}^{2}}\,rN\left(\mathcal{E}\alpha_{\mathfrak{i}}+{\rm P}D_{\mathfrak{i}}\right)D^{\mathfrak{i}}\left(\prod_{n=1}^{\ell}r_{I_{n}}\right)\,.
\end{equation}
Explicitly, the first two terms are
\begin{align}
\dot{\mathtt{p}}^{(\bm{\phi}_{\ell=1})}=\, & -\Phi^{I}\intop_{\mathbb{S}_{r}^{2}}\bm{\epsilon}^{}_{\mathbb{S}^{2}}\,rN\left(\mathcal{E}\alpha_{\mathfrak{i}}+{\rm P}D_{\mathfrak{i}}\right)\mathfrak{B}_{I}^{\mathfrak{i}}\,,\label{eq:p^dot_l1}\\
\dot{\mathtt{p}}^{(\bm{\phi}_{\ell=2})}=\, & -2\Phi^{IJ}\intop_{\mathbb{S}_{r}^{2}}\bm{\epsilon}^{}_{\mathbb{S}^{2}}\,rN\left(\mathcal{E}\alpha_{\mathfrak{i}}+{\rm P}D_{\mathfrak{i}}\right)\left(\mathfrak{B}_{I}^{\mathfrak{i}}r_{J}\right)\,.\label{eq:p^dot_l2}
\end{align}

\subsection{Equation of motion inertial with the background point particle\label{ssec:gralla-wald_pp-inertial}}

\begin{figure}
\noindent \begin{centering}
\includegraphics[scale=0.55]{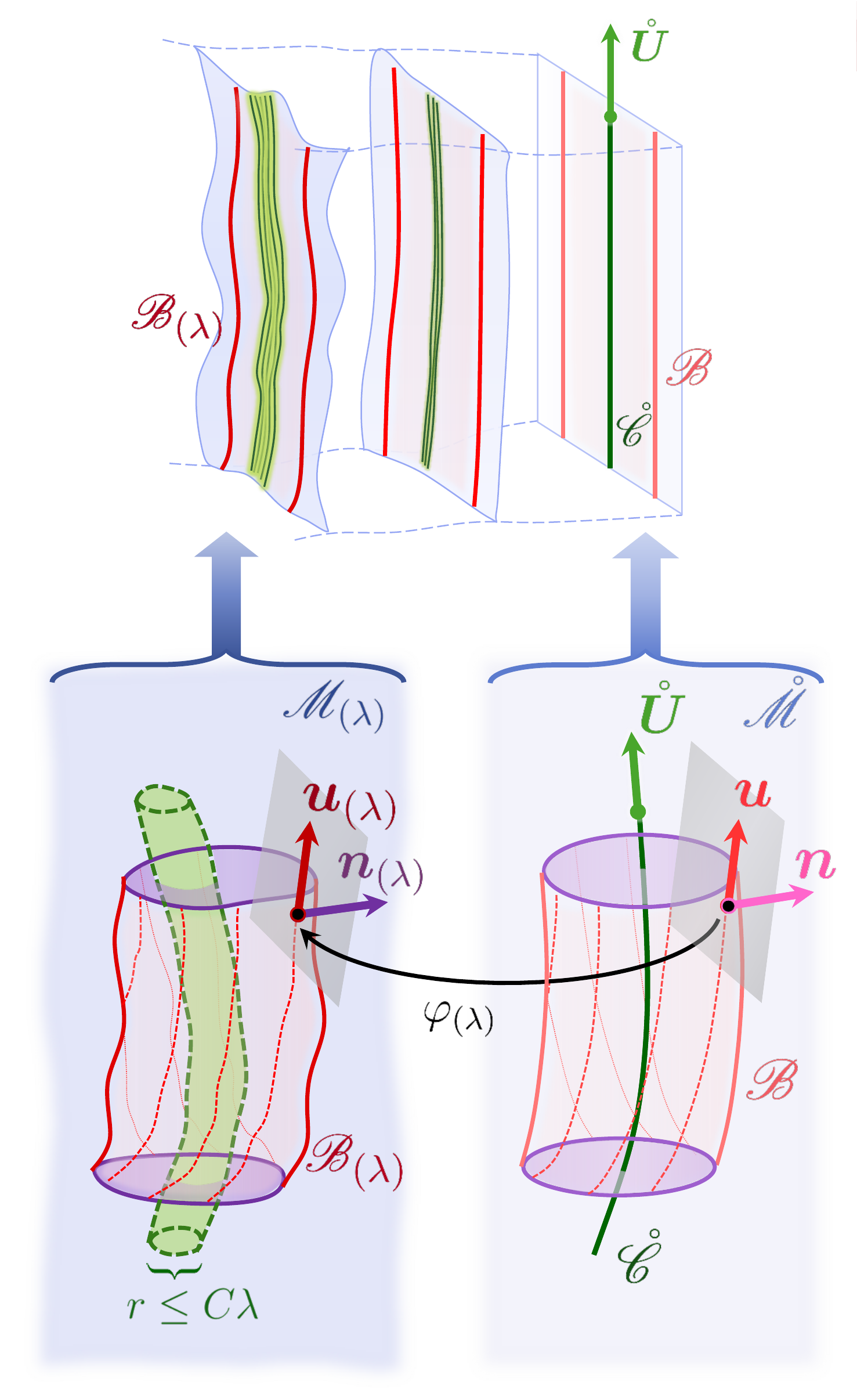}
\par\end{centering}
\caption{A family of rigid quasilocal frames $\{(\mathscr{B}_{(\lambda)};\bm{u}_{(\lambda)})\}$ embedded in the Gralla-Wald family of spacetimes $\{\mathscr{M}_{(\lambda)}\}$ such that the inverse image of any such perturbed quasilocal frame in the background is inertial with the point particle approximation of the moving object, \textit{i.e.} is centered on the geodesic $\mathring{\mathscr{C}}$.}\label{fig-pp}

\end{figure}

Let $\mathscr{G}=\mathring{\mathscr{C}}$. Then $\bm{A}=0$. We will
take this to be the case for the rest of this subsection---corresponding, as discussed, to a rigid quasilocal frame the inverse image in the background of which is inertial with the point particle approximation of the moving object in the background spacetime. This situation is displayed visually in Fig. \ref{fig-pp}.

Let us first compute the zeroth-order (in $\lambda$) part of $\dot{\mathtt{p}}^{(\bm{\phi})}$.
Inserting (\ref{eq:N0})-(\ref{eq:P0}) into the zeroth-order part
of (\ref{eq:p^dot_l1})-(\ref{eq:p^dot_l2}), and making
use of the various properties in Appendix \ref{sec:A}, we find by direct computation:
\begin{align}
\left(\dot{\mathtt{p}}^{(\bm{\phi}_{\ell=1})}\right)_{(0)}=\, & \mathcal{O}\left(r^{2}\right)\,,\label{eq:PP-p^dot_l1_0}\\
\left(\dot{\mathtt{p}}^{(\bm{\phi}_{\ell=2})}\right)_{(0)}=\, & \mathcal{O}\left(r^{2}\right)\,.\label{eq:PP-p^dot_l2_0}
\end{align}
We provide the steps of the calculation in Appendix \ref{sec:B}.

Let us now compute the $\mathcal{O}(\lambda)$, $\ell=1$ part of $\dot{\mathtt{p}}^{(\bm{\phi})}$,
\textit{i.e.} the $\mathcal{O}(\lambda)$ part of Eq. (\ref{eq:p^dot_l1})
which as usual we denote by $\delta\dot{\mathtt{p}}^{(\bm{\phi}_{\ell=1})}$.
One can see that this will involve contributions from five $\mathcal{O}(\lambda)$
terms, respectively containing $\delta N$, $\delta\mathcal{E}$, $\delta\bm{\alpha}$,
$\delta{\rm P}$ and $\delta\bm{D}$. For convenience, we will use
the notation $(\dot{\mathtt{p}}_{(Q)}^{(\bm{\phi}_{\ell})})_{(n)}$
to indicate the term of $\delta ^{n}(\dot{\mathtt{p}}^{(\bm{\phi}_{\ell})})$
that is linear in $Q$, for any $\ell,n$. Thus we write
\begin{equation}
\delta\dot{\mathtt{p}}^{(\bm{\phi}_{\ell=1})}=\sum_{Q\in\{\delta N,\delta\mathcal{E},\delta\bm{\alpha},\delta{\rm P},\delta\bm{D}\}}\delta\dot{\mathtt{p}}_{(Q)}^{(\bm{\phi}_{\ell=1})}\,.
\end{equation}

All of the computational steps are again in Appendix \ref{sec:B}. We find:
\begin{equation}
\delta\dot{\mathtt{p}}_{(\delta N)}^{(\bm{\phi}_{\ell=1})}=-\frac{2}{\kappa}\Phi_{I}\intop_{\mathbb{S}_{r}^{2}}\bm{\epsilon}^{}_{\mathbb{S}^{2}}\,\delta Nr^{I}+\mathcal{O}\left(r^{2}\right)\,.
\end{equation}
If $\delta N$ does not vary significantly over $\mathbb{S}_{r}^{2}$,
the $\mathcal{O}(r^{0})$ part of the above would be negligible owing
to the fact that $\intop_{\mathbb{S}_{r}^{2}}\bm{\epsilon}^{}_{\mathbb{S}^{2}}\,r^{I}=0$.

\begin{figure}
\noindent \begin{centering}
\includegraphics[scale=0.55]{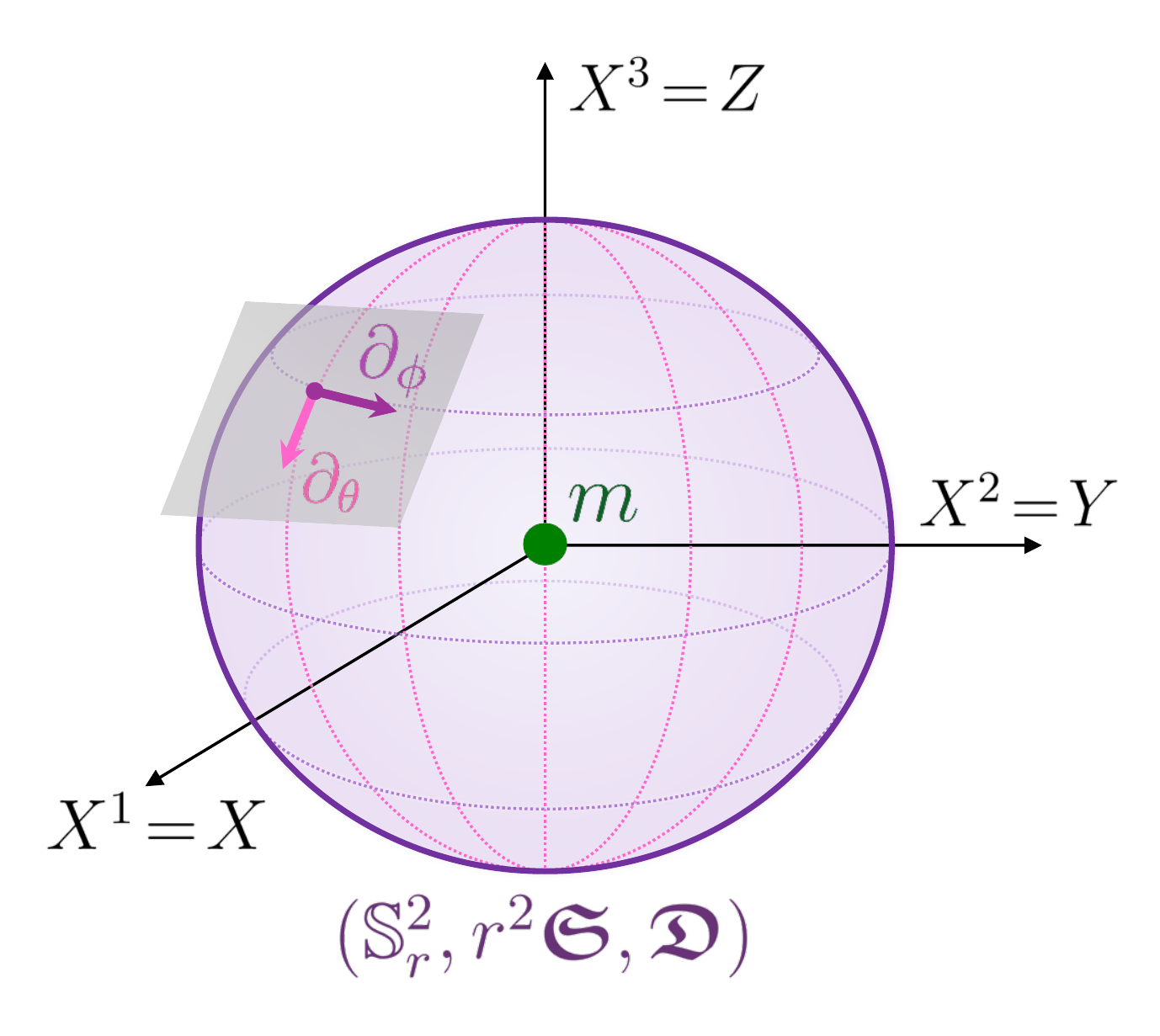}
\par\end{centering}
\caption{An instantaneous rigid quasilocal frame $(\mathbb{S}^{2}_{r},r^{2}\bm{\mathfrak{S}},\bm{\mathfrak{D}})$ (where $\bm{\mathfrak{S}}$ and $\bm{\mathfrak{D}}$ respectively are the metric and derivative compatible with the unit two-sphere) inertial with the background ``point particle''. This means that the latter is located at the center of our Fermi normal coordinate system.}\label{fig-pp-qf}
\end{figure}

Next, let us consider the $\delta\mathcal{E}$ and $\delta{\rm P}$
terms. For this, we find it useful to depict the instantaneous quasilocal frame $(\mathbb{S}^{2}_{r},r^{2}\bm{\mathfrak{S}},\bm{\mathfrak{D}})$ embedded in a constant-time three-slice of $\mathring{\mathscr{M}}$ in Fig. \ref{fig-pp-qf}.

The $\delta\mathcal{E}$ term can be easily determined by realizing that in our
current choice of quasilocal frame, the only background matter is the point particle
which is always at the center of our present coordinate system, \textit{i.e.}
it is always on $\mathring{\mathscr{C}}$ (on which we are here centering
our Fermi normal coordinates). Interpreting the constant $m$ as in the Gralla-Wald approach \cite{gralla_rigorous_2008}
to be the ``mass'' of this point particle, this simply means
that
\begin{equation}
\delta\mathcal{E}=\frac{m}{4\pi r^{2}}\,,\label{eq:C0_deltaE}
\end{equation}
so that when this is integrated (as a surface energy density) over
$\mathbb{S}_{r}^{2}$, we simply recover the mass: $\int_{\mathbb{S}_{r}^{2}}r^{2}\bm{\epsilon}^{}_{\mathbb{S}^{2}}\,\delta\mathcal{E}=m$.
We remark that, by definition, it is possible to express the quasilocal
energy as $\mathcal{E}=u^{a}u^{b}\tau_{ab}=-\tfrac{1}{\kappa}k$ with
$k=\bm{\sigma}:\bm{\Theta}$ the trace of the two-dimensional boundary extrinsic curvature. Notice that the integral of this over
a closed two-surface in the $r\rightarrow\infty$ limit is in fact the same as the usual ADM definition
of the mass/energy; thus $\delta\mathcal{E}=-\tfrac{1}{\kappa}\delta k$,
and so it makes sense to interpret $m$ as the ADM mass of the object.
So now, using Eq. (\ref{eq:C0_deltaE}), we can find that the $\delta\mathcal{E}$
contribution to $\delta\dot{\mathtt{p}}^{(\bm{\phi}_{\ell=1})}$ is
also at most quadratic in $r$:
\begin{equation}
\delta\dot{\mathtt{p}}_{(\delta\mathcal{E})}^{(\bm{\phi}_{\ell=1})}=\mathcal{O}\left(r^{2}\right)\,.
\end{equation}

To compute the $\delta{\rm P}$ term, we now employ the useful identity in Eq.
(\ref{eq:E-P_relation}), which tells us that
\begin{equation}
\delta{\rm P}=\frac{1}{2}\delta\mathcal{E}-\frac{1}{\kappa}\delta a_{\bm{n}}\,.
\end{equation}
Using this, into which we insert the $\delta\mathcal{E}$ from Eq. (\ref{eq:C0_deltaE}),
we find that the $\delta{\rm P}$ contribution to $\delta\dot{\mathtt{p}}^{(\bm{\phi}_{\ell=1})}$
is at most quadratic in $r$ as well,
\begin{equation}
\delta\dot{\mathtt{p}}_{(\delta{\rm P})}^{(\bm{\phi}_{\ell=1})}=\mathcal{O}\left(r^{2}\right)\,.
\end{equation}

Note that the above results may in fact be higher order in $r$ than quadratic.
We have only explicitly checked that they vanish up to linear order
inclusive.

Finally we are left with the $\delta\bm{\alpha}$ and $\delta\bm{D}$
contributions to $\delta\dot{\mathtt{p}}^{(\bm{\phi}_{\ell=1})}$.
By direct computation, it is possible to show that their sum is in
fact precisely what we have referred to as the extended GSF in our general
analysis of the preceding section, \textit{i.e.} it is the $\ell=1$ part of Eq.
(\ref{eq:Delta_p_GSF_general}),
\begin{equation}
\delta\dot{\mathtt{p}}_{(\delta\bm{\alpha})}^{(\bm{\phi}_{\ell=1})}+\delta\dot{\mathtt{p}}_{(\delta\bm{D})}^{(\bm{\phi}_{\ell=1})}=\frac{{\rm d}}{{\rm d}t}\left(\Delta\mathtt{p}_{\textrm{self}}^{(\bm{\phi}_{\ell=1})}\right)\,.
\end{equation}
In particular, they respectively contribute the usual GSF (from $\delta\bm{\alpha}$)
and the gravitational self-pressure force (from $\delta\bm{D}$). 

Thus, we have found that the total $\mathcal{O}(\lambda)$, $\ell=1$
part of the momentum time rate of change is given at leading (zeroth)
order in $r$ by nothing more than the generalized GSF. In other words,
\begin{equation}
\boxed{\delta\dot{\mathtt{p}}^{(\bm{\phi}_{\ell=1})}=-\Phi_{I}\mathtt{F}^{I}+\mathcal{O}\left(r\right)}\,,\label{eq:C0_delta_p^dot_1}
\end{equation}
where we have defined
\begin{equation}
\mathtt{F}^{I}=-\frac{2}{\kappa}\intop_{\mathbb{S}_{r}^{2}}\bm{\epsilon}^{}_{\mathbb{S}^{2}}\,\mathfrak{S}^{\mathfrak{ij}}\mathfrak{B}_{\mathfrak{i}}^{I}\mathcal{F}_{\mathfrak{j}}[\bm{h};\mathring{\bm{u}}]+\mathcal{O}\left(r\right)\,.
\end{equation}

Without loss of generality, let us now pick $\Phi^{I}=(0,0,1)$ to
be the unit vector in the Cartesian $X^{3}=Z$ direction, and denote
its corresponding conformal Killing vector as $\bm{\phi}_{\ell=1}=\bm{\phi}_{\ell=1}^{Z}$.
(Alternately, pick the $Z$-axis to be oriented along $\Phi^{I}$.)
We know $\mathfrak{S}^{\mathfrak{ij}}\mathfrak{B}_{\mathfrak{j}}^{Z}=(-1/\sin\theta,0)$;
moreover, by the coordinate transformation $\mathcal{F}_{\mathfrak{i}}=(\partial x^{J}/\partial x^{\mathfrak{i}})\mathcal{F}_{J}$
we have $\mathcal{F}_{\theta}=\cos\theta(\cos\phi\mathcal{F}_{X}+\sin\phi\mathcal{F}_{Y})-\sin\theta\mathcal{F}_{Z}$.
Inserting these into Eq. (\ref{eq:C0_delta_p^dot_1}) we get
\begin{multline}
\delta\dot{\mathtt{p}}^{(\bm{\phi}_{\ell=1}^{Z})}=\,-\frac{2}{\kappa}\intop_{\mathbb{S}_{r}^{2}}\bm{\epsilon}^{}_{\mathbb{S}^{2}}\,\mathcal{F}_{Z}[\bm{h};\mathring{\bm{u}}]\\
+\frac{2}{\kappa}\intop_{\mathbb{S}_{r}^{2}}{\rm d}\theta\wedge{\rm d}\phi\,\cos\theta\left(\cos\phi\mathcal{F}_{X}+\sin\phi\mathcal{F}_{Y}\right)\,.
\end{multline}
The first line is precisely in the form of the GSF term from the Gralla formula, Eq. (\ref{eq:intro_Gralla}) \cite{gralla_gauge_2011}, except here in the integrand we have (the $Z$-component of) our extended GSF $\bm{\mathcal{F}}$ [Eq. (\ref{eq:main_result_GSF_functional})]: the usual GSF $\bm{F}$ (the only self-force term in Gralla's formula) plus our self-pressure term, $\bm{\wp}$.
The second line contains additional terms involving the extended GSF in the
other two (Cartesian) spatial directions. Notice however that $\intop_{\mathbb{S}_{r}^{2}}{\rm d}\theta\wedge{\rm d}\phi\,\cos\theta\cos\phi=0=\intop_{\mathbb{S}_{r}^{2}}{\rm d}\theta\wedge{\rm d}\phi\,\cos\theta\sin\phi$,
so if $\mathcal{F}_{X}$ and $\mathcal{F}_{Y}$ do not vary significantly
over $\mathbb{S}_{r}^{2}$, their contribution will be subdominant
to that of $\mathcal{F}_{Z}$.

Thus, we have shown that our EoM (\ref{eq:C0_delta_p^dot_1}) always contains Gralla's ``angle average'' of the ``bare'' (usual) GSF. However, the form of (\ref{eq:C0_delta_p^dot_1}) (expressing the perturbative change in the quasilocal momentum) still cannot be \emph{directly} compared, as such, with Gralla's EoM (\ref{eq:Gralla_ange-average}) (expressing the change in a deviation vector representing the perturbative ``correction to the motion''). In the following subsection, we clarify the correspondence by repeating the calculation using a quasilocal frame inertial with the moving extended object in the perturbed spacetime (rather than with the geodesic in the background, as here). Furthermore, we conjecture that a careful imposition of the parity condition on the perturbative gauge---of which we have made no explicit use so far---would make the contribution from our ``self-pressure'' term vanish, but a detailed proof is required and remains to be carried out.

\subsection{Equation of motion inertial with the moving object in the perturbed spacetime\label{ssec:gralla-wald_sco-inertial}}

\begin{figure}
\noindent \begin{centering}
\includegraphics[scale=0.55]{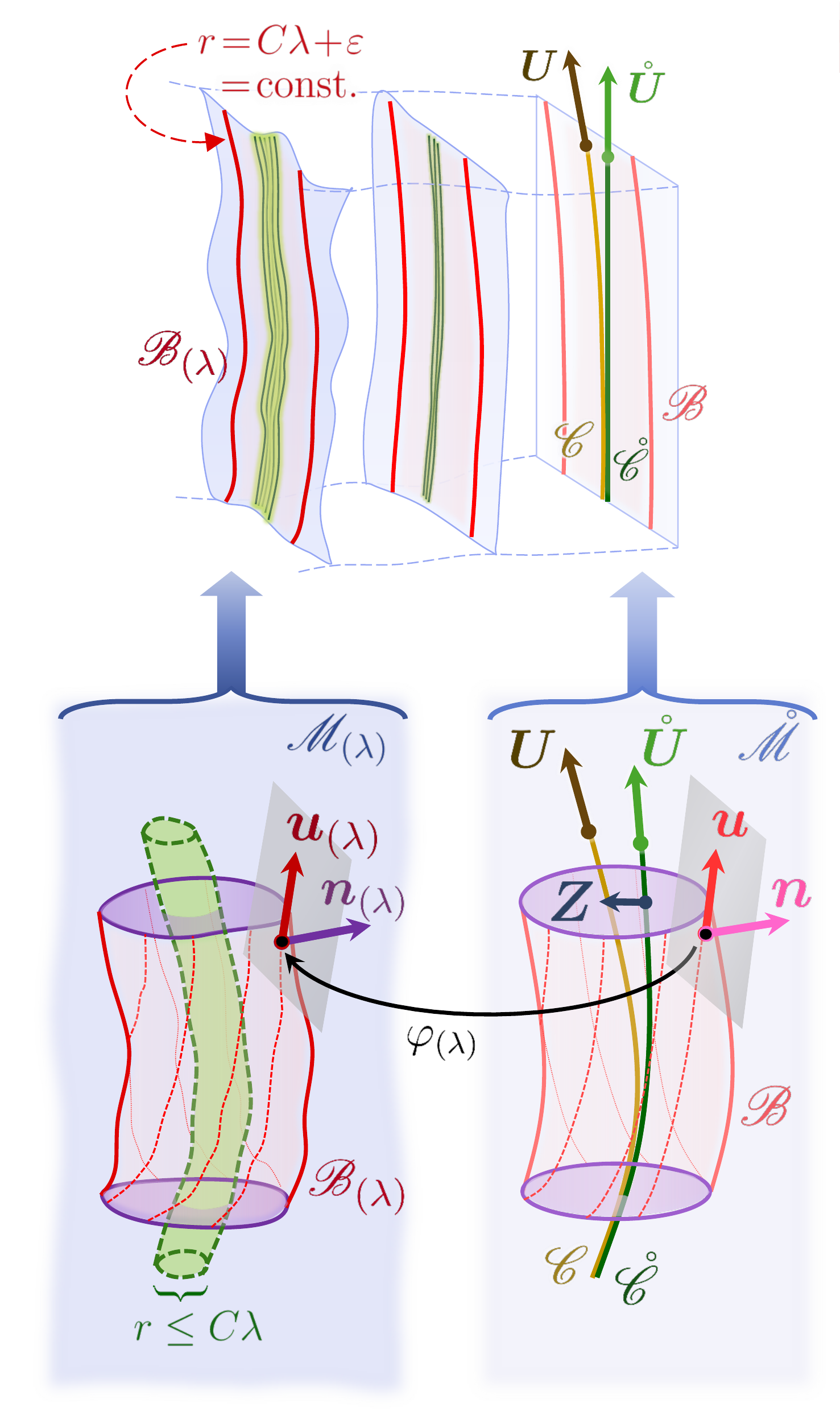}
\par\end{centering}
\caption{A family of rigid quasilocal frames $\{(\mathscr{B}_{(\lambda)};\bm{u}_{(\lambda)})\}$ embedded in the Gralla-Wald family of spacetimes $\{\mathscr{M}_{(\lambda)}\}$ inertial with the moving object in $\mathscr{M}_{(\lambda)}$. This means that $\mathscr{B}_{(\lambda)}$ is defined by the constancy of the Gralla-Wald $r$ coordinate in $\mathscr{M}_{(\lambda)}$, for any $r>C\lambda$. Thus, the inverse image $\mathscr{B}$ of $\mathscr{B}_{(\lambda)}$ in the background $\mathring{\mathscr{M}}$ is centered, in general, \emph{not} on the geodesic $\mathring{\mathscr{C}}$ followed by the point particle background approximation of the object, but on some timelike worldline $\mathscr{C}\neq\mathring{\mathscr{C}}$, with four-velocity $\bm{U}\neq \mathring{\bm{U}}$, which may be regarded as an approximation on $\mathring{\mathscr{M}}$ of the ``true motion'' of the object in $\mathscr{M}_{(\lambda)}$. Between $\mathring{\mathscr{C}}$ and $\mathscr{C}$ there is a deviation vector $\bm{Z}$, which is to be compared with the deviation vector (``correction to the motion'') in the Gralla-Wald approach.}\label{fig-sco}
\end{figure}

Now let $\mathscr{G}=\mathscr{C}\neq\mathring{\mathscr{C}}$ (so $\bm{A}\neq0$
in general) such that the quasilocal frame $(\mathscr{B};\bm{u})$ centered on $\mathscr{C}$
(in $\mathring{\mathscr{M}}$) is the inverse image of the rigid quasilocal frame $(\mathscr{B}_{(\lambda)};\bm{u}_{(\lambda)})$
defined by $r=C\lambda+\varepsilon={\rm const.}$, $\forall\varepsilon>0$,
in $\mathscr{M}_{(\lambda)}$. The meaning of the $r$ coordinate
in the latter is as given in the Gralla-Wald assumptions (Subsection \ref{ssec:gralla-wald_review}). This situation is displayed in Fig. \ref{fig-sco}. 

We now proceed to calculate, in the same way as we did for the point-particle-inertial
case, the various terms in the expansion of the momentum conservation
law, Eqs. (\ref{eq:p^dot_l1})-(\ref{eq:p^dot_l2}). At zeroth order we obtain:
\begin{align}
\left(\dot{\mathtt{p}}^{(\bm{\phi}_{\ell=1})}\right)_{(0)}=\, & \mathcal{O}\left(r^{2}\right)\,,\\
\left(\dot{\mathtt{p}}^{(\bm{\phi}_{\ell=2})}\right)_{(0)}=\, & \mathcal{O}\left(r^{2}\right)\,.
\end{align}
The steps of all these computations are again shown in Appendix \ref{sec:B}.

Let us now compute the $\mathcal{O}(\lambda)$, $\ell=1$ part of $\dot{\mathtt{p}}^{(\bm{\phi})}$.
First, we find that $\delta\dot{\mathtt{p}}_{(\delta N)}^{(\bm{\phi}_{\ell=1})}$
is the same as in the point-particle-inertial case, so if $\delta N$ does not
vary significantly over $\mathbb{S}_{r}^{2}$, the $\mathcal{O}(r^0)$
part thereof is negligible.

\begin{figure}
\noindent \begin{centering}
\includegraphics[scale=0.55]{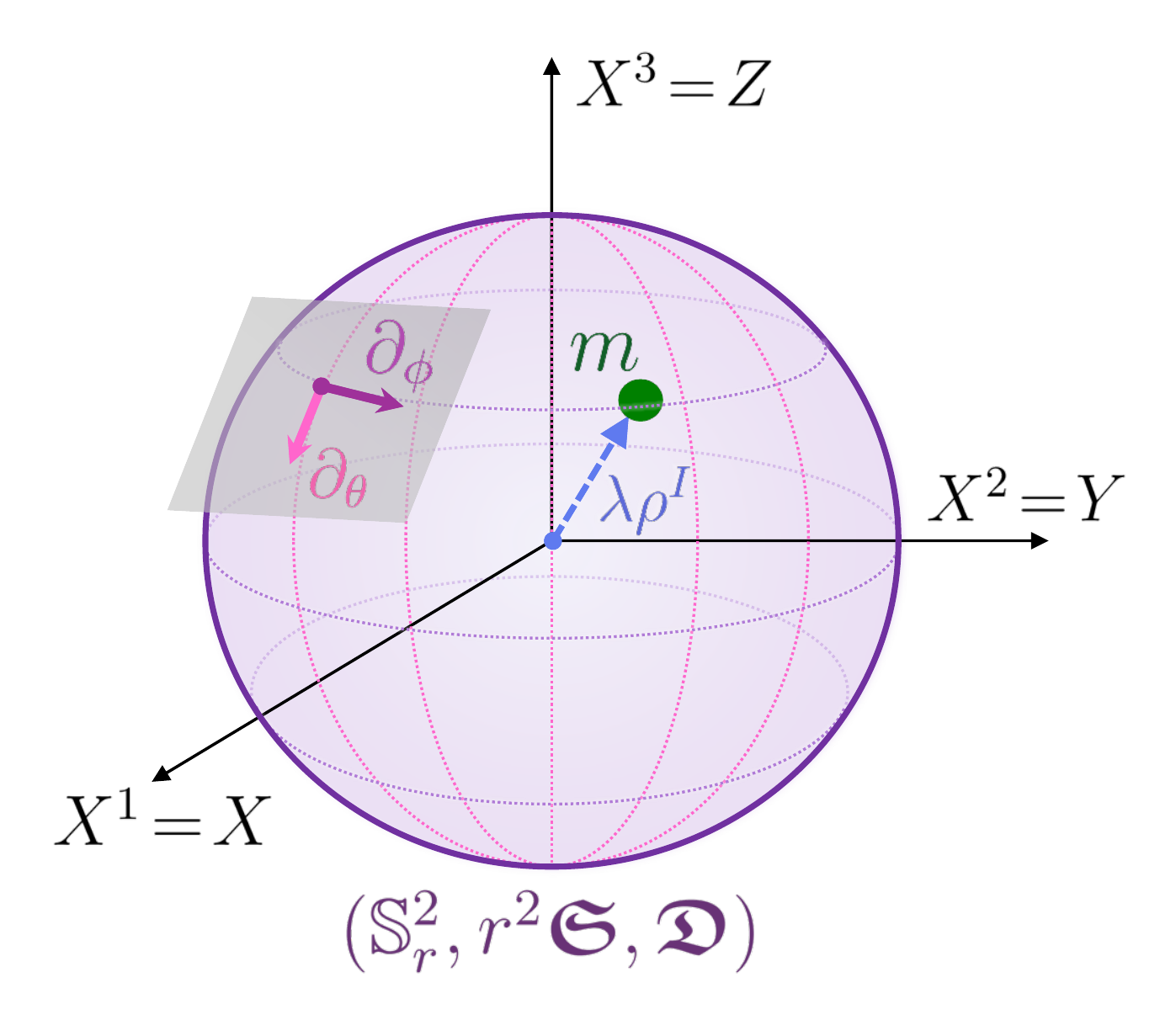}
\par\end{centering}
\caption{An instantaneous rigid quasilocal frame $(\mathbb{S}^{2}_{r},r^{2}\bm{\mathfrak{S}},\bm{\mathfrak{D}})$ (where $\bm{\mathfrak{S}}$ and $\bm{\mathfrak{D}}$ respectively are the metric and derivative compatible with the unit two-sphere) inertial with the moving object in the perturbed spacetime. This means that the point particle approximation of this object in the background spacetime is \emph{not} located at the center of our Fermi normal coordinate system. Instead, it is displaced in some direction $\rho^{I}$, which must be $\mathcal{O}(\lambda)$.}\label{fig-sco-qf}
\end{figure}

Next let us look at the $\delta\mathcal{E}$ and $\delta{\rm P}$
parts. Again, it is useful to consider in this case the visual depiction of the instantaneous quasilocal frame, shown in Fig. \ref{fig-sco-qf}. 

In this case, the particle (delta function) will \emph{not}
be at the center of our coordinate system but instead displaced in
some direction $\rho^{I}$ relative thereto. Nonetheless, we know
that this displacement must itself be $\mathcal{O}(\lambda)$ which
means that it will only contribute $\mathcal{O}(\lambda)$ corrections
to the $\delta\mathcal{E}$ having $m$ exactly at the center, \textit{i.e.} we have
\begin{equation}
\delta\mathcal{E}=\frac{m}{4\pi\left(X^{I}-\lambda\rho^{I}\right)\left(X_{I}-\lambda\rho_{I}\right)}=\frac{m}{4\pi r^{2}}+\mathcal{O}\left(\lambda\right)\,,
\end{equation}
and as before, $\delta{\rm P}=\frac{1}{2}\delta\mathcal{E}-\frac{1}{\kappa}\delta a_{\bm{n}}$.
Using these, we find:
\begin{align}
\delta\dot{\mathtt{p}}_{(\delta\mathcal{E})}^{(\bm{\phi}_{\ell=1})}=\, & -\frac{2}{3}m\Phi_{I}A^{I}+\mathcal{O}\left(r\right)\,,\\
\delta\dot{\mathtt{p}}_{(\delta{\rm P})}^{(\bm{\phi}_{\ell=1})}=\, & +\frac{1}{3}m\Phi_{I}A^{I}+\mathcal{O}\left(r\right)\,,
\end{align}
with the steps shown in Appendix \ref{sec:B}. Thus,
\begin{equation}
\delta\dot{\mathtt{p}}_{(\delta\mathcal{E})}^{(\bm{\phi}_{\ell=1})}+\delta\dot{\mathtt{p}}_{(\delta{\rm P})}^{(\bm{\phi}_{\ell=1})}=-\frac{1}{3}m\Phi_{I}A^{I}+\mathcal{O}\left(r\right)\,.
\end{equation}
Meanwhile, we still have, exactly as in the point-particle-inertial case,
\begin{equation}
\delta\dot{\mathtt{p}}_{(\delta\bm{\alpha})}^{(\bm{\phi}_{\ell=1})}+\delta\dot{\mathtt{p}}_{(\delta\bm{D})}^{(\bm{\phi}_{\ell=1})}=-\Phi_{I}\mathtt{F}^{I}+\mathcal{O}\left(r\right)\,.
\end{equation}

Now, by construction, we know that here $\delta\dot{\mathtt{p}}^{(\bm{\phi}_{\ell=1})}=0$,
as we are inertial with the moving object (in the ``actual'' spacetime $\mathscr{M}_{(\lambda)}$).
Thus summing the above and equating them to zero, we get
\begin{equation}
0=\Phi_{I}\left(-mA^{I}-3\mathtt{F}^{I}\right)+\mathcal{O}\left(r\right)\,.
\end{equation}
Since $\Phi^{I}$ is arbitrary, we thus get the EoM 
\begin{equation}
mA^{I}=-3\mathtt{F}^{I}\,\label{eq:mAI}    
\end{equation}
in the $r\rightarrow0$ limit. 

Finally, to cast this EoM into the same form as Gralla-Wald \cite{gralla_rigorous_2008}, \textit{i.e.}
in terms of a deviation vector $\bm{Z}$ on $\mathring{\mathscr{C}}$
rather than in terms of the proper acceleration $\bm{A}$ of $\mathscr{C}$, we
use the generalized deviation equation (as the name suggests, the deviation equation between arbitrary worldlines, not necessarily geodesics), Eq. (37) of Ref. \cite{puetzfeld_generalized_2016}. In our case, this reads
$\lambda\ddot{Z}^{I}=\lambda A^{I}-\lambda Z^{J}\mathring{E}^{I}\,_{J}+\mathcal{O}(\lambda^{2})$.
Combining this with Eq. (\ref{eq:mAI}), we finally recover the $\mathcal{O}(\lambda)$ EoM
\begin{equation}
\boxed{\lambda m\ddot{Z}^{I}=-3\lambda\mathtt{F}^{I}-\lambda\mathring{E}^{I}\,_{J}Z^{J}+\mathcal{O}\left(\lambda^{2},r\right)}\,.
\end{equation}

Note that the factor of $3$ multiplying the self-force term is in fact present in the EoM in Gralla's Appendix B, that is Eq. (B3) of Ref.  \cite{gralla_gauge_2011}. The latter, in this case, expresses the time evolution not of the deviation vector itself, but of the change in this deviation vector due to a gauge transformation, possibly including extra terms in case that transformation is out of the ``parity-regular'' class. We conjecture that a detailed analysis of the precise correspondence between our deviation vector definition and that of Gralla-Wald (which, while encoding the same intuitive notion of a perturbative ``correction to the motion'', may not be completely identical in general), together with a relation of their gauge transformation properties, would make it possible to relate these EoM's exactly.

\section{Discussion and conclusions\label{sec:concl}}

In this paper, we have used quasilocal conservation laws to develop
a novel formulation of self-force effects in general relativity, one
that is independent of the choice of the perturbative gauge and applicable
to any perturbative scheme designed to describe the correction to
the motion of a localized object. In particular, we have shown that
the correction to the motion of any finite spatial region, due to
any perturbation of any spacetime metric, is dominated when that region
is ``small'' (\textit{i.e.} at zero-th order in a series expansion in its
areal radius) by an \emph{extended }gravitational self-force: this
is the standard gravitational self-force term known up to now plus
a new term, not found in previous analyses and attributable to a gravitational
pressure effect with no analogue in Newtonian gravity, which we have
dubbed the gravitational \emph{self-pressure force}. Mathematically, we have found that the total change
in momentum $\Delta\mathtt{p}^{(\bm{\phi})}=\mathtt{p}_{\textrm{final}}^{(\bm{\phi})}-\mathtt{p}_{\textrm{initial}}^{(\bm{\phi})}$
between an initial and final time of any (gravitational plus matter)
system subject to any metric perturbation $\bm{h}$ is given, in a
direction determined by a conformal Killing vector $\bm{\phi}$ (see Subsection \ref{ssec:qf_cons_laws}), by the following
flux through the portion of the quasilocal frame (worldtube boundary)
$(\mathscr{B};\mathring{\bm{u}})$ delimited thereby:
\begin{equation}
\Delta\mathtt{p}^{(\bm{\phi})}=-\frac{c^{4}}{4\pi G}\intop_{\Delta\mathscr{B}}\bm{\epsilon}_{\mathscr{B}}^{\,}\frac{1}{r}\bm{\phi}\cdot\bm{\mathcal{F}}[\bm{h};\mathring{\bm{u}}]+\mathcal{O}\left(r\right)\,,\label{eq:concl_eom}
\end{equation}
where we have restored units, $r$ is the areal radius, and $\bm{\mathcal{F}}$
is the extended self-force functional. In particular, $\bm{\mathcal{F}}=\bm{F}+\bm{\wp}$
where $\bm{F}$ is the usual ``bare'' self-force [determined by the functional in Eq. (\ref{eq:intro_GSF_functional})] and
$\bm{\wp}$ is our novel self-pressure force [determined by the functional in Eq. (\ref{eq:pressure_functional})].

The most relevant practical application of the self-force is in the
context of modeling EMRIs. Ideally, one would like  to compute
the ``correction to the motion'' at the location of the moving object
(SCO). Yet, once a concrete perturbative procedure is established,
the latter usually ends up being described by a distribution (Dirac
delta function), rendering such a computation ill-defined unless additional
tactics (typically in the form of regularizations or Green's functions
methods) are introduced. However, if one takes a step back from the
exact point denoting the location of the ``particle'' (the distributional
support), and instead considers a flux around it, any singularities
introduced in such a model are avoided by construction.

We have, moreover,
shown that our formulation, when applied in the context of one particular and
very common approach to the self-force---namely that of Gralla and
Wald \cite{gralla_rigorous_2008}---yields equations of motion of the same form as those known up to now; in particular, they always contain, in the appropriate limit, the ``angle
average'' self-force term of Gralla \cite{gralla_gauge_2011}. We conjecture that a more rigorous study of these equations of motion and their gauge transformation properties would prove their exact correspondence under appropriate conditions.

We would like here to offer a concluding discussion on our results
in this paper in Subsection \ref{ssec:discussion}, as well as outlook towards future work
in Subsection \ref{ssec:outlook}.

\subsection{Discussion of results\label{ssec:discussion}}

From a physical point of view, our approach offers a fresh and conceptually
clear perspective on the basic mechanism responsible for the emergence
of self-force effects in general relativity. In particular, we have
demonstrated that the self-force may be regarded as nothing more than the manifestation
of a \emph{physical flux of gravitational momentum} passing through
the boundary enclosing the ``small'' moving object. This gravitational
momentum, and gravitational stress-energy-momentum in general, cannot
be defined locally in general relativity. As we have argued at length
in this paper, such notions must instead be defined quasilocally, \textit{i.e.} as
boundary rather than as a volume densities. This is why the self-force
appears mathematically as a boundary integral
around the moving object [Eq. (\ref{eq:Delta_p_GSF_general})], dominant in the limit where the areal radius
is small.  

The interpretation of the physical meaning of the self-force as a
consequence of conservation principles leads to many interesting implications.
As we have seen, the ``mass'' of the moving object---\textit{e.g.}, the
mass $m$ of the SCO in the EMRI problem---seems to have nothing
to do \emph{fundamentally} with the general existence of a self-force
effect. Indeed, according to our analysis, the self-force is in fact
generically present as a correction to the motion---and dominant
when the moving region is ``small''---whenever one has \emph{any}
perturbation $\boldsymbol{h}$ to the spacetime metric that is non-vanishing
on the boundary of the system.

The usual way to understand the gravitational self-force up
to now has been to regard it as a backreaction of $m$ on the metric,
\textit{i.e.} on the gravitational field, and thus in turn upon its own motion
through that field. Schematically, one thus imagines that the linear
correction to the motion is ``linear in $m$'' (or more generally,
that the full correction is an infinite series in $m$), \textit{i.e.}
that it has the form $\delta\dot{\mathtt{p}}\sim m\delta\mathtt{a}$,
with a ``perturbed acceleration'' $\delta\mathtt{a}$ determined
by $\bm{h}$ (according to some perturbative prescription) causing
a correction to the momentum $\delta\dot{\mathtt{p}}$ by a (linear)
coupling to the mass $m$.

Our analysis, instead, shows that this momentum correction $\delta\dot{\mathtt{p}}$
actually arises fundamentally in the schematic form 
\begin{equation}
\delta\dot{\mathtt{p}}\sim\mathcal{E}_{\textrm{vac}}\delta\mathtt{a}+{\rm P}_{\textrm{vac}}\delta D\,,\label{eq:quasilocal_pdot_schematic}
\end{equation}
where $\mathcal{E}_{\textrm{vac}}$ and ${\rm P}_{\textrm{vac}}$
are the \emph{vacuum} energy and pressure [Eqs. (\ref{eq:E_vac})-(\ref{eq:P_vac}) respectively], and $\delta\mathtt{a}$
and $\delta D$ are perturbed acceleration and gradient terms determined
by $\bm{h}$. Thus it is the vacuum energy (or ``mass'') and vacuum
pressure, \emph{not} the ``mass'' of the moving object, which are
responsible for the backreaction that produces self-force corrections. 

Certainly, the metric perturbation $\bm{h}$ on the system boundary
determining the perturbed acceleration and gradient terms in (\ref{eq:quasilocal_pdot_schematic}) may
in turn be sourced by a ``small mass'' present in the interior of
the system. In fact, if indeed the system is ``small'', there may
well be little physical reason for expecting that (the dominant part
of) $\bm{h}$ would come for anything \emph{other than} the presence
of the ``small'' system itself. Concordantly, the aim of any concrete
self-force analysis is to prescribe exactly \emph{how} $\bm{h}$ is
sourced thereby. Nevertheless, the correction (\ref{eq:concl_eom}) is valid \emph{regardless}
of where $\bm{h}$ comes from, and regardless of the interior description
of the system, which may very well be completely empty of matter or even contain ``exotic'' matter (as long as a well-posed initial value formulation exists). The EMRI problem is just a special case, where $\bm{h}$ is sourced in the background, according to the approach considered here, by a rudimentary point particle of mass $m$.

This opens up many interesting conceptual questions, especially with
regards to the meaning of the quasilocal vacuum energy and pressure.
While traditionally these have often been regarded as unphysical,
to be ``subtracted away'' as reference terms (for the same reason
that a ``reference action'' is often subtracted from the total gravitational
action in Lagrangian formulations of GR), our analysis in this paper
reveals instead that they are absolutely indispensable to accounting for
self-force effects. (Indeed, the initial work \cite{epp_momentum_2013} on the formulation of the quasilocal momentum conservation laws had similarly revealed the necessity of keeping these terms for a proper accounting of gravitational energy-momentum transfer in general.) To put it simply, the vacuum energy is what seems
to play the role of the ``mass'' in the ``mass times acceleration''
of the self-force; the pressure term, leading to what we have called
the self-pressure force, has no Newtonian analogue.

Now let us comment on our results from a more mathematical and technical
point of view. When applied to a specific self-force analysis, namely
that of Gralla and Wald \cite{gralla_rigorous_2008}, we have been able to recover the ``angle
average'' formula of Gralla \cite{gralla_gauge_2011}. The latter was put forward on the basis
of a convenient mathematical argument in a Hamiltonian setting.
As the quasilocal stress-energy-momentum definitions that we have
been working with (namely, as given by the Brown-York tensor) recover
the usual Hamiltonian definitions under appropriate conditions (stationary
asymptotically-flat spacetimes with a parity condition), it is reasonable
that our general equation of motion [Eq. (\ref{eq:Delta_p_GSF_general})]---expressing the physical flux
of gravitational momentum---should thereby recover that of Gralla [Eq. (\ref{eq:Gralla_ange-average})]---expressing
an ``angle average'' in a setting where certain surface integral
definitions of general-relativistic Hamiltonian notions (in particular, a Hamiltonian ``center of mass'') can be well-defined.
The limitation of Gralla's equation of motion (\textit{e.g.} in terms of the
perturbative gauge restriction attached to it) \textit{vis-à-vis} our general
equation of motion is therefore essentially the reflection of the
general limitation of Hamiltonian notions of gravitational stress-energy-momentum
(as defined for a total, asymptotically-flat spacetime with parity
conditions) \textit{vis-à-vis} general quasilocal notions of such concepts---of which the Hamiltonian ones arise simply as a special case.

We conjecture, although a more detailed analysis is needed to show this in detail, that the novel appearance of our self-pressure term and the precise forms of our equations of motion obtained in this paper relative to known results are attributable to our lack of imposition of a perturbative gauge, and possibly a more complicated relationship than direct identification between the simple Hamiltonian center of mass used in past self-force analyses and the choices of the centers of quasilocal frames that we have been working with here. Nevertheless, we maintain that the pressure term in general, though less intuitive than the energy term and having no Newtonian analogue, ought to be viewed as just as fundamental when working with conservation laws based on the Brown-York tensor. Their intimate connection is already implied in the general identity (\ref{eq:E-P_relation}), and we refer the reader to Section 3.3 of Ref. \cite{epp_momentum_2013} for a detailed discussion of why the two vacuum (energy and pressure) terms are logically self-consistent. Essentially, a negative vacuum pressure acts as a positive surface tension on the quasilocal frame which is exactly compensated by the negative vacuum energy.

For carrying out practical EMRI computations, there is a manifest
advantage in formulating the self-force as a closed two-surface integral
around the moving ``particle'' versus standard approaches. In the
latter, one typically attempts to formulate the problem \emph{at}
the ``particle location'', \textit{i.e.} the support of the distributional
matter stress-energy-momentum tensor modeling the moving object (SCO)
in the background spacetime. Of course, due to the distributional
source, $\bm{h}$ actually diverges on its support, and so regularization
or Green's function methods are typically employed in order to make
progress. However, in principle, no such obstacles are encountered
(nor the aforementioned technical solutions needed) if the self-force
is evaluated on a boundary around---very close to, but at a finite
distance away from---the ``particle'', where no formal singularity
is ever encountered: $\bm{h}$ remains everywhere finite over the integration,
and therefore so does the (extended) self-force functional [Eq. (\ref{eq:main_result_GSF_functional})] with it
directly as its argument. 

\subsection{Outlook to future work\label{ssec:outlook}}

A numerical implementation of a concrete self-force computation using
the approach developed in this paper would be arguably the most salient
next step  to take. To our knowledge, no numerical work
has been put forth even using Gralla's ``angle average'' integral
formula \cite{gralla_gauge_2011} (which would further require gauge transformations away from
``parity-regular'' gauges).

We stress here that our proposed equation of motion involving the
gravitational self-force is entirely formulated and in principle valid
in any choice of perturbative gauge. To our knowledge, this is the first such proposal bearing this feature. This may provide a great advantage
for numerical work, as black hole metric perturbations $\bm{h}$ are
often most easily computed (by solving the linearized Einstein equation,
usually with a delta-function source motivated as in or similarly
to the Gralla-Wald approach \cite{gralla_rigorous_2008} described in Subection \ref{ssec:gralla-wald_review}) in gauges that
are \emph{not} in the ``parity-regular'' class restricting Gralla's
formula \cite{gralla_gauge_2011}. In other words, we claim that one may solve the linearized
Einstein equation [Eq. (\ref{eq:first-order_Einstein_equation})] for $\bm{h}^{\bm{\mathsf{X}}}$ in any desired choice
of gauge $\bm{\mathsf{X}}$, insert this $\bm{h}^{\bm{\mathsf{X}}}$
into our extended GSF functional [Eq. (\ref{eq:main_result_GSF_functional})] to obtain $\mathcal{\bm{F}}^{\bm{\mathsf{X}}}[\bm{h}^{\bm{\mathsf{X}}};\mathring{\bm{u}}^{\bm{\mathsf{X}}}]$
(for some choice of background quasilocal frame with four-velocity
$\mathring{\bm{u}}$), and then to integrate this over a ``small
radius'' topological two-sphere surrounding the ``particle'' (so
that $\mathring{\bm{u}}$ can be approximated by the background
geodesic four-velocity of the particle, $\mathring{\bm{U}}$), to obtain the full
\emph{extended} gravitational self-force (or ``correction to the
motion'') \emph{directly in that gauge} $\bm{\mathsf{X}}$. It is
easy to speculate that this may simplify some numerical issues tremendously
\textit{vis-à-vis} current approaches, where much technical machinery is needed
to handle (and to do so in a sufficiently efficient way for future
waveform applications) the necessary gauge transformations involving
distributional source terms.

Nevertheless, further work is needed to bring the relatively abstract
analysis developed in this paper into a form more readily suited for
practical numerics. The most apparent technical issue to be tackled
involves the fact that $\bm{h}$ is usually computed (in some kind
of harmonics) in angular coordinates centered on the MBH, while the
functional $\bm{\mathcal{F}}[\bm{h};\mathring{\bm{u}}]$ is evaluated in
angular coordinates (on a ``small'' topological two-sphere) centered
on the moving ``particle'', \textit{i.e.} the SCO. A detailed understanding
of the transformation between the two sets of angular coordinates
is thus essential to formulate this problem numerically. This issue
is discussed a bit further in Gralla's paper \cite{gralla_gauge_2011}, but a detailed implementation
of such a computation remains to be attempted.

The abstraction and generality of our approach may, on the other hand,
also provide useful ways to address some other technical issues surrounding
the self-force problem. For example, all the calculations in this
paper may be carried on to second order (in the formal expansion parameter
$\lambda$). This is conceptually straightforward given our basic
perturbative setup, but of course  requires a detailed analysis in its
own right. Nonetheless, one may readily see that any higher-order
correction to the motion manifestly remains here in the form of a
boundary flux---only now involving nonlinear terms in the integrand.
Thus any sort of singular behaviour is avoided at the level of the
equations of motion in our approach, up to any order.

As another example, if ever desired (\textit{e.g.} for astrophysical reasons),
linear or any higher-order in $r$ (the areal radius of the SCO boundary)
effects on the correction to the motion can also be computed using
our approach. Moreover, any matter fluxes (described by the usual
matter stress-energy-momentum tensor, $\bm{T}$) can also be accommodated
thanks to our general (gravity plus matter) conservation laws [Eq. (\ref{eq:cons_law_Pa})]. 

Furthermore, while we have applied our ideas in this paper to a specific
self-force approach---that of Gralla and Wald \cite{gralla_rigorous_2008}---our general formulation
(Section \ref{sec:general-analysis}) can just as well be used in any other approach to the gravitational
self-force, \textit{i.e.} any other specification of a perturbative procedure
(of a family of perturbed spacetimes $\{(\mathscr{M}_{(\lambda)},\bm{g}_{(\lambda)}\}$)
for this problem. In other words, our approach permits any alternative
specification of what is meant by a (sufficiently) ``localized source''
in general relativity, as our conservation expressions always involve
fluxes on their boundaries and are not conditioned in any way by the
exact details of their interior modeling. Thus our equation of motion [Eq. (\ref{eq:concl_eom})] could be used not only for a ``self-consistent'' computation (using, \textit{e.g.}, an approach such as that of Refs. \cite{spallicciritter2014,riaospco2016b} for solving the field equations in this context)
within the Gralla-Wald approach, but also, for example, in the context of the (mathematically equivalent) self-consistent formulation of Pound \cite{pound_self-consistent_2010}.

Beyond the gravitational self-force, another avenue to explore from
here---of interest at the very least for conceptual consistency---is
how our approach handles the electromagnetic self-force problem. Although
undoubtedly some conceptual parallels may be drawn between the gravitational
and electromagnetic self-force problems (see \textit{e.g.} Ref. \cite{barack_self-force_2018}), foundationally
they are usually treated as separate problems. Indeed, shortly after
the paper of Gralla and Wald \cite{gralla_rigorous_2008} detailing
the self-force approach used in this work, Gralla, Harte and Wald
\cite{gralla_rigorous_2009} put forth a similar analysis, with an
analogous approach and level of rigour, of the electromagnetic self-force.
It would be of great interest to apply our quasilocal conservation
laws in this setting, as they can be used to account not just for gravitational
but also (and in a consistent way) matter fluxes as well. It may thus prove insightful to study how
the transfer of energy-momentum is actually accounted for (between
the gravitational and the matter sector), as in our approach we are
not restricted to fixing a non-dynamical metric in the spacetime.
In other words, the conservation laws account completely for fluxes
due to a dynamical geometry as well as matter.

\section*{Acknowledgements}

We thank Leor Barack and Abraham Harte for useful comments and  discussions. MO and RBM acknowledge support from the Natural Sciences and Engineering Research Council of Canada. MO and CFS acknowledge support from contracts ESP2013-47637-P, ESP2015-67234-P and ESP2017-90084-P from the Ministry of Economy and Business of Spain (MINECO), and from contract 2017-SGR-1469 from AGAUR (Catalan government). MO and ADAMS acknowledge LISA CNES funding. We also acknowledge networking support by the COST Action GWverse CA16104 (Horizon 2020 Framework Programme of the European Union).


\renewcommand{\thesection}{A}
\section{Conformal killing vectors and the two-sphere}\label{sec:A}

In this appendix we review some basic properties of \emph{conformal Killing vectors} (CKVs), and in particular
CKVs on the two-sphere.

A vector field $\bm{X}$ on any   $n$-dimensional Riemannian manifold $(\mathscr{U},\bm{g}_{\mathscr{U}},\bm{\nabla}_{\mathscr{U}})$
is a CKV if and only if it satisfies the \emph{conformal Killing equation},
\begin{equation}
\mathcal{L}_{\bm{X}}\bm{g}_{\mathscr{U}}=\psi\bm{g}_{\mathscr{U}}\,,\label{eq:A-CKV_equation_general}
\end{equation}
where $\psi\in C^{\infty}(\mathscr{U})$. This function can be determined
uniquely by taking the trace of this equation, yielding
\begin{equation}
\psi=\frac{2}{n}\bm{\nabla}_{\mathscr{U}}\cdot\bm{X}\,.\label{eq:A-CKV_equation_factor}
\end{equation}

Let us now specialize to the $r$-radius two-sphere $(\mathbb{S}_{r}^{2},r^{2}\bm{\mathfrak{S}},\bm{\mathfrak{D}})$,
where we denote our CKV by $\bm{\phi}$. Moreover for ease of notation in this appendix, the two-sphere volume form $\bm{\epsilon}_{\mathbb{S}^{2}}^{\,}$ [Eq. (\ref{eq:S2_volume_form})] is equivalently denoted by
$\bm{\mathfrak{E}}$, \textit{i.e.} $\mathfrak{E}_{\mathfrak{ij}}=\epsilon_{\mathfrak{ij}}^{\mathbb{S}^{2}}$.

In this case, the conformal
Killing equation (\ref{eq:A-CKV_equation_general}) is
\begin{equation}
\mathfrak{D}^{(\mathfrak{i}}\phi^{\mathfrak{j})}=\frac{1}{2r^{2}}\mathfrak{S}^{\mathfrak{ij}}\mathfrak{D}_{\mathfrak{k}}\phi^{\mathfrak{k}}\Leftrightarrow\mathfrak{D}^{\langle\mathfrak{i}}\phi^{\mathfrak{j}\rangle}=0\,,\label{eq:A-CKV_equation_two-sphere}
\end{equation}
where $\langle\cdot\cdot\rangle$ on two indices indicates taking
the STF part. The solution to this equation can be usefully expressed
in the form of a spherical harmonic decomposition in terms of the
standard direction cosines of a radial unit vector in $\mathbb{R}^{3}$,
which we denote by $r^{I}$. In spherical coordinates $\{x^{\mathfrak{i}}\}=\{\theta,\phi\}$,
it is simply given by
\begin{equation}
r^{I}(\theta,\phi)=(\sin\theta\cos\phi,\sin\theta\sin\phi,\cos\theta)\,.\label{eq:A-rI}
\end{equation}
This satisfies the following useful identity:
\begin{equation}
\intop_{\mathbb{S}_{r}^{2}}\bm{\epsilon}^{}_{\mathbb{S}^{2}}\prod_{n=1}^{\ell}r^{I_{\ell}}=\begin{cases}
0\,, & \textrm{for }\ell\textrm{ odd}\,,\\
\tfrac{4\pi}{(\ell+1)!!}\delta^{\{I_{1}I_{2}}\cdots\delta^{I_{\ell-1}I_{\ell}\}}\,, & \textrm{for }\ell\textrm{ even}\,,
\end{cases}
\end{equation}
where $(\ell+1)!!=(\ell+1)(\ell-1)\cdots1$
and the curly brackets on the indices denote the smallest set of permutations
that make the result symmetric. In particular, the $\ell=2$ and $\ell=4$
cases (which suffice for the calculations presented in this paper)
are:
\begin{align}
\intop_{\mathbb{S}_{r}^{2}}\bm{\epsilon}^{}_{\mathbb{S}^{2}}\,r^{I}r^{J}=\, & \frac{4\pi}{3}\delta^{IJ}\,,\\
\intop_{\mathbb{S}_{r}^{2}}\bm{\epsilon}^{}_{\mathbb{S}^{2}}\,r^{I}r^{J}r^{K}r^{L}=\, & \frac{4\pi}{15}\left(\delta^{IJ}\delta^{KL}+\delta^{IK}\delta^{JL}+\delta^{IL}\delta^{KJ}\right)\,.
\end{align}

We can construct from Eq. (\ref{eq:A-rI}) two sets of $\ell=1$ spherical
harmonic forms on $\mathbb{S}_{r}^{2}$, namely the \emph{boost generators},
\begin{multline}
\mathfrak{B}_{\mathfrak{i}}^{I}(\theta,\phi)=\mathfrak{D}_{\mathfrak{i}}r^{I}\\
=\left(\begin{array}{ccc}
\cos\theta\cos\phi & \cos\theta\sin\phi & -\sin\theta\\
-\sin\theta\sin\phi & \sin\theta\cos\phi & 0
\end{array}\right)\,,
\end{multline}
and the \emph{rotation generators},
\begin{multline}
\mathfrak{R}_{\mathfrak{i}}^{I}(\theta,\phi)=\mathfrak{E}_{\mathfrak{i}}\,^{\mathfrak{j}}\mathfrak{B}_{\mathfrak{j}}^{I}=\epsilon^{I}\,_{JK}r^{J}\mathfrak{B}_{\mathfrak{i}}^{K}\\
=\left(\begin{array}{ccc}
-\sin\phi & \cos\phi & 0\\
-\sin\theta\cos\theta\cos\phi & -\sin\theta\cos\theta\sin\phi & \sin^{2}\theta
\end{array}\right)\,,
\end{multline}
where $\epsilon_{IJK}$ is the volume form of $\mathbb{R}^{3}$. We
can obtain from these the vector fields $\mathfrak{B}_{I}^{\mathfrak{i}}=\frac{1}{r^{2}}\delta_{IJ}\mathfrak{S}^{\mathfrak{ij}}\mathfrak{B}_{\mathfrak{j}}^{J}=\mathfrak{D}^{\mathfrak{j}}r_{I}$
and $\mathfrak{R}_{I}^{\mathfrak{i}}=\frac{1}{r^{2}}\delta_{IJ}\mathfrak{S}^{\mathfrak{ij}}\mathfrak{R}_{\mathfrak{j}}^{J}=\mathfrak{E}^{\mathfrak{i}}\,_{\mathfrak{j}}\mathfrak{B}_{I}^{\mathfrak{j}}$,
which satisfy the Lorentz algebra
\begin{align}
\left[\bm{\mathfrak{B}}_{I},\bm{\mathfrak{B}}_{J}\right]=\, & \epsilon_{IJ}\,^{K}\bm{\mathfrak{R}}_{K}\,,\\
\left[\bm{\mathfrak{R}}_{I},\bm{\mathfrak{B}}_{J}\right]=\, & -\epsilon_{IJ}\,^{K}\bm{\mathfrak{B}}_{K}\,,\\
\left[\bm{\mathfrak{R}}_{I},\bm{\mathfrak{R}}_{J}\right]=\, & -\epsilon_{IJ}\,^{K}\bm{\mathfrak{R}}_{K}\,.
\end{align}

From the above, it is possible to derive a number of useful properties:
\begin{align}
r_{I}\mathfrak{B}_{\mathfrak{i}}^{I}=0\,,\quad & \mathfrak{D}_{\mathfrak{i}}\mathfrak{B}_{\mathfrak{j}}^{I}=-\mathfrak{S}_{\mathfrak{ij}}r^{I}\Rightarrow\mathfrak{S}^{\mathfrak{ij}}\mathfrak{D}_{\mathfrak{i}}\mathfrak{B}_{\mathfrak{j}}^{I}=-2r^{I}\,,\\
r_{I}\mathfrak{R}_{\mathfrak{i}}^{I}=0\,,\quad & \mathfrak{D}_{\mathfrak{i}}\mathfrak{R}_{\mathfrak{j}}^{I}=\mathfrak{E}_{\mathfrak{ij}}r^{I}\Rightarrow\mathfrak{S}^{\mathfrak{ij}}\mathfrak{D}_{\mathfrak{i}}\mathfrak{R}_{\mathfrak{j}}^{I}=0\,.
\end{align}
Using these, one can show that the sets of $\ell=1$ vector fields
$\bm{\mathfrak{B}}_{I}$ and $\bm{\mathfrak{R}}_{I}$ all satisfy
the conformal Killing equation, \textit{i.e.}
\begin{equation}
\mathfrak{D}^{\langle\mathfrak{i}}\mathfrak{B}_{I}^{\mathfrak{j}\rangle}=0=\mathfrak{D}^{\langle\mathfrak{i}}\mathfrak{R}_{I}^{\mathfrak{j}\rangle}\,.
\end{equation}
Finally, we give a list of useful relations for various contractions
involving these vector fields:
\begin{align}
\mathfrak{S}^{\mathfrak{ij}}\mathfrak{B}_{\mathfrak{i}}^{I}\mathfrak{B}_{\mathfrak{j}}^{J}=\mathfrak{S}^{\mathfrak{ij}}\mathfrak{R}_{\mathfrak{i}}^{I}\mathfrak{R}_{\mathfrak{j}}^{J}=-\mathfrak{E}^{\mathfrak{ij}}\mathfrak{B}_{\mathfrak{i}}^{I}\mathfrak{R}_{\mathfrak{j}}^{J}=\, & P^{IJ}\,,\\
\mathfrak{S}^{\mathfrak{ij}}\mathfrak{B}_{\mathfrak{i}}^{I}\mathfrak{R}_{\mathfrak{j}}^{J}=\mathfrak{E}_{\mathfrak{ij}}\mathfrak{B}_{I}^{\mathfrak{i}}\mathfrak{B}_{J}^{\mathfrak{j}}=\mathfrak{E}_{\mathfrak{ij}}\mathfrak{R}_{I}^{\mathfrak{i}}\mathfrak{R}_{J}^{\mathfrak{j}}=\, & \epsilon^{IJK}r_{K}\,,\\
\delta_{IJ}\mathfrak{B}_{\mathfrak{i}}^{I}\mathfrak{B}_{\mathfrak{j}}^{J}=\delta_{IJ}\mathfrak{R}_{\mathfrak{i}}^{I}\mathfrak{R}_{\mathfrak{j}}^{J}=\, & \mathfrak{S}_{\mathfrak{ij}}\,,\\
\delta_{IJ}\mathfrak{B}_{\mathfrak{i}}^{I}\mathfrak{R}_{\mathfrak{j}}^{J}=\, & -\mathfrak{E}_{\mathfrak{ij}}\,,\\
\epsilon_{IJK}\mathfrak{B}_{\mathfrak{i}}^{I}\mathfrak{B}_{\mathfrak{j}}^{J}=\epsilon_{IJK}\mathfrak{R}_{\mathfrak{i}}^{I}\mathfrak{R}_{\mathfrak{j}}^{J}=\, & \mathfrak{E}_{\mathfrak{ij}}r_{K}\,,\\
\epsilon_{IJK}\mathfrak{B}_{\mathfrak{i}}^{I}\mathfrak{R}_{\mathfrak{j}}^{J}=\, & \mathfrak{S}_{\mathfrak{ij}}r_{K}\,,
\end{align}
where $P^{IJ}=\delta^{IJ}-r^{I}r^{J}$ projects vectors perpendicular
to the radial direction.

Now we have everything in hand to formulate the general solution to
the conformal Killing equation (\ref{eq:A-CKV_equation_two-sphere})
on $\mathbb{S}_{r}^{2}$; it can be expanded as:
\begin{align}
\phi^{\mathfrak{j}}=\, & \frac{1}{r}\bigg[\mathfrak{D}^{\mathfrak{j}}\left(\sum_{\ell\in\mathbb{N}}\Phi^{I_{1}\cdots I_{\ell}}\prod_{n=1}^{\ell}r_{I_{n}}\right)\nonumber \\
 & +\mathfrak{E}^{\mathfrak{j}}\,_{\mathfrak{k}}\mathfrak{D}^{\mathfrak{k}}\left(\sum_{\ell\in\mathbb{N}}\Psi^{I_{1}\cdots I_{\ell}}\prod_{n=1}^{\ell}r_{I_{n}}\right)\bigg]\,,\label{eq:A-phi_general_solution}\\
=\, & \frac{1}{r}\bigg[\left(\Phi^{I}\mathfrak{B}_{I}^{\mathfrak{j}}+\Psi^{I}\mathfrak{R}_{I}^{\mathfrak{j}}\right)\nonumber \\
 & +\sum_{\ell\geq2}\ell\Big(\Phi^{I_{1}\cdots I_{\ell}}\mathfrak{B}_{I_{1}}^{\mathfrak{j}}+\Psi^{I_{1}\cdots I_{\ell}}\mathfrak{R}_{I_{1}}^{\mathfrak{j}}\Big)\prod_{n=2}^{\ell}r_{I_{n}}\bigg]\,,
\end{align}
where to write the second equality we have used the fact that $\Phi^{I_{1}\cdots I_{\ell}}$
and $\Psi^{I_{1}\cdots I_{\ell}}$ are symmetric in their indices.
One can thus check that\begin{widetext}
\begin{align}
\mathfrak{D}^{\mathfrak{i}}\phi^{\mathfrak{j}}=\, & \chi\bigg[\left(-\Phi^{I}\mathfrak{S}^{\mathfrak{ij}}+\Psi^{I}\mathfrak{E}^{\mathfrak{ij}}\right)r_{I}+2\left(\Phi^{IJ}\left\{ -\mathfrak{S}^{\mathfrak{ij}}r_{I}r_{J}+\mathfrak{B}_{I}^{\mathfrak{i}}\mathfrak{B}_{J}^{\mathfrak{j}}\right\} +\Psi^{IJ}\left\{ \mathfrak{E}^{\mathfrak{ij}}r_{I}r_{J}+\mathfrak{B}_{I}^{\mathfrak{i}}\mathfrak{R}_{J}^{\mathfrak{j}}\right\} \right)\nonumber \\
 & +\sum_{\ell\geq3}\ell\bigg(\Phi^{I_{1}\cdots I_{\ell}}\left\{ -\mathfrak{S}^{\mathfrak{ij}}r_{I_{1}}r_{I_{2}}+\left(\ell-1\right)\mathfrak{B}_{I_{1}}^{\mathfrak{i}}\mathfrak{B}_{I_{2}}^{\mathfrak{j}}\right\} +\Psi^{I_{1}\cdots I_{\ell}}\left\{ \mathfrak{E}^{\mathfrak{ij}}r_{I_{1}}r_{I_{2}}+\left(\ell-1\right)\mathfrak{B}_{I_{1}}^{\mathfrak{i}}\mathfrak{R}_{I_{2}}^{\mathfrak{j}}\right\} \bigg)\prod_{n=2}^{\ell}r_{I_{n}}\bigg]\,.
\end{align}

\end{widetext}

We are interested in working with the $\ell=1$ and $\ell=2$ parts of
$\bm{\phi}$ corresponding to linear momentum only ($\bm{\Psi}=0$):
\begin{align}
\phi_{\ell=1}^{\mathfrak{i}}=\, & \frac{1}{r}\Phi^{I}\mathfrak{B}_{I}^{\mathfrak{i}}\,,\\
\phi_{\ell=2}^{\mathfrak{i}}=\, & \frac{2}{r}\Phi^{IJ}\mathfrak{B}_{I}^{\mathfrak{i}}r_{J}\,.
\end{align}

\begin{widetext}
\renewcommand{\thesection}{B}
\section{Detailed calculations for the equations of motion}\label{sec:B}

\subsection*{Rigid quasilocal frame inertial with the background ``point particle''}

For the background momentum change, the $\ell=1$ and $\ell=2$ parts
are computed respectively as follows:

\begin{align}
\left(\dot{\mathtt{p}}^{(\bm{\phi}_{\ell=1})}\right)_{(0)}=\, & -\intop_{\mathbb{S}_{r}^{2}}\bm{\epsilon}^{}_{\mathbb{S}^{2}}\,r^{2}\mathring{N}\left(\mathring{\mathcal{E}}\mathring{\bm{\alpha}}\cdot\bm{\phi}_{\ell=1}+\mathring{{\rm P}}\bm{\mathfrak{D}}\cdot\bm{\phi}_{\ell=1}\right)\\
=\, & -\intop_{\mathbb{S}_{r}^{2}}\bm{\epsilon}^{}_{\mathbb{S}^{2}}\,r^{2}\left[1+\mathcal{O}\left(r^{2}\right)\right]\Bigg(\left[-\frac{2}{\kappa r}+\mathcal{O}\left(r\right)\right]\frac{1}{r^{2}}\mathfrak{S}^{\mathfrak{ij}}\left[r^{2}\mathring{E}_{JK}\mathfrak{B}_{\mathfrak{i}}^{J}r^{K}+\mathcal{O}\left(r^{3}\right)\right]\left[r\Phi_{I}\mathfrak{B}_{\mathfrak{j}}^{I}\right]\\
 & +\left[-\frac{1}{\kappa r}+\mathcal{O}\left(r\right)\right]\frac{1}{r^{2}}\mathfrak{S}^{\mathfrak{ij}}\mathfrak{D}_{\mathfrak{i}}\left[r\Phi_{I}\mathfrak{B}_{\mathfrak{j}}^{I}\right]\Bigg)\\
=\, & \frac{1}{\kappa}\Phi_{I}\intop_{\mathbb{S}_{r}^{2}}\bm{\epsilon}^{}_{\mathbb{S}^{2}}\,\left[1+\mathcal{O}\left(r^{2}\right)\right]\left(\left[2+\mathcal{O}\left(r^{2}\right)\right]\left[\mathcal{O}\left(r^{2}\right)\right]+\left[1+\mathcal{O}\left(r^{2}\right)\right]\mathfrak{S}^{\mathfrak{ij}}\mathfrak{D}_{\mathfrak{i}}\mathfrak{B}_{\mathfrak{j}}^{I}\right)\\
=\, & \frac{1}{\kappa}\Phi_{I}\intop_{\mathbb{S}_{r}^{2}}\bm{\epsilon}^{}_{\mathbb{S}^{2}}\,\left(-2r^{I}\right)+\mathcal{O}\left(r^{2}\right)\\
=\, & \mathcal{O}\left(r^{2}\right)\,,
\end{align}
and
\begin{align}
\left(\dot{\mathtt{p}}^{(\bm{\phi}_{\ell=2})}\right)_{(0)}=\, & -\intop_{\mathbb{S}_{r}^{2}}\bm{\epsilon}^{}_{\mathbb{S}^{2}}\,r^{2}\mathring{N}\left(\mathring{\mathcal{E}}\mathring{\bm{\alpha}}\cdot\bm{\phi}_{\ell=2}+\mathring{{\rm P}}\bm{\mathfrak{D}}\cdot\bm{\phi}_{\ell=2}\right)\\
=\, & -\intop_{\mathbb{S}_{r}^{2}}\bm{\epsilon}^{}_{\mathbb{S}^{2}}\,r^{2}\left[1+\mathcal{O}\left(r^{2}\right)\right]\Bigg(\left[-\frac{2}{\kappa r}+\mathcal{O}\left(r\right)\right]\frac{1}{r^{2}}\mathfrak{S}^{\mathfrak{ij}}\left[r^{2}\mathring{E}_{JK}\mathfrak{B}_{\mathfrak{i}}^{J}r^{K}+\mathcal{O}\left(r^{3}\right)\right]\left[2r\Phi_{IJ}r^{I}\mathfrak{B}_{\mathfrak{j}}^{J}\right]\\
 & +\left[-\frac{1}{\kappa r}+\mathcal{O}\left(r\right)\right]\frac{1}{r^{2}}\mathfrak{S}^{\mathfrak{ij}}\mathfrak{D}_{\mathfrak{i}}\left[2r\Phi_{IJ}r^{I}\mathfrak{B}_{\mathfrak{j}}^{J}\right]\Bigg)\\
=\, & \frac{1}{\kappa}\intop_{\mathbb{S}_{r}^{2}}\bm{\epsilon}^{}_{\mathbb{S}^{2}}\,\left[1+\mathcal{O}\left(r^{2}\right)\right]\left(\left[2+\mathcal{O}\left(r^{2}\right)\right]\left[\mathcal{O}\left(r^{2}\right)\right]+\left[1+\mathcal{O}\left(r^{2}\right)\right]\mathfrak{S}^{\mathfrak{ij}}\mathfrak{D}_{\mathfrak{i}}\left[2\Phi_{IJ}r^{I}\mathfrak{B}_{\mathfrak{j}}^{J}\right]\right)\\
=\, & \frac{2}{\kappa}\Phi_{IJ}\intop_{\mathbb{S}_{r}^{2}}\bm{\epsilon}^{}_{\mathbb{S}^{2}}\,\left(\mathfrak{S}^{\mathfrak{ij}}\mathfrak{D}_{\mathfrak{i}}\left[r^{I}\mathfrak{B}_{\mathfrak{j}}^{J}\right]\right)+\mathcal{O}\left(r^{2}\right)\\
=\, & \frac{2}{\kappa}\Phi_{IJ}\intop_{\mathbb{S}_{r}^{2}}\bm{\epsilon}^{}_{\mathbb{S}^{2}}\,\left(\mathfrak{S}^{\mathfrak{ij}}\mathfrak{B}_{\mathfrak{i}}^{I}\mathfrak{B}_{\mathfrak{j}}^{J}+r^{I}\mathfrak{S}^{\mathfrak{ij}}\mathfrak{D}_{\mathfrak{i}}\mathfrak{B}_{\mathfrak{j}}^{J}\right)+\mathcal{O}\left(r^{2}\right)\\
=\, & \frac{2}{\kappa}\Phi_{IJ}\intop_{\mathbb{S}_{r}^{2}}\bm{\epsilon}^{}_{\mathbb{S}^{2}}\,\left(P^{IJ}-2r^{I}r^{J}\right)+\mathcal{O}\left(r^{2}\right)\\
=\, & \mathcal{O}\left(r^{2}\right)\,.
\end{align}
For the perturbed momentum change, we have
\begin{align}
\delta\dot{\mathtt{p}}_{(\delta N)}^{(\bm{\phi}_{\ell=1})}=\, & -\intop_{\mathbb{S}_{r}^{2}}\bm{\epsilon}^{}_{\mathbb{S}^{2}}\,r^{2}\delta N\left(\mathring{\mathcal{E}}\mathring{\bm{\alpha}}\cdot\bm{\phi}_{\ell=1}+\mathring{{\rm P}}\bm{\mathfrak{D}}\cdot\bm{\phi}_{\ell=1}\right)\\
=\, & -\intop_{\mathbb{S}_{r}^{2}}\bm{\epsilon}^{}_{\mathbb{S}^{2}}\,r^{2}\delta N\Bigg(\left[-\frac{2}{\kappa r}+\mathcal{O}\left(r\right)\right]\frac{1}{r^{2}}\mathfrak{S}^{\mathfrak{ij}}\left[r^{2}\mathring{E}_{JK}\mathfrak{B}_{\mathfrak{i}}^{J}r^{K}+\mathcal{O}\left(r^{3}\right)\right]\left[r\Phi_{I}\mathfrak{B}_{\mathfrak{j}}^{I}\right]\\
 & +\left[-\frac{1}{\kappa r}+\mathcal{O}\left(r\right)\right]\frac{1}{r^{2}}\mathfrak{S}^{\mathfrak{ij}}\mathfrak{D}_{\mathfrak{i}}\left[r\Phi_{I}\mathfrak{B}_{\mathfrak{j}}^{I}\right]\Bigg)\\
=\, & \frac{1}{\kappa}\Phi_{I}\intop_{\mathbb{S}_{r}^{2}}\bm{\epsilon}^{}_{\mathbb{S}^{2}}\,\delta N\left(\left[2+\mathcal{O}\left(r^{2}\right)\right]\left[r^{2}\mathring{E}_{JK}r^{K}P^{JI}+\mathcal{O}\left(r^{3}\right)\right]+\left[1+\mathcal{O}\left(r^{2}\right)\right]\left[-2r^{I}\right]\right)\\
=\, & -\frac{2}{\kappa}\Phi_{I}\intop_{\mathbb{S}_{r}^{2}}\bm{\epsilon}^{}_{\mathbb{S}^{2}}\,\delta Nr^{I}+\mathcal{O}\left(r^{2}\right)\,,
\end{align}
\begin{align}
\delta\dot{\mathtt{p}}_{(\delta\mathcal{E})}^{(\bm{\phi}_{\ell=1})}=\, & -\intop_{\mathbb{S}_{r}^{2}}\bm{\epsilon}^{}_{\mathbb{S}^{2}}\,r^{2}\mathring{N}\delta\mathcal{E}\mathring{\bm{\alpha}}\cdot\bm{\phi}\\
=\, & -\intop_{\mathbb{S}_{r}^{2}}\bm{\epsilon}^{}_{\mathbb{S}^{2}}\,r^{2}\left[1+\mathcal{O}\left(r^{2}\right)\right]\left[\frac{m}{4\pi r^{2}}\right]\frac{1}{r^{2}}\mathfrak{S}^{\mathfrak{ij}}\left[r^{2}\mathring{E}_{JK}\mathfrak{B}_{\mathfrak{i}}^{J}r^{K}+\mathcal{O}\left(r^{3}\right)\right]\left[r\Phi_{I}\mathfrak{B}_{\mathfrak{j}}^{I}\right]\\
=\, & -\frac{m}{4\pi}r\Phi_{I}\mathring{E}_{JK}\intop_{\mathbb{S}_{r}^{2}}\bm{\epsilon}^{}_{\mathbb{S}^{2}}\,\left[1+\mathcal{O}\left(r^{2}\right)\right]\left[r^{K}P^{JI}\right]+\mathcal{O}\left(r^{2}\right)\\
=\, & \mathcal{O}\left(r^{2}\right)\,,
\end{align}
and
\begin{align}
\delta\dot{\mathtt{p}}_{(\delta{\rm P})}^{(\bm{\phi}_{\ell=1})}=\, & -\intop_{\mathbb{S}_{r}^{2}}\bm{\epsilon}^{}_{\mathbb{S}^{2}}\,r^{2}\mathring{N}\delta{\rm P}\bm{\mathfrak{D}}\cdot\bm{\phi}_{\ell=1}\\
=\, & -\intop_{\mathbb{S}_{r}^{2}}\bm{\epsilon}^{}_{\mathbb{S}^{2}}\,r^{2}\left[1+\frac{1}{2}r^{2}\mathring{E}_{IJ}r^{I}r^{J}+\mathcal{O}\left(r^{3}\right)\right]\left[\frac{m}{8\pi r^{2}}-\frac{1}{\kappa}\delta a_{\bm{n}}\right]\frac{1}{r^{2}}\mathfrak{S}^{\mathfrak{ij}}\mathfrak{D}_{\mathfrak{i}}\left[r\Phi_{I}\mathfrak{B}_{\mathfrak{j}}^{I}\right]\\
=\, & -r\Phi_{I}\intop_{\mathbb{S}_{r}^{2}}\bm{\epsilon}^{}_{\mathbb{S}^{2}}\,\left[1+\frac{1}{2}r^{2}\mathring{E}_{IJ}r^{I}r^{J}+\mathcal{O}\left(r^{3}\right)\right]\left[\frac{m}{8\pi r^{2}}-\frac{1}{\kappa}\delta a_{\bm{n}}\right]\left(-2r^{I}\right)\\
=\, & \mathcal{O}\left(r^{2}\right)\,.
\end{align}

\subsection*{Rigid quasilocal frame inertial with the moving object in the perturbed spacetime}

The $\ell=1$ background momentum change is given by $(\dot{\mathtt{p}}^{(\bm{\phi}_{\ell=1})})_{(0)}=(\dot{\mathtt{p}}_{(\mathring{\mathcal{E}})}^{(\bm{\phi}_{\ell=1})})_{(0)}+(\dot{\mathtt{p}}_{(\mathring{{\rm P}})}^{(\bm{\phi}_{\ell=1})})_{(0)}$.
We have:
\begin{align}
\left(\dot{\mathtt{p}}_{(\mathring{\mathcal{E}})}^{(\bm{\phi}_{\ell=1})}\right)_{(0)}=\, & -\intop_{\mathbb{S}_{r}^{2}}\bm{\epsilon}^{}_{\mathbb{S}^{2}}\,r^{2}\mathring{N}\mathring{\mathcal{E}}\mathring{\bm{\alpha}}\cdot\bm{\phi}_{\ell=1}\\
=\, & -\intop_{\mathbb{S}_{r}^{2}}\bm{\epsilon}^{}_{\mathbb{S}^{2}}\,r^{2}\left[1+\mathcal{O}\left(r\right)\right]\left[-\frac{2}{\kappa r}+\mathcal{O}\left(r\right)\right]\frac{1}{r^{2}}\mathfrak{S}^{\mathfrak{ij}}\left[rA_{K}\mathfrak{B}_{\mathfrak{i}}^{K}+\mathcal{O}\left(r^{2}\right)\right]\left[r\Phi_{I}\mathfrak{B}_{\mathfrak{j}}^{I}\right]\\
=\, & \frac{2}{\kappa}\Phi_{I}\intop_{\mathbb{S}_{r}^{2}}\bm{\epsilon}^{}_{\mathbb{S}^{2}}\,\left[1+\mathcal{O}\left(r\right)\right]\left[1+\mathcal{O}\left(r^{2}\right)\right]\mathfrak{S}^{\mathfrak{ij}}\left[rA_{J}\mathfrak{B}_{\mathfrak{i}}^{J}\mathfrak{B}_{\mathfrak{j}}^{I}+\mathcal{O}\left(r^{2}\right)\right]\\
=\, & \frac{2r}{\kappa}\Phi_{I}A_{J}\intop_{\mathbb{S}_{r}^{2}}\bm{\epsilon}^{}_{\mathbb{S}^{2}}\,\mathfrak{S}^{\mathfrak{ij}}\mathfrak{B}_{\mathfrak{i}}^{J}\mathfrak{B}_{\mathfrak{j}}^{I}+\mathcal{O}\left(r^{2}\right)\\
=\, & \frac{2r}{\kappa}\Phi_{I}A_{J}\intop_{\mathbb{S}_{r}^{2}}\bm{\epsilon}^{}_{\mathbb{S}^{2}}\,P^{IJ}+\mathcal{O}\left(r^{2}\right)\\
=\, & \frac{16\pi}{3\kappa}r\Phi^{I}A_{I}+\mathcal{O}\left(r^{2}\right)\,,
\end{align}
using the fact that $\intop_{\mathbb{S}_{r}^{2}}\bm{\epsilon}^{}_{\mathbb{S}^{2}}\,P_{IJ}=\intop_{\mathbb{S}_{r}^{2}}\bm{\epsilon}^{}_{\mathbb{S}^{2}}\,(\delta_{IJ}-r_{I}r_{J})=4\pi(1-\frac{1}{3})\delta_{IJ}=\frac{8\pi}{3}\delta_{IJ}$,
and
\begin{align}
\left(\dot{\mathtt{p}}_{(\mathring{{\rm P}})}^{(\bm{\phi}_{\ell=1})}\right)_{(0)}=\, & -\intop_{\mathbb{S}_{r}^{2}}\bm{\epsilon}^{}_{\mathbb{S}^{2}}\,r^{2}\mathring{N}\mathring{{\rm P}}\bm{\mathfrak{D}}\cdot\bm{\phi}_{\ell=1}\\
=\, & -\intop_{\mathbb{S}_{r}^{2}}\bm{\epsilon}^{}_{\mathbb{S}^{2}}\,r^{2}\left[1+rA_{J}r^{J}+\mathcal{O}\left(r^{2}\right)\right]\left[-\frac{1}{\kappa r}-\frac{1}{\kappa}A_{K}r^{K}+\mathcal{O}\left(r\right)\right]\frac{1}{r^{2}}\mathfrak{S}^{\mathfrak{ij}}\mathfrak{D}_{\mathfrak{i}}\left[r\Phi_{I}\mathfrak{B}_{\mathfrak{j}}^{I}\right]\\
=\, & \frac{1}{\kappa}\Phi_{I}\intop_{\mathbb{S}_{r}^{2}}\bm{\epsilon}^{}_{\mathbb{S}^{2}}\,\left[1+rA_{J}r^{J}+\mathcal{O}\left(r^{2}\right)\right]\left[1+rA_{K}r^{K}+\mathcal{O}\left(r^{2}\right)\right]\mathfrak{S}^{\mathfrak{ij}}\mathfrak{D}_{\mathfrak{i}}\mathfrak{B}_{\mathfrak{j}}^{I}\\
=\, & -\frac{2}{\kappa}\Phi_{I}\intop_{\mathbb{S}_{r}^{2}}\bm{\epsilon}^{}_{\mathbb{S}^{2}}\,\left[1+rA_{J}r^{J}+\mathcal{O}\left(r^{2}\right)\right]\left[1+rA_{K}r^{K}+\mathcal{O}\left(r^{2}\right)\right]r^{I}\\
=\, & -\frac{4}{\kappa}r\Phi_{I}A_{J}\intop_{\mathbb{S}_{r}^{2}}\bm{\epsilon}^{}_{\mathbb{S}^{2}}\,r^{J}r^{I}+\mathcal{O}\left(r^{2}\right)\\
=\, & -\frac{16\pi}{3\kappa}r\Phi^{I}A_{I}+\mathcal{O}\left(r^{2}\right)\,,
\end{align}
where in the fifth equality, the fact that $\intop_{\mathbb{S}_{r}^{2}}\bm{\epsilon}^{}_{\mathbb{S}^{2}}\,r_{I}=0$
leads to the vanishing of the $\mathcal{O}(r^{0})$ term. Adding the
two results, $(\dot{\mathtt{p}}^{(\bm{\phi}_{\ell=1})})_{(0)}=\mathcal{O}(r^{2})$.
For the $\ell=2$ part, we similarly have $(\dot{\mathtt{p}}^{(\bm{\phi}_{\ell=2})})_{(0)}=(\dot{\mathtt{p}}_{(\mathring{\mathcal{E}})}^{(\bm{\phi}_{\ell=2})})_{(0)}+(\dot{\mathtt{p}}_{(\mathring{{\rm P}})}^{(\bm{\phi}_{\ell=2})})_{(0)}$.
First,
\begin{align}
\left(\dot{\mathtt{p}}_{(\mathring{\mathcal{E}})}^{(\bm{\phi}_{\ell=2})}\right)_{(0)}=\, & -\intop_{\mathbb{S}_{r}^{2}}\bm{\epsilon}^{}_{\mathbb{S}^{2}}\,r^{2}\mathring{N}\mathring{\mathcal{E}}\mathring{\bm{\alpha}}\cdot\bm{\phi}_{\ell=2}\\
=\, & -\intop_{\mathbb{S}_{r}^{2}}\bm{\epsilon}^{}_{\mathbb{S}^{2}}\,r^{2}\left[1+rA_{N}r^{N}+\mathcal{O}\left(r^{2}\right)\right]\left[-\frac{2}{\kappa r}+\mathcal{O}\left(r\right)\right]\\
 & \times\frac{1}{r^{2}}\mathfrak{S}^{\mathfrak{ij}}\left[rA_{K}\mathfrak{B}_{\mathfrak{i}}^{K}+r^{2}\left(\mathring{E}_{LM}-A_{L}A_{M}\right)\mathfrak{B}_{\mathfrak{i}}^{L}r^{M}+\mathcal{O}\left(r^{3}\right)\right]\left[2r\Phi_{IJ}r^{I}\mathfrak{B}_{\mathfrak{j}}^{J}\right]\\
=\, & \frac{4}{\kappa}r\Phi_{IJ}\intop_{\mathbb{S}_{r}^{2}}\bm{\epsilon}^{}_{\mathbb{S}^{2}}\,\left[1+rA_{N}r^{N}+\mathcal{O}\left(r^{2}\right)\right]\left[1+\mathcal{O}\left(r^{2}\right)\right]\\
 & \times\mathfrak{S}^{\mathfrak{ij}}\left[A_{K}\mathfrak{B}_{\mathfrak{i}}^{K}+r\left(\mathring{E}_{LM}-A_{L}A_{M}\right)\mathfrak{B}_{\mathfrak{i}}^{L}r^{M}+\mathcal{O}\left(r^{2}\right)\right]\left[r^{I}\mathfrak{B}_{\mathfrak{j}}^{J}\right]\\
=\, & \frac{4}{\kappa}r^{2}\Phi_{IJ}\intop_{\mathbb{S}_{r}^{2}}\bm{\epsilon}^{}_{\mathbb{S}^{2}}\,r^{I}\mathfrak{S}^{\mathfrak{ij}}\mathfrak{B}_{\mathfrak{j}}^{J}\left[\left(\mathring{E}_{LM}-A_{L}A_{M}\right)\mathfrak{B}_{\mathfrak{i}}^{L}r^{M}+A_{N}r^{N}A_{K}\mathfrak{B}_{\mathfrak{i}}^{K}\right]+\mathcal{O}\left(r^{3}\right)\\
=\, & \frac{4}{\kappa}r^{2}\Phi_{IJ}\mathring{E}_{LM}\intop_{\mathbb{S}_{r}^{2}}\bm{\epsilon}^{}_{\mathbb{S}^{2}}\,r^{I}\mathfrak{S}^{\mathfrak{ij}}\mathfrak{B}_{\mathfrak{j}}^{J}\mathfrak{B}_{\mathfrak{i}}^{L}r^{M}+\mathcal{O}\left(r^{3}\right)\\
=\, & \frac{4}{\kappa}r^{2}\Phi_{IJ}\mathring{E}_{LM}\intop_{\mathbb{S}_{r}^{2}}\bm{\epsilon}^{}_{\mathbb{S}^{2}}\,P^{LJ}r^{I}r^{M}+\mathcal{O}\left(r^{3}\right)\\
=\, & \frac{16\pi}{5\kappa}r^{2}\Phi^{IJ}\mathring{E}_{IJ}+\mathcal{O}\left(r^{3}\right)
\end{align}
where in the fourth equality, the $\mathcal{O}(r)$ term vanishes
upon integration since it is an $\ell=3$ spherical harmonic, and
in the seventh, we have used the fact that
\begin{equation}
\intop_{\mathbb{S}_{r}^{2}}\bm{\epsilon}^{}_{\mathbb{S}^{2}}\,P^{LJ}r^{I}r^{M}=\intop_{\mathbb{S}_{r}^{2}}\bm{\epsilon}^{}_{\mathbb{S}^{2}}\,(\delta^{LJ}-r^{L}r^{J})r^{I}r^{M}=\frac{4\pi}{15}(4\delta^{LJ}\delta^{IM}-\delta^{LI}\delta^{JM}-\delta^{LM}\delta^{JI})\,,
\end{equation}
along with the fact that $\Phi_{IJ}$ and $\mathring{E}_{IJ}$ are
STF terms. Next,
\begin{align}
\left(\dot{\mathtt{p}}_{(\mathring{{\rm P}})}^{(\bm{\phi}_{\ell=2})}\right)_{(0)}=\, & -\intop_{\mathbb{S}_{r}^{2}}\bm{\epsilon}^{}_{\mathbb{S}^{2}}\,r^{2}\mathring{N}\mathring{{\rm P}}\bm{\mathfrak{D}}\cdot\bm{\phi}_{\ell=2}\\
=\, & -\intop_{\mathbb{S}_{r}^{2}}\bm{\epsilon}^{}_{\mathbb{S}^{2}}\,r^{2}\left[1+rA_{L}r^{L}+\mathcal{O}\left(r^{2}\right)\right]\left[-\frac{1}{\kappa r}-\frac{1}{\kappa}A_{K}r^{K}+\mathcal{O}\left(r\right)\right]\frac{1}{r^{2}}\mathfrak{S}^{\mathfrak{ij}}\mathfrak{D}_{\mathfrak{i}}\left[2r\Phi_{IJ}r^{I}\mathfrak{B}_{\mathfrak{j}}^{J}\right]\\
=\, & \frac{2}{\kappa}\Phi_{IJ}\intop_{\mathbb{S}_{r}^{2}}\bm{\epsilon}^{}_{\mathbb{S}^{2}}\,\left[1+rA_{L}r^{L}+\mathcal{O}\left(r^{2}\right)\right]\left[1+rA_{K}r^{K}+\mathcal{O}\left(r^{2}\right)\right]\mathfrak{S}^{\mathfrak{ij}}\left[\mathfrak{B}_{\mathfrak{i}}^{I}\mathfrak{B}_{\mathfrak{j}}^{J}+r^{I}\mathfrak{D}_{\mathfrak{i}}\mathfrak{B}_{\mathfrak{j}}^{J}\right]\\
=\, & \frac{2}{\kappa}\Phi_{IJ}\intop_{\mathbb{S}_{r}^{2}}\bm{\epsilon}^{}_{\mathbb{S}^{2}}\,\left[1+rA_{L}r^{L}+\mathcal{O}\left(r^{2}\right)\right]\left[1+rA_{K}r^{K}+\mathcal{O}\left(r^{2}\right)\right]\left[P^{IJ}+r^{I}\left(-2r^{J}\right)\right]\\
=\, & \frac{2}{\kappa}\Phi_{IJ}\intop_{\mathbb{S}_{r}^{2}}\bm{\epsilon}^{}_{\mathbb{S}^{2}}\,\left[1+rA_{L}r^{L}+\mathcal{O}\left(r^{2}\right)\right]\left[1+rA_{K}r^{K}+\mathcal{O}\left(r^{2}\right)\right]\left[P^{IJ}-2r^{I}r^{J}\right]\\
=\, & \mathcal{O}\left(r^{2}\right)\,.
\end{align}
Here we would need the next (linear in $r$) term in the expansion
of $\mathring{{\rm P}}$ (the one for $\mathring{N}$ is easy to obtain)
in order to explicitly calculate the $\mathcal{O}\left(r^{2}\right)$
term above. In any case, we have found, just as in the PP-inertial
case, $(\dot{\mathtt{p}}^{(\bm{\phi}_{\ell=2})})_{(0)}=\mathcal{O}(r^{2})$.

For the $\mathcal{O}(\lambda)$ part, as in the point-particle-inertial case,
we will have
\begin{equation}
\delta\dot{\mathtt{p}}^{(\bm{\phi}_{\ell=1})}=\sum_{Q\in\{\delta N,\delta\mathcal{E},\delta\bm{\alpha},\delta{\rm P},\delta\bm{D}\}}\delta\dot{\mathtt{p}}_{(Q)}^{(\bm{\phi}_{\ell=1})}\,.
\end{equation}
We compute these terms one by one. First,
\begin{align}
\delta\dot{\mathtt{p}}_{(\delta N)}^{(\bm{\phi}_{\ell=1})}=\, & -\intop_{\mathbb{S}_{r}^{2}}\bm{\epsilon}^{}_{\mathbb{S}^{2}}\,r^{2}\delta N\left(\mathring{\mathcal{E}}\mathring{\bm{\alpha}}\cdot\bm{\phi}_{\ell=1}+\mathring{{\rm P}}\bm{\mathfrak{D}}\cdot\bm{\phi}_{\ell=1}\right)\\
=\, & -\intop_{\mathbb{S}_{r}^{2}}\bm{\epsilon}^{}_{\mathbb{S}^{2}}\,r^{2}\delta N\Bigg(\left[-\frac{2}{\kappa r}+\mathcal{O}\left(r\right)\right]\frac{1}{r^{2}}\mathfrak{S}^{\mathfrak{ij}}\left[rA_{K}\mathfrak{B}_{\mathfrak{i}}^{K}+\mathcal{O}\left(r^{2}\right)\right]\left[r\Phi_{I}\mathfrak{B}_{\mathfrak{j}}^{I}\right]\\
 & +\left[-\frac{1}{\kappa r}+\mathcal{O}\left(1\right)\right]\frac{1}{r^{2}}\mathfrak{S}^{\mathfrak{ij}}\mathfrak{D}_{\mathfrak{i}}\left[r\Phi_{I}\mathfrak{B}_{\mathfrak{j}}^{I}\right]\Bigg)\\
=\, & \frac{1}{\kappa}\Phi_{I}\intop_{\mathbb{S}_{r}^{2}}\bm{\epsilon}^{}_{\mathbb{S}^{2}}\,\delta N\left(\left[2+\mathcal{O}\left(r^{2}\right)\right]\mathfrak{S}^{\mathfrak{ij}}\left[rA_{K}\mathfrak{B}_{\mathfrak{i}}^{K}\mathfrak{B}_{\mathfrak{j}}^{I}+\mathcal{O}\left(r^{2}\right)\right]+\left[1+\mathcal{O}\left(r\right)\right]\mathfrak{S}^{\mathfrak{ij}}\mathfrak{D}_{\mathfrak{i}}\mathfrak{B}_{\mathfrak{j}}^{I}\right)\\
=\, & -\frac{2}{\kappa}\Phi_{I}\intop_{\mathbb{S}_{r}^{2}}\bm{\epsilon}^{}_{\mathbb{S}^{2}}\,\delta Nr^{I}+\mathcal{O}\left(r\right)\,.
\end{align}
Next, we have
\begin{align}
\delta\dot{\mathtt{p}}_{(\delta\mathcal{E})}^{(\bm{\phi}_{\ell=1})}=\, & -\intop_{\mathbb{S}_{r}^{2}}\bm{\epsilon}^{}_{\mathbb{S}^{2}}\,r^{2}\mathring{N}\delta\mathcal{E}\mathring{\bm{\alpha}}\cdot\bm{\phi}\\
=\, & -\intop_{\mathbb{S}_{r}^{2}}\bm{\epsilon}^{}_{\mathbb{S}^{2}}\,r^{2}\left[1+\mathcal{O}\left(r\right)\right]\left[\frac{m}{4\pi r^{2}}\right]\frac{1}{r^{2}}\mathfrak{S}^{\mathfrak{ij}}\left[rA_{K}\mathfrak{B}_{\mathfrak{i}}^{K}+\mathcal{O}\left(r^{2}\right)\right]\left[r\Phi_{I}\mathfrak{B}_{\mathfrak{j}}^{I}\right]\\
=\, & -\frac{m}{4\pi}\Phi_{I}A_{K}\intop_{\mathbb{S}_{r}^{2}}\bm{\epsilon}^{}_{\mathbb{S}^{2}}\,P^{KI}+\mathcal{O}\left(r\right)\\
=\, & -\frac{m}{4\pi}\Phi_{I}A_{K}\intop_{\mathbb{S}_{r}^{2}}\bm{\epsilon}^{}_{\mathbb{S}^{2}}\,P^{KI}+\mathcal{O}\left(r\right)\\
=\, & -\frac{2}{3}m\Phi_{I}A^{I}+\mathcal{O}\left(r\right)\,,
\end{align}
and
\begin{align}
\delta\dot{\mathtt{p}}_{(\delta{\rm P})}^{(\bm{\phi}_{\ell=1})}=\, & -\intop_{\mathbb{S}_{r}^{2}}\bm{\epsilon}^{}_{\mathbb{S}^{2}}\,r^{2}\mathring{N}\delta{\rm P}\bm{\mathfrak{D}}\cdot\bm{\phi}_{\ell=1}\\
=\, & -\intop_{\mathbb{S}_{r}^{2}}\bm{\epsilon}^{}_{\mathbb{S}^{2}}\,r^{2}\left[1+rA_{L}r^{L}+\mathcal{O}\left(r^{2}\right)\right]\left[\frac{m}{8\pi r^{2}}-\frac{1}{\kappa}\delta a_{\bm{n}}\right]\frac{1}{r^{2}}\mathfrak{S}^{\mathfrak{ij}}\mathfrak{D}_{\mathfrak{i}}\left[r\Phi_{I}\mathfrak{B}_{\mathfrak{j}}^{I}\right]\\
=\, & -\Phi_{I}\intop_{\mathbb{S}_{r}^{2}}\bm{\epsilon}^{}_{\mathbb{S}^{2}}\,\left[1+rA_{L}r^{L}\right]\left[\frac{m}{8\pi r^{2}}\right]r\left(-2r^{I}\right)+\mathcal{O}\left(r\right)\\
=\, & \frac{m}{4\pi}\Phi_{I}A_{L}\intop_{\mathbb{S}_{r}^{2}}\bm{\epsilon}^{}_{\mathbb{S}^{2}}\,r^{L}r^{I}+\mathcal{O}\left(r\right)\\
=\, & \frac{1}{3}m\Phi_{I}A^{I}+\mathcal{O}\left(r\right)\,.
\end{align}

\end{widetext}







%

\end{document}